\documentclass[11pt,a4paper]{article}
\pdfoutput=1
\usepackage{amsfonts}
\usepackage{amssymb}
\usepackage[centertags]{amsmath}
\usepackage{graphicx}
\usepackage{hyperref}
\usepackage{booktabs}
\usepackage{theorem}
\usepackage{array}
\usepackage{epsfig}
\usepackage{colordvi}
\usepackage{color}
\usepackage{ulem}
\usepackage[small,bf]{caption}
\usepackage{cite}
\usepackage{color}
\usepackage{subfigure}
\usepackage{pdflscape}

\def\1{\mathbf{1}}
\def\3{\mathbf{3}}
\def\2{\mathbf{2}}

\def\th{\theta}

\newcommand{\mefff}{\mbox{$ \langle\!\, m \,\!\rangle $}}
\newcommand{\meff}{\mbox{$\left|  \langle \!\,  m  \,\!  \rangle \right| $}}
\newcommand{\betabeta}{\mbox{$(\beta \beta)_{0 \nu}  $}}
\def\ltap{\ \raisebox{-.4ex}{\rlap{$\sim$}} \raisebox{.4ex}{$<$}\ }
\def\gtap{\ \raisebox{-.4ex}{\rlap{$\sim$}} \raisebox{.4ex}{$>$}\ }

\allowdisplaybreaks[1]

\usepackage{a4wide}

\newcommand{\bec}{\begin{cases}}
\newcommand{\eec}{\end{cases}}
\newcommand{\beq}{\begin{equation*}}
\newcommand{\eeq}{\end{equation*}}
\newcommand{\be}{\begin{equation}}
\newcommand{\ee}{\end{equation}}
\newcommand{\ba}{\begin{eqnarray}}
\newcommand{\ea}{\end{eqnarray}}

\DeclareMathOperator{\diag}{diag}

\makeatletter

\newcommand{\Rmnum}[1]{\expandafter\@slowromancap\romannumeral #1@}
\makeatother

\begin{document}

\begin{titlepage}

\vspace*{-15mm}
\begin{flushright}
SISSA 26/2016/FISI\\
IPMU16-0064
\end{flushright}
\vspace*{0.7cm}

\begin{center}
{\bf\Large {Predictions for the Majorana CP Violation Phases in the}} \\
[4mm]
{\bf\Large {Neutrino Mixing Matrix and Neutrinoless Double Beta Decay}}\\
[8mm]
\vspace{0.4cm} I. Girardi$\mbox{}^{a)}$, S. T. Petcov$\mbox{}^{a,b)}$~%
\footnote{Also at: Institute of Nuclear Research and Nuclear Energy,
Bulgarian Academy of Sciences, 1784 Sofia, Bulgaria.}
and A. V. Titov$\mbox{}^{a)}$ 
\\[1mm]
\end{center}
\vspace*{0.50cm}
\centerline{$^{a}$ \it SISSA/INFN, Via Bonomea 265, 34136 Trieste, Italy}
\vspace*{0.2cm}
\centerline{$^{b}$ \it Kavli IPMU (WPI), University of Tokyo,
5-1-5 Kashiwanoha, 277-8583 Kashiwa, Japan}
\vspace*{1.20cm}

\begin{abstract}
\noindent
We obtain predictions for the Majorana phases 
$\alpha_{21}/2$ and $\alpha_{31}/2$ 
of the $3\times 3$ unitary neutrino mixing
matrix $U = U_e^{\dagger} \, U_{\nu}$, $U_e$ and $U_{\nu}$ being 
the $3\times 3$ unitary matrices 
resulting from the diagonalisation 
of the charged lepton and neutrino Majorana 
mass matrices, respectively.
We focus on forms of $U_e$ and $U_{\nu}$ 
permitting to express 
$\alpha_{21}/2$ and $\alpha_{31}/2$   
in terms of the Dirac phase $\delta$ 
and the three
neutrino mixing angles of the standard 
parametrisation of $U$, 
and the angles and the two Majorana-like 
phases $\xi_{21}/2$ and $\xi_{31}/2$ 
present, in general,  in $U_{\nu}$.
The concrete forms of $U_{\nu}$ considered 
are fixed by, or associated with, symmetries 
(tri-bimaximal, bimaximal, etc.), 
so that the angles in $U_{\nu}$ are 
fixed. For each of these forms and forms of $U_e$ that 
allow to reproduce the measured 
values of the three neutrino
mixing angles $\theta_{12}$,  $\theta_{23}$ 
and $\theta_{13}$, we derive predictions for 
phase differences $(\alpha_{21}/2 - \xi_{21}/2)$,
$(\alpha_{31}/2 - \xi_{31}/2)$, etc., which 
are completely determined by the values of the 
mixing angles. We show that the requirement of generalised 
CP invariance of the neutrino Majorana mass term 
implies $\xi_{21} = 0$ or $\pi$ and  $\xi_{31} = 0$ or $\pi$.
For these values of $\xi_{21}$ and  $\xi_{31}$ 
and the best fit values of $\theta_{12}$,  $\theta_{23}$ 
and $\theta_{13}$, we present predictions 
for the effective Majorana mass 
in neutrinoless double beta decay for both 
neutrino mass spectra with normal and inverted 
ordering.
\end{abstract}

\vspace{0.5cm}

\end{titlepage}
\setcounter{footnote}{0}

%
\section{Introduction}
%
%

Determining the status of the CP symmetry in the lepton sector, 
discerning the type of spectrum the neutrino masses obey,  
identifying the nature~---~Dirac or Majorana~---~of 
massive neutrinos and 
determining the absolute neutrino mass scale
are among the highest priority goals of the programme of 
future research in neutrino physics (see, e.g., \cite{PDG2014}). 
The results obtained within this ambitious research programme can 
shed light, in particular, on the origin of the observed pattern 
of neutrino mixing. Comprehending the origin of the patterns 
of neutrino masses and mixing is one of the most 
challenging problems in neutrino physics. It is an integral 
part of the more general fundamental problem 
in particle physics of deciphering the origins of flavour, i.e., 
of the patterns of quark, charged lepton and neutrino 
masses and of the quark and neutrino mixing. 

   In refs. 
\cite{Petcov:2014laa,Girardi:2014faa,Girardi:2015vha,Girardi:2015rwa} (see also \cite{Marzocca:2013cr}), working in the framework of the 
reference 3-neutrino mixing scheme (see, e.g., \cite{PDG2014}), 
we have derived predictions for the
Dirac CP violation (CPV) phase in the 
Pontecorvo, Maki, Nakagawa and Sakata (PMNS) 
neutrino mixing matrix   
within the discrete flavour symmetry approach to 
neutrino mixing. This approach provides a natural 
explanation of the observed pattern of neutrino mixing 
and is widely explored at present (see, e.g., 
\cite{King:2013eh,King:2014nza} and  references therein).
In the present article, using the method 
developed and utilised in \cite{Petcov:2014laa}, 
we derive predictions for the Majorana  CPV phases in the PMNS 
matrix  \cite{Bilenky:1980cx} within the same 
approach based on discrete flavour symmetries.
Our study is a natural continuation of 
the studies performed in 
\cite{Petcov:2014laa,Girardi:2014faa,Girardi:2015vha,Girardi:2015rwa,Marzocca:2013cr}. 

 As is well known, the PMNS matrix  will contain 
physical CPV Majorana phases if the massive 
neutrinos are Majorana particles \cite{Bilenky:1980cx}.
The massive neutrinos are predicted to be Majorana fermions 
by a large number of theories of neutrino mass generation 
(see, e.g., \cite{King:2013eh,BiPet87,Mohapatra:2005xxx}), 
most notably, by the theories based on 
the seesaw mechanism \cite{seesaw}.  
The flavour neutrino oscillation probabilities do 
not depend on the Majorana phases \cite{Bilenky:1980cx,Lang87}.
The Majorana phases play particularly important role 
in processes involving real or virtual 
neutrinos, which are characteristic 
of Majorana nature of massive neutrinos
and in which the total lepton charge $L$
changes by two units, $|\Delta L| = 2$ 
(see, e.g., \cite{Petcov:2013poa}).
One widely discussed and experimentally relevant 
example is neutrinoless double beta
($\betabeta$-) decay of even-even nuclei 
(see, e.g., \cite{BiPet87,BPP1,bb0nuth})  
$^{48}Ca$, $^{76}Ge$, $^{82}Se$,
$^{100}Mo$, $^{130}Te$, $^{136}Xe$, etc.: 
$(A,Z) \rightarrow (A,Z+2) + e^- + e^-$.
The predictions for the 
rates of the lepton flavour violating processes, 
$\mu \rightarrow e + \gamma$ and 
$\mu \rightarrow 3e$ decays, $\mu - e$ conversion in nuclei, etc., 
in theories of neutrino mass generation with massive Majorana 
neutrinos (e.g., TeV scale type I seesaw model, the Higgs 
triplet model, etc.) depend on the Majorana phases 
(see, e.g., \cite{PPY03,Dinh:2012bp}). 
And the Majorana phases in the PMNS matrix can provide 
the CP violation necessary for the generation of the 
observed baryon asymmetry of the Universe \cite{Pascoli:2006ci}~%
\footnote{This possibility can be realised within  
the leptogenesis scenario of the 
baryon asymmetry generation \cite{LGFY,kuzmin}, 
which is based on the type I seesaw mechanism  
of neutrino mass generation \cite{seesaw}.}.

 In the reference case of 3-neutrino mixing, which we are 
going to consider in the present article, 
there can be two physical Majorana CPV phases in 
the PMNS neutrino mixing matrix in addition to 
the Dirac CPV phase  \cite{Bilenky:1980cx}. 
The PMNS matrix in this case is given by
\begin{equation}
U = VQ\,, \quad
Q = \diag\left(1, e^{i \frac{\alpha_{21}}{2}}, e^{i \frac{\alpha_{31}}{2}}\right)\,,
\label{eq:VQ}
\end{equation}
%
where $\alpha_{21,31}$ are the two Majorana CPV phases 
and $V$ is a CKM-like matrix containing the Dirac CPV phase. 
The matrix $V$ has 
the following form in the standard 
parametrisation of the PMNS matrix \cite{PDG2014}, 
which we are going to employ in what follows: 
\begin{equation}
\begin{array}{c}
\label{eq:Vpara}
V = \left(\begin{array}{ccc}
 c_{12} c_{13} & s_{12} c_{13} & s_{13} e^{-i \delta}  \\[0.2cm]
 -s_{12} c_{23} - c_{12} s_{23} s_{13} e^{i \delta}
 & c_{12} c_{23} - s_{12} s_{23} s_{13} e^{i \delta}
 & s_{23} c_{13} 
\\[0.2cm]
 s_{12} s_{23} - c_{12} c_{23} s_{13} e^{i \delta} &
 - c_{12} s_{23} - s_{12} c_{23} s_{13} e^{i \delta}
 & c_{23} c_{13} 
\\
  \end{array}
\right)\,.
\end{array}
\end{equation}
%
Here
$0 \leq \delta \leq 2\pi$ is the Dirac CPV phase and
we have used the standard notation
$c_{ij} = \cos\theta_{ij}$,
$s_{ij} = \sin\theta_{ij}$ with
$0 \leq  \theta_{ij} \leq \pi/2$.  
In the case of CP invariance 
we have $\delta =0$, $\pi$, $2\pi$, 
$0$ and $2\pi$ being physically
indistinguishable, and \cite{LW81}
$\alpha_{21} = k\pi$, $\alpha_{31} = k'\pi$,
$k,k'=0,1,2$~%
\footnote{If the neutrino masses are generated 
via the  type I seesaw mechanism, 
the interval in which  $\alpha_{21}$ 
and $\alpha_{31}$ vary is 
$[0,4\pi)$ \cite{Molinaro:2008rg}. Thus, 
in this case  $\alpha_{21}$ and $\alpha_{31}$ 
have CP-conserving values for $k,k'=0,1,2,3,4$.}.

  The neutrino mixing parameters
$\sin^2\theta_{12}$, $\sin^2\theta_{23}$ and $\sin^2\theta_{13}$ 
play important role in our further considerations. They 
were determined with relatively small uncertainties 
in the most recent analysis of the global neutrino 
oscillation data performed in \cite{Capozzi:2016rtj}
(for earlier analyses see, e.g., 
\cite{Capozzi:2013csa,Gonzalez-Garcia:2014bfa}).
The  authors of ref. \cite{Capozzi:2016rtj}, 
using, in particular, the first NO$\nu$A (LID) data 
on $\nu_{\mu} \rightarrow \nu_e$ oscillations  
from \cite{Adamson:2016tbq},  
find the following best fit values and  
3$\sigma$ allowed ranges of 
$\sin^2\theta_{12}$, $\sin^2\theta_{23}$ and $\sin^2\theta_{13}$:
\begin{eqnarray}
\label{th12values}
(\sin^2 \theta_{12})_{\rm BF} = 0.297\,,~~~~
 0.250 \leq \sin^2 \theta_{12} \leq 0.354\,,\\ [0.30cm]
\label{th23values}
(\sin^2\theta_{23})_{\rm BF} = 0.437~(0.569)\,,~~~~
 0.379~(0.383) \leq \sin^2\theta_{23} \leq 0.616~(0.637)\,,\\[0.30cm]
\label{th13values}
(\sin^2\theta_{13})_{\rm BF} = 0.0214~(0.0218)\,,~~~~
0.0185~(0.0186) \leq \sin^2\theta_{13} \leq 0.0246~(0.0248)\,.
\end{eqnarray}
%
The values (values in brackets)
correspond to neutrino mass spectrum with normal ordering
(inverted ordering) (see, e.g., \cite{PDG2014}),
denoted further as the NO (IO) spectrum.
Note, in particular, that  
$\sin^2\theta_{23}$ can differ significantly from 0.5 and that 
$\sin^2\theta_{23} = 0.5$ lies in the $2\sigma$ interval of 
allowed values. Using the same set of data the authors of 
 \cite{Capozzi:2016rtj} find also the following best fit value 
and $2\sigma$ allowed range of the Dirac phase $\delta$:
\begin{equation}
\delta = 1.35\,\pi~(1.32\,\pi)\,,~~~~
0.92\,\pi~(0.83\,\pi)\leq \delta \leq 1.99\,\pi\,.
\label{deltaexp}
\end{equation}
%
 
 The discrete flavour symmetry approach 
to neutrino mixing is based on the observation that the 
PMNS neutrino mixing angles $\theta_{12}$,  $\theta_{23}$ and 
$\theta_{13}$ have values which differ from those of 
specific symmetry forms of the mixing matrix by 
subleading perturbative corrections (see further).
The fact that the PMNS matrix 
in the case of 3-neutrino mixing 
is a product of two $3\times 3$ unitary matrices $U_{e}$ and $U_{\nu}$, 
originating from the diagonalisation of the charged 
lepton and neutrino mass matrices,
\begin{equation}
U = U_e^{\dagger}\, U_{\nu}\,,
\label{PMNS1}
\end{equation}
%
is also widely exploited.
In terms of the parameters of $U_{e}$ and $U_{\nu}$, 
in the absence of constraints
the PMNS matrix can be parametrised  
as \cite{Frampton:2004ud} 
\begin{equation}
U = U_e^{\dagger}\, U_{\nu} = 
(\tilde{U}_{e})^\dagger\, \Psi\, \tilde{U}_{\nu} \, Q_0\,.
\label{PMNS2}
\end{equation}
%
Here $\tilde{U}_e$ and $\tilde{U}_\nu$ 
are CKM-like $3\times 3$ unitary matrices, 
and  $\Psi$ and $Q_0$ are given by
\begin{equation} 
\Psi =
{\rm diag} \left(1,\text{e}^{-i \psi}, \text{e}^{-i \omega} \right)\,, \quad
Q_0 = {\rm diag} \left(1,\text{e}^{i \frac{\xi_{21}}{2}}, 
\text{e}^{i \frac{\xi_{31}}{2}} \right)\,,
\label{PsieQ0}
\end{equation}
%
where $\psi$, $\omega$, $\xi_{21}$ and $\xi_{31}$  are phases 
which contribute to physical CPV phases.
The phases in $Q_0$ result from the diagonalisation of 
the neutrino Majorana mass term and 
contribute to the Majorana phases 
in the PMNS matrix.  

 In the approach of interest one assumes the existence 
at certain energy scale of a (lepton) flavour  
symmetry corresponding to a non-Abelian 
discrete group $G_f$.  
The symmetry group $G_f$ can be broken, in general, 
to different symmetry subgroups, or ``residual symmetries'',
$G_e$ and $G_{\nu}$ of the charged lepton and 
neutrino mass terms, respectively. 
Given a discrete symmetry $G_f$, there are more than one 
(but still a finite number of) possible 
residual symmetries $G_e$ and $G_{\nu}$.
The subgroup $G_e$, in particular, can be trivial.
Non-trivial residual symmetries $G_e$ and $G_{\nu}$ 
(of a given $G_f$) constrain the forms of the 
matrices $U_{e}$ and $U_{\nu}$, and thus 
the form~of~$U$. 

 Among the widely considered symmetry forms of $U$ are: 
i) the tri-bimaximal (TBM) form 
\cite{TBM,Xing:2002sw}, 
ii) the bimaximal (BM) form~%
\footnote{Bimaximal mixing can also be 
a consequence of 
the conservation of the lepton charge
$L' = L_e - L_{\mu} - L_{\tau}$ (LC) \cite{SPPD82}, 
supplemented by a $\mu - \tau$ symmetry.} 
\cite{BM},
iii) the golden ratio type A (GRA) form 
\cite{GRAM,GRAM2}, 
iv) the golden ratio type B (GRB) form \cite{GRBM},
and v) the hexagonal (HG) form
\cite{Albright:2010ap,HGM}.
It is typically assumed that the matrix $\tilde U_{\nu}$ 
in eq.~(\ref{PMNS2}), and not  $\tilde U_{e}$,    
has a symmetry form and, in particular,
has one of the forms discussed above.
For all these forms we have 
\begin{equation}
\tilde{U}_\nu = R_{23}(\theta^{\nu}_{23}) \, R_{12}(\theta^{\nu}_{12})\,,
\label{tU2312}
\end{equation}
%
with $\theta^\nu_{23} = -\,\pi/4$, $R_{23}$ and $R_{12}$ 
being $3\times 3$ orthogonal matrices 
describing rotations in the 2-3 and 1-2 planes:
\begin{equation}
R_{12}\left( \theta^\nu_{12} \right) = \begin{pmatrix}
\cos \theta^\nu_{12} & \sin \theta^\nu_{12} & 0\\
- \sin \theta^\nu_{12} & \cos \theta^\nu_{12} & 0\\
0 & 0 & 1 \end{pmatrix} \,,
\quad
R_{23}\left( \theta^\nu_{23} \right) = \begin{pmatrix}
1 & 0 & 0\\
0 & \cos \theta^\nu_{23} & \sin \theta^\nu_{23} \\
0 & - \sin \theta^\nu_{23}  & \cos \theta^\nu_{23} \\
\end{pmatrix} \,.
\label{R1223}
\end{equation}
%
The value of the angle  $\theta^{\nu}_{12}$, and thus of  
$\sin^2\theta^{\nu}_{12}$, depends on the form of $\tilde{U}_{\nu}$.
For the TBM, BM, GRA, GRB and HG forms we have: 
i) $\sin^2\theta^{\nu}_{12} = 1/3$ (TBM),
ii)  $\sin^2\theta^{\nu}_{12} = 1/2$ (BM),
iii)  $\sin^2\theta^{\nu}_{12} =  (2 + r)^{-1} \cong 0.276$ (GRA),
$r$ being the golden ratio, $r = (1 +\sqrt{5})/2$,
iv) $\sin^2\theta^{\nu}_{12} = (3 - r)/4 \cong 0.345$ (GRB), and
v) $\sin^2\theta^{\nu}_{12} = 1/4$ (HG).

  The TBM form of $\tilde{U}_{\nu}$, for example,  
can be obtained from a $G_f = A_4$ symmetry,
when the residual symmetry is $G_\nu = Z_2$.
In this case there is an additional accidental 
$\mu-\tau$ symmetry, which together with the $Z_2$ symmetry 
leads to the TBM form of $\tilde{U}_{\nu}$ 
(see, e.g., \cite{Altarelli:2010gt}).
The TBM  form can also be derived from 
$G_f = T'$ with $G_\nu = Z_2$, provided  
the left-handed (LH) charged lepton and neutrino fields each 
transform as triplets  of $T^{\prime}$~%
\footnote{When working with 3-dimensional and 1-dimensional
representations of $T^{\prime}$, there is no way to distinguish 
$T^{\prime}$ from $A_4$ \cite{Feruglio:2007uu}.}.
One can obtain the BM form from, e.g., 
the $G_f = S_4$ symmetry, when 
$G_{\nu} = Z_2$. 
There is an accidental $\mu-\tau$ symmetry in this case as well
\cite{Altarelli:2009gn}.
The $A_5$ symmetry group can be utilised to generate GRA mixing, 
while the groups $D_{10}$ and $D_{12}$ can lead to  
the GRB and HG mixing forms, respectively. 

 The symmetry forms of $\tilde{U}_\nu$ considered above do not 
include rotation in the 1-3 plane, i.e., $\theta^{\nu}_{13} = 0$.
However, forms of  $\tilde{U}_\nu$ of the type 
\begin{equation}
\tilde{U}_\nu = R_{23}(\theta^{\nu}_{23}) \, R_{13}(\theta^{\nu}_{13})\, 
R_{12}(\theta^{\nu}_{12})\,,
\label{tU231312}
\end{equation}
%
with non-zero values of $\theta^{\nu}_{13}$ are
inspired by certain types of flavour symmetries
(see, e.g., 
\cite{Bazzocchi:2011ax,Rodejohann:2014xoa,Toorop:2011jn,King:2012in}).
In \cite{Bazzocchi:2011ax}, for example, 
the so-called tri-permuting pattern, corresponding to  
$\theta^{\nu}_{12} = \theta^{\nu}_{23} = - \pi/4$ 
and $\theta^{\nu}_{13} = \sin^{-1} (1 / 3)$, 
was proposed and investigated.
In the study we will perform we will consider 
also the form in eq.~(\ref{tU231312})
for three 
representative values of $\theta^{\nu}_{13}$ 
discussed in the literature:
$\theta^{\nu}_{13} = \pi/20$, $\pi/10$ and $\sin^{-1} (1 / 3)$. 

   The symmetry values of the angles in the matrix $\tilde{U}_{\nu}$  
typically, and in all cases considered above, differ 
by relatively small perturbative corrections 
from the experimentally determined values of at least 
some of the angles $\theta_{12}$, $\theta_{23}$ and $\theta_{13}$.
The requisite corrections are provided by the 
matrix $U_e$, or equivalently, by $\tilde{U}_e$.
In the approach followed in 
\cite{Petcov:2014laa,Girardi:2014faa,Marzocca:2013cr,Girardi:2015vha}
we are going to adopt, the matrix $\tilde{U}_e$ is unconstrained 
and was chosen on phenomenological grounds.  
This corresponds to the case of trivial 
subgroup $G_e$, i.e., of the charged lepton mass term 
breaking the symmetry $G_f$ completely.
The matrix $\tilde{U}_e$ in the general case 
depends on three angles and one phase \cite{Frampton:2004ud}. 
However, in a class of theories of (lepton) flavour 
and neutrino mass generation, 
based on a GUT and/or a discrete symmetry (see, e.g., 
\cite{Gehrlein:2014wda,Meroni:2012ty,Marzocca:2011dh,
Antusch:2012fb,Girardi:2013sza,Chen:2009gf}), 
$\tilde{U}_e$ is an orthogonal matrix which describes  
one rotation in the 1-2 plane, 
\begin{align} 
\label{Re12}
\tilde{U}_e &=  R^{-1}_{12}(\theta^{e}_{12})\,,
\end{align}
%
or two rotations in 
the planes 1-2 and 2-3,
\begin{align} 
\label{Re12Re23}
\tilde{U}_e &= R^{-1}_{23}(\theta^{e}_{23})\,
R^{-1}_{12}(\theta^{e}_{12})\,,
\end{align}
%
$\theta^{e}_{12}$ and $\theta^{e}_{23}$ being the 
 corresponding rotation angles.
Other possibilities include $\tilde{U}_e$ being 
an orthogonal matrix which 
describes  i) one rotation in the 1-3 plane~%
\footnote{The case of $\tilde{U}_e$ representing a rotation 
in the 2-3 plane is ruled out for the five symmetry 
forms of $\tilde{U}_{\nu}$ listed above, since 
in this case a realistic value of 
$\theta_{13} \neq 0$ cannot be generated.},
\begin{align} 
\label{Re13}
\tilde{U}_e & =  R^{-1}_{13}(\theta^{e}_{13})\,,
\end{align}
%
or ii) two rotations in 
any other two of the three planes, e.g., 
\begin{align} 
\label{Re13Re23}
\tilde{U}_e &=  R^{-1}_{23}(\theta^{e}_{23})\, 
R^{-1}_{13}(\theta^{e}_{13})\,,~~{\rm or} \\ 
\label{Re13Re12}
\tilde{U}_e &= R^{-1}_{13}(\theta^{e}_{13})\,  
R^{-1}_{12}(\theta^{e}_{12})\,. 
\end{align}
%
We use the inverse matrices in 
eqs.~(\ref{Re12})~--~(\ref{Re13Re12})
for convenience of the notations in 
expressions that will appear further in our analysis. 

In refs. \cite{Petcov:2014laa,Girardi:2015vha} 
sum rules for the cosine of the Dirac phase 
$\delta$ of the PMNS matrix, by which 
$\cos\delta$ is expressed in terms of the three 
measured neutrino
angles $\theta_{12}$, $\theta_{23}$ and $\theta_{13}$,    
were derived in the cases of the following forms of 
$\tilde{U}_e$ and  $\tilde{U}_\nu$: 
\begin{itemize}
\item[A.] $\tilde{U}_\nu = R_{23}(\theta^{\nu}_{23})R_{12}(\theta^{\nu}_{12})$~~%
and~~%
i)  $\tilde{U}_e =  R^{-1}_{12}(\theta^{e}_{12})$,~~%
ii)  $\tilde{U}_e =  R^{-1}_{13}(\theta^{e}_{13})$,~~%
iii) $\tilde{U}_e =  R^{-1}_{23}(\theta^{e}_{23})R^{-1}_{12}(\theta^{e}_{12})$,~~%
iv) $\tilde{U}_e =  R^{-1}_{23}(\theta^{e}_{23})R^{-1}_{13}(\theta^{e}_{13})$,~~%
v) $\tilde{U}_e = R^{-1}_{13}(\theta^{e}_{13}) R^{-1}_{12}(\theta^{e}_{12})$; 
\item[B.] $\tilde{U}_\nu = R_{23}(\theta^{\nu}_{23})R_{13}(\theta^{\nu}_{13}) 
 R_{12}(\theta^{\nu}_{12})$~~%
and~~%
vi) $\tilde{U}_e = R^{-1}_{12}(\theta^{e}_{12})$,~~%
vii) $\tilde{U}_e = R^{-1}_{13}(\theta^{e}_{13})$.
\end{itemize}
The sum rules thus found allowed us 
in the cases of the TBM, BM (LC), GRA, GRB and HG 
mixing forms of $\tilde U_\nu$ in item A and 
for certain fixed values of $\theta^\nu_{ij}$ in item B 
to obtain predictions for $\cos\delta$ (see refs. 
\cite{Petcov:2014laa,Girardi:2014faa,Girardi:2015vha,Marzocca:2013cr})
 as well as for the rephasing invariant  
\begin{equation}
J_{\rm CP} = {\rm Im} \left\{ U^*_{e1} U^*_{\mu 3} U_{e3} U_{\mu 1} \right\}
= \frac{1}{8} \sin \delta \sin 2\theta_{13} \sin 2\theta_{23}
\sin 2\theta_{12} \cos \theta_{13} \,,
\label{JCP}
\end{equation}
%
on which the magnitude of CP-violating 
effects in neutrino oscillations 
depends \cite{PKSP3nu88}.
The results of these studies showed that the 
predictions for $\cos\delta$ exhibit strong 
dependence on the symmetry form of 
 $\tilde{U}_\nu$. This led to the conclusion that 
a sufficiently precise measurement of $\cos\delta$ 
combined with high precision measurements of 
$\sin^2\theta_{12}$, $\sin^2\theta_{23}$ and 
 $\sin^2\theta_{13}$ can allow to test 
critically the idea of existence 
of an underlying discrete symmetry form 
of the PMNS matrix and, thus, of existence of a new 
symmetry in particle physics. 

 In ref. \cite{Petcov:2014laa} predictions for the Majorana phases 
of the PMNS matrix $\alpha_{21}$ and $\alpha_{31}$
in the case of 
$\tilde{U}_\nu = R_{23}(\theta^{\nu}_{23})R_{12}(\theta^{\nu}_{12})$,   
corresponding to the TBM, BM (LC), GRA, GRB and HG 
symmetry forms,  
and  $\tilde{U}_e =  R^{-1}_{23}(\theta^{e}_{23})R^{-1}_{12}(\theta^{e}_{12})$ 
were derived under the assumption that the phases 
$\xi_{21}$ and $\xi_{31}$ in eqs.~(\ref{PMNS2}) and  
(\ref{PsieQ0}), which originate 
from the diagonalisation of the neutrino Majorana mass term, 
are known (i.e., are fixed by symmetry or other arguments).
In the present article we extend the analysis performed in 
\cite{Petcov:2014laa} to obtain predictions for the phases 
 $\alpha_{21}$ and $\alpha_{31}$ in the cases of the forms 
of the matrices  $\tilde{U}_\nu$ and   $\tilde{U}_e$ 
listed in items A and B above.
This allows us to obtain predictions 
for the phase differences $(\alpha_{21} - \xi_{21})$ and 
$(\alpha_{31}-\xi_{31})$. 
We further employ the generalised CP symmetry constraint 
in the neutrino sector 
\cite{Branco:1986gr,Feruglio:2012cw,Holthausen:2012dk}, 
which allows us to fix the values 
of the phases $\xi_{21}$ and $\xi_{31}$, and thus to predict 
the values of  $\alpha_{21}$ and $\alpha_{31}$.
We use these results together with 
the sum rule results on $\cos\delta$ 
to derive (in graphic form) 
predictions for the dependence of the 
absolute value of the 
$\betabeta$-decay effective Majorana mass 
(see, e.g., \cite{BiPet87}), $\meff$, on the lightest 
neutrino mass in all cases considered 
for both the NO and IO spectra. 

 Our article is organised as follows.
In Section~\ref{sec:ije23nu12nu} we obtain 
sum rules for 
$(\alpha_{21} - \xi_{21})$ and $(\alpha_{31}-\xi_{31})$
in schemes containing one rotation from the charged lepton sector, 
i.e., $\tilde{U}_e = R^{-1}_{12}(\theta^e_{12})$, or 
$\tilde{U}_e = R^{-1}_{13}(\theta^e_{13})$,
and two rotations from the neutrino sector:
$\tilde{U}_\nu = R_{23}(\theta^{\nu}_{23}) \, R_{12}(\theta^{\nu}_{12})$.
In these schemes the PMNS matrix has the form
\begin{align}
U = R_{ij}(\theta^e_{ij}) \, \Psi \, R_{23}(\theta^{\nu}_{23}) \,
R_{12}(\theta^{\nu}_{12}) \, Q_0 \,, 
\label{eq:Uij}
\end{align}
%
with $(ij) = (12)$, $(13)$. 
We obtain results in the general case
of arbitrary fixed values of $\theta^{\nu}_{23}$ 
and  $\theta^{\nu}_{12}$.  
In Section~\ref{sec:ijekle23nu12nu} we analyse schemes 
with $\tilde{U}_e = R^{-1}_{23}(\theta^e_{23}) \, R^{-1}_{12}(\theta^e_{12})$,
$\tilde{U}_e = R^{-1}_{23}(\theta^e_{23}) \, R^{-1}_{13}(\theta^e_{13})$,
or $\tilde{U}_e = R^{-1}_{13}(\theta^e_{13}) \, R^{-1}_{12}(\theta^e_{12})$,
and~%
\footnote{We consider only the ``standard'' ordering 
of the two rotations in $\tilde{U}_e$, see 
\cite{Marzocca:2013cr}.  The case with 
$\tilde{U}_e =  R^{-1}_{23}(\theta^e_{23}) \, R^{-1}_{12}(\theta^e_{12})$
has been investigated in \cite{Petcov:2014laa} and 
we consider it here briefly for completeness.} 
two rotations from the neutrino sector, i.e.,
\begin{align}
&U = R_{ij}(\theta^e_{ij}) \,  R_{kl}(\theta^e_{kl})  \, \Psi \, 
R_{23}(\theta^{\nu}_{23}) \, R_{12}(\theta^{\nu}_{12}) \, Q_0 \,,
\label{eq:Uijkl} 
\end{align}
%
with $(ij)-(kl) = (12)-(23)$, $(13)-(23)$, $(12)-(13)$. 
Again we provide results for arbitrary fixed values of 
$\theta^{\nu}_{23}$ and $\theta^{\nu}_{12}$.
Further, in Section \ref{sec:ije23nu13nu12nu}, 
we extend the analysis performed in Section~\ref{sec:ije23nu12nu} 
to the case of a third rotation matrix present in 
$\tilde{U}_{\nu}$:
\begin{align}
U = R_{ij}(\theta^e_{ij}) \, \Psi \, R_{23}(\theta^{\nu}_{23})\,
R_{13}(\theta^{\nu}_{13}) \, R_{12}(\theta^{\nu}_{12}) \, Q_0 \,,
\label{eq:Uija} 
\end{align}
%
with $(ij) = (12)$, $(13)$, $(23)$.
Section~\ref{sec:sumrules} contains a brief summary of 
the sum rules for the Majorana phases 
$\alpha_{21}/2$ and $\alpha_{31}/2$ derived in Sections~%
\ref{sec:ije23nu12nu}~--~\ref{sec:ije23nu13nu12nu}.
Using the sum rules, we present in Section~\ref{sec:predictions} 
predictions for phase differences  
$(\alpha_{21}/2 - \xi_{21}/2)$, $(\alpha_{31}/2 - \xi_{31}/2)$, etc.,
 involving the Majorana phases 
$\alpha_{21}/2$ and $\alpha_{31}/2$,
which are determined just by the values of the three 
neutrino mixing angles $\theta_{12}$,  $\theta_{23}$ and 
$\theta_{13}$, 
and of the fixed angles $\theta^\nu_{ij}$.
In the cases listed in item A we give results for values of
$\theta^{\nu}_{23}$ ($ = -\,\pi/4$) and $\theta^{\nu}_{12}$,
corresponding to the TBM, BM~(LC), GRA, GRB and HG symmetry 
forms of $\tilde{U}_{\nu}$. In each of the two cases 
given in item B 
the reported results are for $\theta^{\nu}_{23} = -\,\pi/4$ 
and five sets of values of $\theta^{\nu}_{13}$ and $\theta^{\nu}_{12}$ 
associated with symmetries. 
We then set $(\xi_{21},\xi_{31})=(0,0)$, 
$(0,\pi)$, $(\pi,0)$ and $(\pi,\pi)$ 
and use the resulting values of 
$\alpha_{21}/2$ and $\alpha_{31}/2$ 
to derive graphical predictions for 
the absolute value of 
the effective Majorana 
mass in $\betabeta$-decay, $\meff$,
as a function of the lightest neutrino mass 
in the schemes of mixing studied.
We show in Section~\ref{sec:GCP} that 
the requirement of generalised CP invariance 
of the neutrino Majorana mass term  
in the cases of $S_4$, $A_4$, $T^\prime$ and $A_5$ lepton flavour symmetries 
leads indeed to $\xi_{21} =0$ or $\pi$, $\xi_{31} = 0$ or $\pi$. 
In the first two cases (third case)
studied in Section~\ref{sec:ijekle23nu12nu},  B1 and B2 (B3), 
the phase  $\alpha_{31}/2$ 
(the phases $\delta$, $\alpha_{21}/2$ and $\alpha_{31}/2$)
depends (depend) on an additional phase, 
$\beta$ ($\omega$), which, in general, is not 
constrained. For schemes B1 and B2, 
the predictions for $\meff$ are obtained 
in Section~\ref{sec:predictions} by varying $\beta$ in 
the interval $[0,\pi]$.
In the case of scheme B3 the results for 
the Majorana phases and $\meff$ are derived for the value of $\omega = 0$, 
for which the Dirac phase $\delta$ has a value 
in its $2\sigma$ allowed interval 
quoted in eq.~(\ref{deltaexp}). 
Section~\ref{sec:summary} contains 
summary of the results of the present study and conclusions.

We note finally that the titles of Sections~%
\ref{sec:ije23nu12nu}~--~\ref{sec:ije23nu13nu12nu} and of their subsections 
reflect the rotations contained in the corresponding parametrisation,
eqs.~(\ref{eq:Uij})~--~(\ref{eq:Uija}).

%
\section{The Cases of $\boldsymbol{\theta^e_{ij} - (\theta^\nu_{23}, \theta^\nu_{12})}$ 
 Rotations }
\label{sec:ije23nu12nu}
%
%

In this section we  derive the 
sum rules for $\alpha_{21}$ and $\alpha_{31}$ of interest 
in the case when the matrix $\tilde{U}_\nu = R_{23}(\theta^{\nu}_{23})\,
R_{12}(\theta^{\nu}_{12})$ with fixed (e.g., symmetry) values 
of the angles $\theta^{\nu}_{23}$ and 
$\theta^{\nu}_{12}$, 
gets correction only due to
one rotation from the charged lepton sector.
The neutrino mixing matrix $U$ has the 
form given in eq.~(\ref{eq:Uij}).
We do not consider the case of eq.~(\ref{eq:Uij}) 
with $(ij) = (23)$, because in this case the reactor angle
$\theta_{13}=0$ and thus the measured value of $\theta_{13} \cong 0.15$
cannot be reproduced.

\subsection{The Scheme with $\boldsymbol{\theta^e_{12} - (\theta^\nu_{23}, \theta^\nu_{12})}$ 
 Rotations (Case A1)}
\label{sec:12e23nu12nu}

 In the present subsection we consider the parametrisation
of the neutrino mixing matrix given in eq.~(\ref{eq:Uij}) 
with $(ij) = (12)$. In this parametrisation the PMNS 
matrix has the form
\begin{align}
&U = R_{12}(\theta^e_{12}) \, \Psi \, 
R_{23}(\theta^{\nu}_{23}) \, R_{12}(\theta^{\nu}_{12}) \, Q_0 \,.
\label{eq:U12e23nu12nu} 
\end{align}
%
The phase $\omega$ in the phase matrix $\Psi$ is unphysical.

 We are interested in deriving analytic expressions for 
the Majorana phases $\alpha_{21}$ and  $\alpha_{31}$
i) in terms of the parameters of the parametrisation in 
eq.~(\ref{eq:U12e23nu12nu}), 
$\theta^e_{12}$, $\psi$, $\theta^{\nu}_{23}$, $\theta^{\nu}_{12}$, 
$\xi_{21}$ and $\xi_{31}$, 
and possibly 
ii) in terms of the angles $\theta_{12}$, $\theta_{13}$, $\theta_{23}$ 
and the Dirac phase $\delta$ of the standard parametrisation 
of the PMNS matrix, 
the fixed angles $\theta^\nu_{23}$ and $\theta^\nu_{12}$,
and the phases $\xi_{21}$ and $\xi_{31}$.
The values of the phases 
$\alpha_{21}$ and  $\alpha_{31}$ in the latter case, as we will see,
indeed depend on the value of the Dirac phase $\delta$.
Thus, we first recall the sum rule satisfied 
by the Dirac phase $\delta$
in the case under study,   
by which $\cos\delta$  is expressed in terms of 
the angles  $\theta_{12}$, $\theta_{13}$ and $\theta_{23}$. 
The sum rule of interest reads \cite{Petcov:2014laa}:
\begin{equation}
\cos\delta =  \frac{\tan\theta_{23}}{\sin2\theta_{12}\sin\theta_{13}}\,
\left [\cos2\theta^{\nu}_{12} +
\left (\sin^2\theta_{12} - \cos^2\theta^{\nu}_{12} \right )\,
 \left (1 - \cot^2\theta_{23}\,\sin^2\theta_{13}\right )\right ]\,.
\label{cosdthnu}
\end{equation}
%
Although the expression in eq.~(\ref{cosdthnu}) was derived in \cite{Petcov:2014laa} 
for $\theta^{\nu}_{23} = -\pi/4$, it
was shown in \cite{Girardi:2015vha} to be valid for 
arbitrary $\theta^{\nu}_{23}$. 
The dependence of $\cos\delta$ on  $\theta^{\nu}_{23}$ is ``hidden'', 
in particular, in the specific relation between 
$\theta_{23}$ and  $\theta^{\nu}_{23}$:
\begin{align}
\sin^2 \theta_{23} & = \frac{|U_{\mu3}|^2}{1-|U_{e3}|^2} = 
\frac{\sin^2 \theta^{\nu}_{23}-\sin^2 \theta_{13}}{1 - \sin^2 \theta_{13}} 
\label{eq:th23A0}\,.
\end{align}
%
We give also the expressions of $\sin^2 \theta_{13}$ 
and  $\sin^2 \theta_{12}$ in terms of the  parameters 
of the parametrisation of the PMNS matrix 
given in eq.~(\ref{eq:U12e23nu12nu}), which  will be used further 
in the analysis performed in this subsection:
\begin{align}
\label{eq:th13A0}
\sin^2 \theta_{13} & = |U_{e3}|^2  = \sin^2 \theta^e_{12} \sin^2 \theta^{\nu}_{23}\,,\\
\sin^2 \theta_{12} &  = \frac{|U_{e2}|^2}{1-|U_{e3}|^2}  = \frac{1}{1-\sin^2 \theta_{13}} 
\bigg[ \cos^2 \theta^{\nu}_{23} \sin^2 \theta^e_{12} \cos^2 \theta^{\nu}_{12} +  
\cos^2 \theta^e_{12} \sin^2 \theta^{\nu}_{12}   \nonumber \\
& + \dfrac{1}{2} \sin 2 \theta^e_{12} \sin 2 \theta^{\nu}_{12} \cos \theta^{\nu}_{23} \cos \psi  \bigg] 
\label{eq:th12A0}\,.
\end{align}
%
The parameters  $\sin^2 \theta^\nu_{23}$ and $\sin^2 \theta^e_{12}$ 
can be expressed in terms  of $\sin^2\theta_{13}$ and $\sin^2\theta_{23}$
using eqs.~(\ref{eq:th23A0}) and ~(\ref{eq:th13A0}).

 From eqs.~(\ref{eq:th13A0}) and (\ref{eq:th12A0}) we get the
following expression for $\cos \psi$:
\begin{align}
& \cos \psi = 
\dfrac{\sin^2 \theta^{\nu}_{23} \left( \cos^2\theta_{13}\sin^2 \theta_{12} 
- \sin^2 \theta^{\nu}_{12} \right) 
+ \sin^2 \theta_{13} \left( \cos^2 \theta^{\nu}_{12} \sin^2 \theta^{\nu}_{23} 
- \cos 2 \theta^{\nu}_{12} \right)}
{{\rm sgn}(\sin2\theta^e_{12})\, \sin 2 \theta^{\nu}_{12} \cos \theta^{\nu}_{23} \sin \theta_{13} 
\left(\sin^2 \theta^{\nu}_{23} - \sin^2 \theta_{13}\right)^{1/2}} \,. 
\label{eq:thpsiA0}
\end{align}
%
The sign of $\sin2\theta^e_{12}$ is supposed to be fixed 
in the underlying theory leading to the neutrino mixing 
given in eq.~(\ref{eq:U12e23nu12nu}). In what follows we will 
account for both possibilities of  $\sin2\theta^e_{12} > 0$ 
and  $\sin2\theta^e_{12} < 0$. 
Using eq.~(\ref{eq:th23A0}) and setting
$\sin^2 \theta^{\nu}_{23} = \sin^2\theta_{23} \cos^2\theta_{13} 
+ \sin^2\theta_{13}$,  
$\cos\theta^{\nu}_{23} = \cos\theta_{23} \cos\theta_{13}$ 
and ${\rm sgn}(\sin2\theta^e_{12}) = 1$
in eq.~(\ref{eq:thpsiA0}) leads to an expression 
for $\cos\psi$ in terms of $\theta^\nu_{12}$ and 
the standard parametrisation mixing angles 
$\theta_{12}$, $\theta_{13}$ and $\theta_{23}$, 
which coincides with the expression 
for $\cos\phi$ given in eq.~(22) in \cite{Petcov:2014laa}.
For $\theta^{\nu}_{23} = -\pi/4$,
eq.~(\ref{eq:thpsiA0}) reduces to the expression 
for $\cos\phi$ in eq. (46) in ref. \cite{Petcov:2014laa}
and in eq.~(37) in ref. \cite{Girardi:2014faa}.

The cosine of the phase $\psi$ can be determined 
uniquely using eq. (\ref{eq:thpsiA0}), i.e., using as input 
${\rm sgn}(\sin2\theta^e_{12})$, 
the symmetry values of $\theta^{\nu}_{12}$ and $\theta^{\nu}_{23}$
(of  $\theta^{\nu}_{12}$) and the measured value of 
$\theta_{12}$ and $\theta_{13}$  
($\theta_{12}$, $\theta_{13}$ and $\theta_{23}$).
However, the sign of $\sin\psi$ in this case remains unfixed 
if no additional information allowing to fix it is available.
This in turn leads to an ambiguity in the determination of 
the phase $\psi$ from the value of $\cos\psi$: 
in the interval $[0,2\pi]$, two values of $\psi$ 
will be possible. 

  Sum rules for the Majorana phases 
$\alpha_{21}$ and  $\alpha_{31}$ of the type we are 
interested in were  derived in \cite{Petcov:2014laa}.
The sum rules for $\alpha_{21}$ and  $\alpha_{31}$ 
we are aiming to obtain in this subsection 
turn out to be a particular case of the 
sum rules derived in \cite{Petcov:2014laa}.
This becomes clear from a comparison 
of eq.~(18)  in \cite{Petcov:2014laa}, which fixes 
the parametrisation of $U$ used in \cite{Petcov:2014laa},  
and the expression for $U$ in eq.~(\ref{eq:U12e23nu12nu}). 
It shows that to get the sum rules for 
$\alpha_{21}$ and  $\alpha_{31}$ of interest, 
one has formally to set $\hat \theta_{23} = \theta^\nu_{23}$, 
$\phi = -\psi$ and $\beta = 0$ in the sum rules  
for $\alpha_{21}$ and  $\alpha_{31}$ derived in 
eq.~(102) in \cite{Petcov:2014laa} and 
to take into account the two possible
signs of the product  
$c^e_{12}c^\nu_{23}s^\nu_{23}
\equiv \cos\theta^e_{12}\cos\theta^\nu_{23}\sin\theta^\nu_{23}$:
\begin{equation}
\frac{\alpha_{21}}{2} = \beta_{e2} - \beta_{e1} +\frac{\xi_{21}}{2}\,,~~~  
\label{Majph21}
\end{equation}
%
\begin{equation}
\frac{\alpha_{31}}{2} = \beta_{e2}  + \tilde{\varphi} + \frac{\xi_{31}}{2}\,, 
\quad
e^{i\tilde{\varphi}} = {\rm sgn}(c^e_{12}c^\nu_{23}s^\nu_{23}) 
= +1~{\rm or}~(-1)\,.
\label{Majph31}
\end{equation}
%
Thus, $\tilde{\varphi} = 0$ or $\pi$. 
The results in eqs.~(\ref{Majph21}) and (\ref{Majph31}) 
can be obtained formally from eqs. (88), (89) and (95) 
in \cite{Petcov:2014laa} by setting $\hat \theta_{23} = \theta^\nu_{23}$, 
$\phi = -\psi$, 
$Q_1 = {\rm diag}(1,1,1)$ and 
$Q_2 = {\rm diag}(1,e^{i(\beta_{e2}-\beta_{e1})},
{\rm sgn}(c^e_{12}c^\nu_{23}s^\nu_{23})\, e^{i\beta_{e2}})$. 
We note that in the case considered of arbitrary fixed signs of 
$c^e_{12}$, $s^e_{12} \equiv \sin\theta^e_{12}$, 
$c^\nu_{23}$ and $s^\nu_{23}$, 
the $U_{e3}$ element of the PMNS matrix in eq.~(95) in 
\cite{Petcov:2014laa} must also be replaced by 
$U_{e3} \, {\rm sgn}(c^e_{12}s^e_{12}c^\nu_{23})$.
Correspondingly, in terms of the parametrisation in 
eq.~(\ref{eq:U12e23nu12nu}) of the PMNS matrix, 
the phases  $\beta_{e2}$ and $\beta_{e1}$ 
are given by eqs.~(90) and (91) in  \cite{Petcov:2014laa}: 
\begin{align}
\label{be1}
\beta_{e1} & = {\rm arg}(U_{e 1}) =  
{\rm arg}\left (c^e_{12} c^\nu_{12} - 
 s^e_{12}c^{\nu}_{23} s^\nu_{12} e^{-i \psi} \right )\,,
\\[0.25cm]
\label{be2}
\beta_{e2} & =  {\rm arg}(U_{e 2}\,e^{-i\frac{\xi_{21}}{2}}) =  
{\rm arg} \left ( c^e_{12} s^\nu_{12} 
+ s^e_{12}c^\nu_{23} c^\nu_{12} e^{-i \psi} \right )\,,
\end{align}
%
where  $c^\nu_{12} \equiv \cos\theta^\nu_{12}$ and 
$s^\nu_{12} \equiv \sin\theta^\nu_{12}$. 
For  $\tilde{\varphi} = 0$, 
eq.~(\ref{Majph31}) reduces to the expression for 
$\alpha_{31}/2$ in eq.~(102) in \cite{Petcov:2014laa}.
By using eq.~(\ref{eq:th13A0}), $s^e_{12}$ and $c^e_{12}$
in eqs. (\ref{be1}) and (\ref{be2})
can be expressed (given their signs) 
in terms of $\sin\theta_{13}$ and  
$\sin\theta^\nu_{23}$, while the phase $\psi$ is determined via  
eq.~(\ref{eq:thpsiA0}) by the values of $\theta_{12}$,  
$\theta_{13}$, $\theta^\nu_{12}$ and $\theta^\nu_{23}$ 
(up to an ambiguity of the sign of $\sin\psi$). 
The phases  $\beta_{e2}$ and $\beta_{e1}$ 
in this case will be given in terms of 
$\theta_{12}$, $\theta_{13}$, $\theta^\nu_{12}$ and 
$\theta^\nu_{23}$, i.e., in terms of mixing angles 
which are measured or fixed by symmetry arguments.
It is often convenient to express $\sin\theta^\nu_{23}$ and  
$\cos\theta^\nu_{23}$ in terms of the measured 
angles $\theta_{13}$ and $\theta_{23}$
of the standard parametrisation of the PMNS matrix 
using the relation in eq.~(\ref{eq:th23A0}).
 
 As can be shown employing the formalism developed in
\cite{Petcov:2014laa} and taking into account 
the possibility of negative signs of 
$c^e_{12}s^\nu_{12}$ and $c^e_{12}c^\nu_{12}$,
the expressions for the phases $\beta_{e2}$ and $\beta_{e1}$ 
in terms of the angles $\theta_{12}$, $\theta_{13}$, $\theta_{23}$ 
and the Dirac phase $\delta$ of the standard parametrisation 
of the PMNS matrix have the form:
\begin{align}
\label{be2tau1ph}
\beta_{e2} & = 
{\rm arg}\left (U_{\tau 1}{\rm sgn}(c^e_{12}s^\nu_{12})\right) = 
{\rm arg} \left[\left (s_{12} s_{23} 
- c_{12} c_{23} s_{13} e^{i \delta}\right){\rm sgn}(c^e_{12}s^\nu_{12})\right]\,,
\\[0.25cm]
 \label{be1tau2ph}
\beta_{e 1} & = 
{\rm arg}\left (U_{\tau 2}\,e^{i\pi}{\rm sgn}(c^e_{12}c^\nu_{12})
\,e^{-i\frac{\alpha_{21}}{2}}\right) =   
{\rm arg}\left[ \left (c_{12} s_{23} 
+ s_{12} c_{23} s_{13} e^{i \delta} \right){\rm sgn}(c^e_{12}c^\nu_{12})
\right ]\,.
\end{align}
%
For ${\rm sgn}(c^e_{12}s^\nu_{12}) =1$ and 
${\rm sgn}(c^e_{12}c^\nu_{12}) =1$, eqs.~(\ref{be2tau1ph}) 
and (\ref{be1tau2ph}) reduce respectively to eqs.~(100) and 
(101) in ref. \cite{Petcov:2014laa}.

 It follows from eqs.~(\ref{be2tau1ph}) and (\ref{be1tau2ph}) 
that the phases $\beta_{e1}$ and $\beta_{e 2}$ are determined by 
the values of the standard parametrisation mixing angles 
$\theta_{12}$, $\theta_{13}$, $\theta_{23}$  
and of the Dirac phase $\delta$. The phase  $\delta$ is also determined 
(up to a sign ambiguity of $\sin\delta$) by the values of 
``standard'' angles $\theta_{12}$, $\theta_{13}$, $\theta_{23}$ via the sum 
rule given in eq.~(\ref{cosdthnu}). Since the relations 
in eqs.~(\ref{Majph21}) and  (\ref{Majph31}) 
between the Majorana phases $\alpha_{21}$ and $\alpha_{31}$ 
and the phases $\beta_{e1}$ and $\beta_{e 2}$ involve 
the phases  $\xi_{21}$ and $\xi_{31}$ originating 
from the diagonalisation of the neutrino Majorana mass term, 
$\alpha_{21}$ and $\alpha_{31}$ will be determined by the values of the 
``standard'' neutrino mixing angles $\theta_{12}$, $\theta_{13}$, $\theta_{23}$ 
(up to the mentioned ambiguity related to the undetermined so far sign of 
$\sin\delta$), provided the values of $\xi_{21}$ and $\xi_{31}$ are 
known. Thus, predictions for the Majorana phases $\alpha_{21}$ 
and $\alpha_{31}$ can be obtained when the phases 
$\xi_{21}$ and $\xi_{31}$ are fixed 
by additional considerations of, e.g.,  generalised CP invariance, 
symmetries, etc. In theories with discrete lepton flavour symmetries 
the phases  $\xi_{21}$ and $\xi_{31}$ are often determined 
by the employed symmetries of the theory 
(see, e.g., \cite{Girardi:2013sza,Chen:2009gf,Gehrlein:2014wda,Ballett:2015wia,Turner:2015uta} and references quoted therein).
We will show in Section~\ref{sec:GCP} how the phases 
$\xi_{21}$ and $\xi_{31}$ are fixed 
by the requirement of generalised CP invariance 
of the neutrino Majorana mass term 
in the cases of the non-Abelian discrete 
flavour symmetries $S_4$, $A_4$, $T^\prime$ and $A_5$. 
In all these cases the generalised CP invariance    
constraint fixes the values of  
$\xi_{21}$ and $\xi_{31}$, which allows us 
to obtain predictions for 
the Majorana phases $\alpha_{21}$ and $\alpha_{31}$.

 The phases 
$\beta_{e1}$, $\beta_{e2}$, $\psi$ and  $\delta$
can be shown to satisfy the relation: 
\begin{equation}
\delta = \psi + \beta_{e1} + \beta_{e2}  + \varphi\,, \quad
e^{i\varphi} = {\rm sgn}(c^e_{12}s^e_{12}c^\nu_{23}) = +1~{\rm or}~(-1)\,. 
\label{dphibeta}
\end{equation}
%
For $\varphi = 0$ (${\rm sgn}(c^e_{12}s^e_{12}c^\nu_{23}) = +1$),
this relation reduces to eq.~(94) in ref. \cite{Petcov:2014laa} 
by setting $ \psi = -\phi$.
From eqs.~(\ref{Majph21}), (\ref{Majph31}) and (\ref{dphibeta}) 
we get further
\begin{equation}
(\alpha_{31} - \xi_{31}) - \frac{1}{2}\, (\alpha_{21} - \xi_{21}) 
= \beta_{e1} + \beta_{e2} + 2\tilde{\varphi} 
= \delta - \psi - \varphi\,, \quad \varphi = 0~{\rm or}~\pi\,,
\label{Majxidpsi}
\end{equation}
%
where we took into account that $2\tilde{\varphi} = 0$ or $2\pi$.
 
The Dirac phase $\delta$ and the phase $\psi$ 
are related \cite{Petcov:2014laa}. We will give below only the relation 
between $\sin\delta$ and $\sin\psi$. It can be obtained from 
eq.~(28) in \cite{Petcov:2014laa} by setting~%
\footnote{The relation between $\cos\delta$ and $\cos\psi$
can be deduced from eq.~(29) in \cite{Petcov:2014laa}.
}
$\phi = - \psi$ and by taking into account 
that in the case considered both signs of
$\sin2\theta^e_{12}\cos\theta^\nu_{23}$ 
are, in principle, allowed~%
\footnote{In \cite{Petcov:2014laa} both $\sin2\theta^e_{12}$ 
and $\cos\theta^\nu_{23}$ could be and were 
considered to be positive without loss of generality.
}:
\begin{align}
\label{sindsinphi}
\sin\delta =
&\; {\rm sgn}\,(\sin2\theta^e_{12}\cos\theta^\nu_{23})\,
 \frac{\sin2\theta^{\nu}_{12}}{\sin2\theta_{12}}\,\sin\psi\,.
\end{align}
%

 We note that within the approach employed in our analysis,
the results presented in eqs.~(\ref{Majph21})~--~(\ref{sindsinphi})
are exact and are valid for arbitrary fixed values of 
$\theta^{\nu}_{12}$ and $\theta^{\nu}_{23}$ and for arbitrary 
signs of $\sin\theta^e_{12}$ and  $\cos\theta^e_{12}$ 
($|\sin\theta^e_{12}|$ and  $|\cos\theta^e_{12}|$
can be expressed in terms of $\theta_{13}$ and 
$\theta^{\nu}_{23}$).

 Although the sum rules  
 derived above allow to determine 
the values of the Majorana phases 
$\alpha_{21}$ and $\alpha_{31}$ 
(up to a two-fold ambiguity related 
to the ambiguity of ${\rm sgn}(\sin\delta)$  
or of  ${\rm sgn}(\sin\psi)$) 
if the phases $\xi_{21}$ and $\xi_{31}$ are 
known, we will present 
below an alternative method of 
determination of $\alpha_{21}$ and 
$\alpha_{31}$,  which can be used 
in the cases when the method developed 
in \cite{Petcov:2014laa} cannot be applied.
The alternative method makes use of the 
rephasing invariants associated with the two 
Majorana phases of the PMNS matrix.

In the case of 3-neutrino mixing 
under discussion there are, 
in principle, three independent 
CPV rephasing invariants.
The first is associated with the Dirac 
phase $\delta$ and is given 
by the well-known expression 
in eq.~(\ref{JCP}), where we have shown 
also the expression of the $J_{\rm CP}$ 
factor in the standard parametrisation.
The other two,  $I_1$ and $I_2$,
are related to the two Majorana 
CPV phases in the PMNS matrix and 
can be chosen as \cite{JMaj87,ASBranco00,BPP1}~%
\footnote{The expressions for the invariants
$I_{1,2}$ we give and will use further 
correspond to Majorana conditions 
satisfied by the fields of the 
light massive Majorana 
neutrinos, which do not contain 
phase factors, see, e.g., \cite{BPP1}.}:
\beq
I_1 = {\rm Im}\left\{ U_{e 1}^\ast \, U_{e 2} \right\}\,, \quad
I_2 = {\rm Im}\left\{ U_{e 1}^\ast \, U_{e 3} \right\}\,.
\label{eq:SCP}
\eeq
%
The rephasing invariants associated 
with the Majorana phases are 
not uniquely determined.
Instead of $I_1$ defined above we could have chosen, e.g.,
$I'_1 = {\rm Im}\left\{U_{\tau 1}^\ast \, U_{\tau 2}\right\}$ 
or $I''_1 = {\rm Im}\left\{ U_{\mu 1} U_{\mu 2}^\ast \right\}$, 
while instead of $I_2$ we could have used 
$I'_2 = {\rm Im}\left\{  U_{\tau 2}^\ast \, U_{\tau 3} \right\}$, 
or $I''_2 = {\rm Im}\left\{ U_{\mu 2} \, U_{\mu 3}^\ast \right\}$.
However, the three invariants~---~%
$J_{\rm CP}$ and any two chosen Majorana phase 
invariants~---~form a complete set in the case of 
3-neutrino mixing: any other two rephasing invariants associated 
with the Majorana phases can be expressed in terms of the 
two chosen Majorana phase invariants and the $J_{\rm CP}$ factor 
\cite{JMaj87}.
We note also that CP violation due to the Majorana phase
$\alpha_{21}$ requires that both
$I_1 = {\rm Im}\left\{ U^\ast_{e 1} U_{e 2} \right\}\neq 0$ and
${\rm Re}\left\{ U^\ast_{e 1} U_{e 2} \right\}\neq 0$ \cite{ASBranco00}.
Similarly,
$I_2 = {\rm Im}\left\{ U_{e 1}^\ast U_{e 3} \right\} \neq 0$
would imply violation of the CP symmetry only 
if in addition ${\rm Re}\left\{ U_{e 1}^\ast U_{e 3} \right\} \neq 0$.

 In the standard parametrisation of the PMNS matrix $U$, 
the rephasing invariants $I_1$ and $I_2$ are given by
\begin{align}
I_1 & =  
\cos\theta_{12} \sin \theta_{12} \cos^2 \theta_{13} \sin (\alpha_{21}/2) \,, 
\label{eq:StI1}\\
I_2 & = 
\cos\theta_{12} \sin\theta_{13} \cos\theta_{13} \sin (\alpha_{31}/2 - \delta) \,.
\label{eq:StI2}
\end{align}
%
Comparing these expressions with the expressions for 
$I_1$ and $I_2$ in the parametrisation of $U$  
defined in eq.~(\ref{eq:U12e23nu12nu}),
we obtain sum rules for $\sin (\alpha_{21}/2)$ and 
$\sin (\alpha_{31}/2 - \delta)$ in terms 
of $\theta^e_{12}$, $\psi$, $\theta^\nu_{12}$, $\theta^\nu_{23}$ 
and the standard parametrisation mixing angles $\theta_{12}$ and 
$\theta_{13}$:
\begin{align}
\label{eq:alpha21A0gen}
\sin (\alpha_{21}/2) & = \dfrac{1}{\cos^2 \theta_{13}  \sin 2 \theta_{12}}
\bigg[ \sin 2 \theta^e_{12} \cos \theta^{\nu}_{23}
\big(\sin(\xi_{21}/2 - \psi)
 - 2\sin^2\theta^\nu_{12}\cos \psi \sin(\xi_{21}/2)\big) 
\nonumber 
\\
&  +  \sin 2 \theta^\nu_{12}\sin(\xi_{21}/2)
\big(\cos^2 \theta^e_{12} -  
\sin^2 \theta^e_{12} \cos^2 \theta^{\nu}_{23} 
  \big) 
\bigg] \,, 
\\
\sin (\alpha_{31}/2 - \delta) & = 
\dfrac{2\sin \theta^e_{12} \sin \theta^{\nu}_{23}}
{\cos\theta_{12} \sin2\theta_{13}} 
\bigg[
\cos \theta^e_{12}  \cos \theta^{\nu}_{12} \sin(\xi_{31}/2 - \psi) 
-  \cos\theta^{\nu}_{23}\sin \theta^e_{12}
\sin\theta^{\nu}_{12}\sin(\xi_{31}/2) \bigg] \,.
\label{eq:alpha31dA0gen}
\end{align}
%
The result in eq.~(\ref{eq:alpha31dA0gen}) 
can be derived also from eqs.~(\ref{Majph31}) and (\ref{dphibeta}), 
which lead to 
\begin{equation}
\frac{\alpha_{31}}{2} - \delta 
=  -\,\psi -\,\beta_{e1} + \,\tilde\varphi - \,\varphi + \frac{\xi_{31}}{2}\,,
\label{eq:alph31delta}
\end{equation}
%
and by using further eq. (\ref{be1}) for $\beta_{e1}$.
The expression for $\sin (\alpha_{31}/2)$, which can be obtained from 
eqs. (\ref{Majph31}) and (\ref{be2}),
has a form similar to that of $\sin (\alpha_{31}/2 - \delta)$:
\begin{align}
\sin (\alpha_{31}/2) & = \dfrac{{\rm sgn}(c^e_{12}c^\nu_{23}s^\nu_{23})}
{\sin \theta_{12}  \cos \theta_{13}} 
\bigg[\sin \theta^e_{12}
\cos \theta^\nu_{12}  \cos \theta^{\nu}_{23} \sin(\xi_{31}/2 - \psi) 
+ \cos \theta^e_{12} \sin \theta^{\nu}_{12} \sin (\xi_{31}/2) \bigg] \,.
\label{eq:alpha31A0gen}
\end{align}
%

The angles $\theta^\nu_{12}$ and $\theta^\nu_{23}$ in 
eqs.~(\ref{eq:alpha21A0gen}), (\ref{eq:alpha31dA0gen}) 
and (\ref{eq:alpha31A0gen}), 
as we have already emphasised, 
are assumed to be fixed by symmetry arguments,  
$\theta^e_{12}$ can be expressed in terms of 
$\theta_{13}$ and $\theta^\nu_{23}$ using eq.~(\ref{eq:th13A0}), 
while eq.~(\ref{eq:thpsiA0}) allows to express $\psi$ in terms of 
$\theta_{12}$, $\theta_{13}$, $\theta^\nu_{12}$ and $\theta^\nu_{23}$.
The formulae for $\cos (\alpha_{21}/2)$ and 
$\cos (\alpha_{31}/2 - \delta)$, which 
enter into the expression for the absolute 
value of the effective Majorana mass 
in $\betabeta$-decay (see, e.g., \cite{BPP1}), $\meff$, 
can be obtained from eqs.~(\ref{eq:alpha21A0gen}) and 
(\ref{eq:alpha31dA0gen}) by changing $\xi_{21}$ to 
$\xi_{21} + \pi$ and $\xi_{31}$ to $\xi_{31} + \pi$, respectively.

 In terms of the standard parametrisation mixing angles 
$\theta_{12}$, $\theta_{13}$, $\theta_{23}$ and the Dirac phase $\delta$, and 
the angles $\theta^\nu_{12}$ and $\theta^\nu_{23}$, 
the expressions for $\sin (\alpha_{21}/2)$ and $\sin (\alpha_{31}/2)$ read:
\begin{align}
\sin (\alpha_{21}/2) & = \dfrac{1}{\sin^2\theta^\nu_{23}\sin2\theta^\nu_{12}}
\bigg[ \sin2\theta_{23} \sin\theta_{13}
\big(\sin(\xi_{21}/2 - \delta)
 - 2\cos^2\theta_{12}\cos \delta\sin(\xi_{21}/2)\big) 
\nonumber 
\label{eq:alpha21A0gensp}\\
&  + \sin(\xi_{21}/2) \sin 2 \theta_{12}\big(\sin^2 \theta_{23} -  
\cos^2 \theta_{23} \sin^2\theta_{13}
  \big) \bigg] \,, \\
\sin (\alpha_{31}/2) & = \dfrac{{\rm sgn}(c^\nu_{23})}
{\sin \theta^\nu_{12}\sin\theta^\nu_{23}} 
\bigg[
\sin\theta_{12}\sin\theta_{23} \sin(\xi_{31}/2) 
- \cos\theta_{12}\cos\theta_{23}\sin\theta_{13} 
\sin (\xi_{31}/2+\delta) \bigg] \,,
\label{eq:alpha31A0gensp}
\end{align}
%
where, we recall, 
$\sin^2\theta^\nu_{23} = 1 - \cos^2\theta_{23} \cos^2\theta_{13}$.

 The phases  $\xi_{21}$ and $\xi_{31}$, as we have already 
discussed, are supposed to be fixed by symmetry arguments.
Thus, it proves convenient to have analytic expressions  
which allow to calculate the phase differences 
$(\alpha_{21}/2 - \xi_{21}/2)$, $(\alpha_{31}/2 - \delta - \xi_{31}/2)$
and $(\alpha_{31}/2 - \xi_{31}/2)$. 
We find for $\sin (\alpha_{21}/2 - \xi_{21}/2)$, 
$\sin (\alpha_{31}/2 - \delta - \xi_{31}/2)$ and 
$\sin (\alpha_{31}/2 - \xi_{31}/2)$:
\begin{align}
\sin (\alpha_{21}/2 - \xi_{21}/2) 
& = - \,\dfrac{\sin 2 \theta^e_{12}}{\cos^2 \theta_{13}
\sin 2 \theta_{12}} \cos \theta^{\nu}_{23} \sin \psi 
= -\, \dfrac{\sin2\theta_{23} \sin\theta_{13}}
{\sin^2\theta^\nu_{23}\sin2\theta^\nu_{12}}\,\sin\delta\,,
\label{eq:alpha21A0d}\\
\sin (\alpha_{31}/2 - \delta - \xi_{31}/2) & = - \,\dfrac{\sin 2\theta^e_{12}}
{\cos \theta_{12}\sin 2 \theta_{13}}\, \cos \theta^{\nu}_{12}  
\sin \theta^{\nu}_{23}\, \sin \psi \,
\nonumber
\\
 & = - \, {\rm sgn}(\cos\theta^\nu_{23})\,\dfrac{\sin\theta_{12}\sin\theta_{23}}
{\sin\theta^\nu_{12}\sin\theta^\nu_{23}}\,\sin\delta\,,
\label{eq:alpha31A0d}\\
\sin (\alpha_{31}/2 - \xi_{31}/2) & = 
-\, \dfrac{{\rm sgn}(c^e_{12}c^\nu_{23}s^\nu_{23})}
{\sin \theta_{12}  \cos \theta_{13}}\, 
\sin \theta^e_{12} \cos \theta^\nu_{12}\cos \theta^{\nu}_{23}\sin\psi 
\nonumber
\\
 & = - \dfrac{{\rm sgn}(\cos\theta^\nu_{23})}
{\sin\theta^\nu_{12}  \sin \theta^\nu_{23}}\, 
\sin \theta_{13} \cos \theta_{12}\cos \theta_{23}\sin\delta\,. 
\label{eq:alpha31A0d}
\end{align}
%
It follows from eqs. (\ref{eq:alpha21A0d}) and (\ref{eq:alpha31A0d}) that 
$|\sin(\alpha_{21(31)}/2 - \xi_{21(31)}/2 )|\propto \sin\theta_{13}$. 
Using the results  given, 
e.g., in eqs. (\ref{Majph21}), (\ref{Majph31}), (\ref{be2tau1ph}), 
(\ref{be1tau2ph}), (\ref{cosdthnu}), and the best fit values of the 
neutrino oscillation parameters 
quoted in eqs.~(\ref{th12values})~--~(\ref{th13values}),  
we can obtain predictions for the values of the phases 
$(\alpha_{21}/2 - \xi_{21}/2)$ and $(\alpha_{31}/2 - \xi_{31}/2)$
for the symmetry forms of $\tilde{U}_{\nu}$ (TBM, BM (LC), GRA, etc.) 
considered. These predictions as well as predictions for the 
values of $(\alpha_{21}/2 - \xi_{21}/2)$ and $(\alpha_{31}/2 - \xi_{31}/2)$ 
in the cases investigated in the next subsection and in Sections~%
\ref{sec:ijekle23nu12nu} and \ref{sec:ije23nu13nu12nu} will be presented in 
Section~\ref{sec:predictions}.

\subsection{The Scheme with $\boldsymbol{\theta^e_{13} - (\theta^\nu_{23}, \theta^\nu_{12})}$ 
 Rotations (Case A2)}
\label{sec:13e23nu12nu}

 In the present subsection we consider the parametrisation
of the neutrino mixing matrix given in eq.~(\ref{eq:Uij}) with $(ij) = (13)$.
In this parametrisation the PMNS matrix has the form
\begin{align}
&U = R_{13}(\theta^e_{13}) \, \Psi \, 
R_{23}(\theta^{\nu}_{23}) \, R_{12}(\theta^{\nu}_{12}) \, Q_0 \,.
\label{eq:U13e23nu12nu} 
\end{align}
%
Now the phase $\psi$ in the phase matrix $\Psi$ is unphysical.
We employ the approaches used in the preceding 
subsection, which are based on the method developed in 
\cite{Petcov:2014laa} and on the relevant rephasing invariants,
for determining the Majorana phases 
$\alpha_{21}$ and $\alpha_{31}$.
 
We first give the expressions for  $\sin^2 \theta_{13}$, 
$\sin^2 \theta_{23}$ and $\sin^2 \theta_{12}$
in terms of the parameters of the parametrisation in 
eq.~(\ref{eq:U13e23nu12nu}), 
which will be used in our analysis:
\begin{align}
\sin^2 \theta_{13} & = |U_{e3}|^2  = \sin^2 \theta^e_{13} \cos^2 \theta^{\nu}_{23} \label{eq:th13B0gen}\,,\\
\sin^2 \theta_{23} & = \frac{|U_{\mu3}|^2}{1-|U_{e3}|^2} = \frac{\sin^2 \theta^{\nu}_{23}}{1 - \sin^2 \theta_{13}} \label{eq:th23B0gen}\,,\\
\sin^2 \theta_{12} & = \frac{|U_{e2}|^2}{1-|U_{e3}|^2} = \frac{1}{1-\sin^2 \theta_{13}}
\bigg[ \sin^2 \theta^{\nu}_{23} \sin^2 \theta^e_{13} \cos^2 \theta^{\nu}_{12}  \nonumber \\
& +  \cos^2 \theta^e_{13} \sin^2 \theta^{\nu}_{12} - \dfrac{1}{2} \sin 2 \theta^e_{13} \sin 2 \theta^{\nu}_{12} \sin \theta^{\nu}_{23} \cos \omega \bigg]  \label{eq:th12B0gen}\,.
\end{align}
%
The formulae 
for $\sin^2 \theta_{13}$ and $\sin^2 \theta_{23}$ given above 
have been derived in \cite{Girardi:2015vha}.
The expression for $\sin^2 \theta_{12}$ 
is a generalisation to  arbitrary fixed values of $\theta^{\nu}_{23}$ 
of that derived in \cite{Girardi:2015vha} for $\theta^{\nu}_{23} = -\pi/4$. 

 From eqs.~(\ref{eq:th13B0gen}) and (\ref{eq:th12B0gen}) we  
obtain an expression for $\cos \omega$ in terms of
the measured mixing angles $\theta_{12}$ and $\theta_{13}$, 
and the known $\theta^{\nu}_{12}$ and $\theta^{\nu}_{23}$:
\begin{align}
\cos \omega & = 
- \dfrac{\cos^2 \theta^{\nu}_{23} \left( \cos^2 \theta_{13} \sin^2 \theta_{12} - \sin^2 \theta^{\nu}_{12} \right) + \sin^2 \theta_{13} \left( \sin^2 \theta^{\nu}_{12} - \cos^2 \theta^{\nu}_{12} \sin^2 \theta^{\nu}_{23} \right)}
{ {\rm sgn} (\sin 2 \theta^e_{13}) \sin 2 \theta^{\nu}_{12} \sin \theta^{\nu}_{23} \sin \theta_{13} \left(\cos^2 \theta^{\nu}_{23} - \sin^2 \theta_{13}\right)^{1/2}} \,. 
\label{eq:omega1gen}
\end{align}
%
For $\theta^{\nu}_{23} = -\pi/4$ and ${\rm sgn} (\sin 2 \theta^e_{13}) = 1$, 
this sum rule reduces to 
the sum rule for $\cos \omega$ given in eq.~(25) in 
\cite{Girardi:2015vha}.

 As we will see, the expressions for the Majorana phases 
$\alpha_{21}$ and $\alpha_{31}$ we will obtain 
depend on the Dirac phase $\delta$. Therefore we give also 
the sum rule for the Dirac phase $\delta$ in the considered 
case by which $\cos\delta$ is expressed in terms of the 
measured angles $\theta_{12}$ and $\theta_{13}$
of the standard parametrisation of the PMNS matrix 
\cite{Girardi:2015vha}:
\begin{align} 
\cos\delta & = -\,\frac{(\cos 2 \theta_{13} 
+ \cos 2 \theta^{\nu}_{23})^{\frac{1}{2}}}
{\sqrt{2}\sin2\theta_{12}\sin\theta_{13}|\sin\theta^{\nu}_{23}|}\,
\bigg[ \cos2\theta^{\nu}_{12} \nonumber \\ 
& + \left (\sin^2\theta_{12} - \cos^2\theta^{\nu}_{12} \right )\,
\frac{2\cos^2 \theta^{\nu}_{23} - (3-\cos 2 \theta^{\nu}_{23}) \sin^2\theta_{13}}
{\cos 2 \theta_{13} + \cos 2 \theta^{\nu}_{23}} \bigg]\,.
\label{eq:cosdelta13e23nu12nugen}
\end{align}
%

 Equating the expressions for the rephasing invariant 
associated with the Dirac phase in the PMNS matrix, $J_{\rm CP}$,
obtained in the standard parametrisation and in the 
parametrisation given in eq.~(\ref{eq:U13e23nu12nu}) 
allows us to get a relation between $\sin\delta$ 
and $\sin\omega$:
\be
\sin \delta = {\rm sgn} (\sin 2\theta^e_{13} \sin \theta^\nu_{23}) 
\frac{\sin 2\theta^\nu_{12}}{\sin 2\th_{12}} \sin \omega\,.
\ee
%

 As can be shown using the method developed in 
\cite{Petcov:2014laa} and employed in the preceding subsection, 
the phases $\delta$, $\alpha_{21}/2$ and $\alpha_{31}/2$ 
are related with the phase $\omega$ and the phases 
$\beta_{e1}$ and $\beta_{e2}$, 
\begin{align}
\label{betae1Re13}
\beta_{e1} &= \arg\left(U_{e1}\right) = 
\arg\left(c^e_{13} c^\nu_{12} + s^e_{13} s^\nu_{23} s^\nu_{12} e^{-i \omega}\right)\,,\\
\beta_{e2} &= \arg(U_{e2}\, e^{-i\frac{\xi_{21}}{2}}) = 
\arg\left(c^e_{13} s^\nu_{12} - s^e_{13} s^\nu_{23} c^\nu_{12} e^{-i \omega}\right)\,, 
\label{betae2Re13}
\end{align}
in the following way:
\begin{align}
\delta &= \omega + \beta_{e1} + \beta_{e2} + \arg\left(s^e_{13} c^e_{13} s^\nu_{23}\right)\,, \label{deltaRe13}\\
\frac{\alpha_{21}}{2} &= \beta_{e2} - \beta_{e1} + 
\frac{\xi_{21}}{2}\,, \label{alpha21Re13}\\
\frac{\alpha_{31}}{2} &= \beta_{e2} + 
\frac{\xi_{31}}{2} + \arg\left(c^e_{13} s^\nu_{23} c^\nu_{23}\right)\,. 
\label{alpha31Re13}
\end{align}
%
From eqs.~(\ref{deltaRe13})\,--\,(\ref{alpha31Re13}) we get 
a relation analogous to that in eq.~(\ref{Majxidpsi}) 
in the preceding subsection:
\begin{equation}
\left(\alpha_{31} - \xi_{31}\right) - \frac{1}{2} \left(\alpha_{21} - \xi_{21}\right) = 
\beta_{e1} + \beta_{e2} = \delta - \omega - \arg\left(s^e_{13} c^e_{13} s^\nu_{23}\right)\,,
\end{equation}
%
where we took into account that 
$2\arg\left(c^e_{13} s^\nu_{23} c^\nu_{23}\right) = 0$ or $2\pi$. 

Equation~(\ref{eq:th13B0gen}) allows one to express $s^e_{13}$ and $c^e_{13}$ 
(given their signs) in terms of $\sin\theta_{13}$ and $\cos\theta^\nu_{23}$.
The phase $\omega$ is determined by the angles $\theta_{12}$, $\theta_{13}$, 
$\theta^\nu_{12}$ and $\theta^\nu_{23}$ via eq.~(\ref{eq:omega1gen}) 
(up to an ambiguity of the sign of $\sin\omega$). 
Thus, using eqs.~(\ref{betae1Re13}) and (\ref{betae2Re13}), 
the phases $\beta_{e1}$ and $\beta_{e2}$ 
can be expressed in terms of the measured mixing 
angles $\theta_{12}$ and $\theta_{13}$ and the 
angles $\theta^\nu_{12}$ and $\theta^\nu_{23}$ fixed by symmetry 
arguments.

It is not difficult to derive expressions for $\beta_{e1}$ and $\beta_{e2}$ 
in terms of the angles $\theta_{12}$, $\theta_{13}$, $\theta_{23}$ and 
the phase $\delta$ of the standard parametrisation of the PMNS matrix.
They read:
\begin{align}
\beta_{e1} &= \arg\left(U_{\mu2}\, {\rm sgn}\left(c^e_{13} c^\nu_{12}\right) 
e^{-i \frac{\alpha_{21}}{2}}\right) = 
\arg\left[\left(c_{12} c_{23} - s_{12} s_{23} s_{13} e^{i \delta}\right)
{\rm sgn}\left(c^e_{13} c^\nu_{12}\right)\right]\,,\\
\beta_{e2} &= \arg\left(U_{\mu1}\, e^{i \pi} {\rm sgn}\left(c^e_{13} s^\nu_{12}\right)\right) =
\arg\left[\left(s_{12} c_{23} + c_{12} s_{23} s_{13} e^{i \delta}\right)
{\rm sgn}\left(c^e_{13} s^\nu_{12}\right)\right]\,.
\end{align}
%

From eqs.~(\ref{betae1Re13})~--~(\ref{alpha31Re13}) 
it is not difficult to get the  following results for, e.g.,  
$\sin (\alpha_{21}/2)$,  $\sin (\alpha_{31}/2 - \delta)$ and 
$\sin (\alpha_{31}/2)$ in the case of arbitrary fixed values of 
the phases $\xi_{21}$ and $\xi_{31}$  in $Q_0$:
\begin{align}
\sin (\alpha_{21}/2) & = \dfrac{1}{\cos^2 \theta_{13}  \sin 2 \theta_{12}}
\bigg[ -\,\sin 2 \theta^e_{13} \sin \theta^{\nu}_{23} 
\big( \sin(\xi_{21}/2 - \omega) - 
2\sin^2\theta^{\nu}_{12}\sin(\xi_{21}/2)\cos\omega \big ) 
\nonumber 
\label{eq:alpha21B0gen} \\
&  + \sin(\xi_{21}/2) \sin 2 \theta^{\nu}_{12} 
\big(\cos^2 \theta^e_{13} -  \sin^2 \theta^{\nu}_{23} \sin^2 \theta^e_{13} \big) 
\bigg] \,, 
\\
\sin (\alpha_{31}/2 - \delta) & = \dfrac{2 \sin \theta^e_{13} \cos \theta^{\nu}_{23}}{\cos \theta_{12}  \sin 2 \theta_{13}} \bigg[
\cos \theta^e_{13}  \cos \theta^{\nu}_{12} \sin(\xi_{31}/2 - \omega) +  \sin \theta^{\nu}_{23} \sin \theta^e_{13} \sin \theta^{\nu}_{12} \sin (\xi_{31}/2) \bigg] \,,
\label{eq:alpha31dB0gen}
\end{align}
%
\begin{equation}
\sin(\alpha_{31}/2) = \frac{{\rm sgn}\left(c^e_{13} s^\nu_{23} c^\nu_{23}\right)}
{\sin\theta_{12} \cos\theta_{13}}
\bigg[-\sin\theta^e_{13} \cos\theta^\nu_{12} \sin\theta^\nu_{23} \sin(\xi_{31}/2 
- \omega) + 
\cos\theta^e_{13} \sin\theta^\nu_{12} \sin(\xi_{31}/2)\bigg]\,,
\label{eq:alpha31B0gen}
\end{equation}
%
where $\theta^e_{13}$ and $\omega$ are given in 
eqs.~(\ref{eq:th13B0gen}) and (\ref{eq:omega1gen}), respectively.
The formulae for $\cos (\alpha_{21}/2)$, $\cos (\alpha_{31}/2 - \delta)$ 
and  $\cos (\alpha_{31}/2)$ can be obtained
from eqs.~(\ref{eq:alpha21B0gen})~--~(\ref{eq:alpha31B0gen}) by changing 
$\xi_{21}$ to $\xi_{21} + \pi$ and $\xi_{31}$ to $\xi_{31} + \pi$, respectively.
The results for $\sin (\alpha_{21}/2)$ and 
$\sin (\alpha_{31}/2 - \delta)$ can also be obtained by 
equating the expressions for the rephasing invariants $I_1$
and $I_2$ related to the Majorana phases, 
derived in the parametrisation of the PMNS matrix 
in eq. (\ref{eq:U13e23nu12nu}), with those given respectively 
in  eqs.~(\ref{eq:StI1}) and (\ref{eq:StI2}).

  It proves convenient for the calculation of the Majorana phases 
to use expressions of  $\sin(\alpha_{21}/2)$ and $\sin(\alpha_{31}/2)$ 
in terms of  the standard parametrisation mixing angles 
$\theta_{12}$, $\theta_{13}$, $\theta_{23}$, the Dirac phase $\delta$, 
and the angles $\theta^\nu_{12}$ and $\theta^\nu_{23}$ fixed by 
symmetries. The expressions of interest 
are not difficult to derive and they read: 
\begin{align}
\sin(\alpha_{21}/2) &= \frac{1}{\sin2\theta^\nu_{12} \cos^2\theta^\nu_{23}}
\bigg[-\sin2\theta_{23} \sin\theta_{13} 
\left(\sin(\xi_{21}/2 - \delta) -2 \cos^2\theta_{12} \cos\delta \sin(\xi_{21}/2)\right) \nonumber\\
&+ \sin2\theta_{12} \left(\cos^2\theta_{23} - \sin^2\theta_{23} \sin^2\theta_{13}\right)
\sin(\xi_{21}/2)\bigg]\,,\\
\sin(\alpha_{31}/2) &= \frac{{\rm sgn}(s^\nu_{23})}
{\sin\theta^\nu_{12} \cos\theta^\nu_{23}}
\bigg[\sin\theta_{12} \cos\theta_{23} \sin(\xi_{31}/2) + 
\cos\theta_{12} \sin\theta_{23} \sin\theta_{13} \sin(\xi_{31}/2 + \delta)\bigg]\,.
\end{align}
%
We recall that $\sin^2\theta^\nu_{23} = \cos^2\theta_{13} \sin^2\theta_{23}$.

The expressions for 
$\sin(\alpha_{21}/2 -\xi_{21})$, 
$\sin(\alpha_{31}/2 - \xi_{31})$ and 
$\sin(\alpha_{31}/2 - \delta - \xi_{31}/2)$ take the simple forms:
\begin{align}
\sin(\alpha_{21}/2 - \xi_{21}/2) &= 
\dfrac{\sin 2 \theta^e_{13}}{\cos^2 \theta_{13}  \sin 2 \theta_{12}} 
\sin \theta^{\nu}_{23} \sin \omega
= \frac{\sin2\theta_{23} \sin\theta_{13}}{\sin2\theta^\nu_{12} \cos^2\theta^\nu_{23}}
\sin\delta\,, 
\label{eq:alpha21B0}
\\
\sin(\alpha_{31}/2 - \xi_{31}/2) &= \frac{{\rm sgn}\left(c^e_{13} s^\nu_{23} c^\nu_{23}\right)}
{\sin\theta_{12} \cos\theta_{13}}
\sin\theta^e_{13} \cos\theta^\nu_{12} \sin\theta^\nu_{23} \sin\omega \nonumber\\
&= \frac{{\rm sgn}(s^\nu_{23})}{\sin\theta^\nu_{12} \cos\theta^\nu_{23}}
\cos\theta_{12} \sin\theta_{23} \sin\theta_{13} \sin\delta\,, 
\label{eq:alpha31B0*}
\\
\sin(\alpha_{31}/2 - \delta - \xi_{31}/2) &= 
- \dfrac{\sin 2 \theta^e_{13}}{\cos \theta_{12} \sin 2 \theta_{13}} 
\cos \theta^{\nu}_{12}  \cos \theta^{\nu}_{23} \sin \omega =
- {\rm sgn}(s^\nu_{23}) \frac{\sin\theta_{12} \cos\theta_{23}}
{\sin\theta^\nu_{12} \cos\theta^\nu_{23}} \sin\delta\,.
\label{eq:alpha31B0}
\end{align}
%

Equations~(\ref{eq:alpha21B0}), (\ref{eq:alpha31B0*}) and 
(\ref{eq:alpha31B0}) do not allow one
to obtain unique predictions for 
$\sin (\alpha_{21}/2)$, $\sin (\alpha_{31}/2)$ 
and $\sin (\alpha_{31}/2 - \delta)$ because of the 
ambiguity in determining
the sign of $\sin \omega$ ($\sin\delta$).
As in the case discussed in the preceding subsection, 
we have $|\sin(\alpha_{21}/2 - \xi_{21}/2)| \propto \sin\theta_{13}$ and 
$|\sin(\alpha_{31}/2 - \xi_{31}/2)| \propto \sin\theta_{13}$.
Predictions for $(\alpha_{21}/2 - \xi_{21}/2)$ and 
$(\alpha_{31}/2 - \xi_{31}/2)$ in the case studied in this 
subsection will be given in Section~\ref{sec:predictions}.

%
\section{The Cases of $\boldsymbol{(\theta^e_{ij},\theta^e_{kl}) - (\theta^\nu_{23}, \theta^\nu_{12})}$ 
 Rotations }
\label{sec:ijekle23nu12nu}
%
%

As it follows from eqs.~(\ref{eq:th23A0}) and (\ref{eq:th23B0gen}) 
in the preceding Section,
in the cases when the matrix 
$\tilde{U}_e$ originating from the charged lepton sector
contains one rotation angle ($\theta^e_{12}$ or $\theta^e_{13}$)
and  $\theta^{\nu}_{23} = -\pi/4$, 
the mixing angle $\theta_{23}$ cannot deviate 
significantly from $\pi/4$ due to the smallness of 
the angle $\theta_{13}$.  If the matrix 
$\tilde{U}_{\nu}$ has one of the symmetry forms 
considered in this study,
the matrix $\tilde{U}_e$ has to contain at least
two rotation angles in order to be possible to 
reproduce the current best fit values
of the neutrino mixing parameters 
quoted in eqs.~(\ref{th12values})~--~(\ref{th13values}), 
or more generally, in order to be possible 
to account for deviations of 
$\sin^2\theta_{23}$ from 0.5 which are bigger 
than $\sin^2\theta_{13}$, i.e., for 
$\sin^2\theta_{23}\neq 0.5(1 \mp \sin^2\theta_{13})$.
In this Section we consider the 
determination of the Majorana phases 
$\alpha_{21}$ and $\alpha_{31}$ 
in the cases when the matrix $\tilde{U}_e$ contains 
two rotation angles.
%
\subsection{The Scheme with $\boldsymbol{(\theta^e_{12},\theta^e_{23}) - 
(\theta^\nu_{23}, \theta^\nu_{12})}$ Rotations (Case B1)}
\label{sec:12e23e23nu12nu}
%
%

The PMNS matrix in this scheme has the form
\begin{equation}
U = 
R_{12}(\theta^e_{12}) \,  R_{23}(\theta^e_{23})  \, \Psi \, 
R_{23}(\theta^{\nu}_{23}) \, R_{12}(\theta^{\nu}_{12}) \, Q_0 \,.
\label{eq:Ue1223Unu2312} 
\end{equation}
%
The scheme has been analysed in detail in \cite{Petcov:2014laa}, 
where a sum rule for $\cos \delta$ and analytic expressions for 
$\alpha_{21}$ and  $\alpha_{31}$ were derived 
for $\theta^{\nu}_{23} = - \pi/4$. As was shown in \cite{Girardi:2015vha}, 
the sum rule for $\cos \delta$ found in \cite{Petcov:2014laa}  
holds for an arbitrary fixed value of $\theta^{\nu}_{23}$.
The sum rule under discussion,  eq.~(30) in \cite{Petcov:2014laa}, 
coincides with the sum rule given in eq.~(\ref{cosdthnu}) in 
subsection~\ref{sec:12e23nu12nu}. However, in contrast to the case 
considered in subsection~\ref{sec:12e23nu12nu}, 
the PMNS mixing angle $\theta_{23}$ in the scheme 
under discussion can differ significantly from $\theta^\nu_{23}$ 
and from $\pi/4$: 
\begin{align}
\sin^2 \theta_{23} & = \frac{|U_{\mu3}|^2}{1-|U_{e3}|^2} = 
\frac{\sin^2 \hat \theta_{23}-\sin^2 \theta_{13}}{1 - \sin^2 \theta_{13}} 
\label{eq:th23B1}\,,
\end{align}
%
where
\begin{align}
\sin\hat \theta_{23} &= 
\left |\,\text{e}^{ -i \psi} \cos\theta^e_{23} \sin\theta^{\nu}_{23}
+ \text{e}^{-i \omega}\sin\theta^e_{23} \cos\theta^{\nu}_{23} \right |\,,
\nonumber
\\
\cos\hat \theta_{23} &= 
\left |\,\text{e}^{ -i \psi} \cos\theta^e_{23} \cos\theta^{\nu}_{23}
- \text{e}^{-i \omega}\sin\theta^e_{23} \sin\theta^{\nu}_{23} \right |\,.
\label{eq:hattheta23genB1}
\end{align}
%
In the preceding equations $\sin \hat \theta_{23}$ and 
$\cos\hat \theta_{23}$ 
are expressed in terms of the parameters of the scheme considered,   
defined in eq.~(\ref{eq:Ue1223Unu2312}) for the PMNS matrix.
Obviously, $\sin\hat \theta_{23} > 0$ and $\cos\hat \theta_{23} > 0$.
The parameter $\sin^2\hat\theta_{23}$ enters also into 
the expression for $\sin^2\theta_{13}$:
\be
\sin^2\theta_{13} = |U_{e3}|^2 = 
\sin^2 \theta^e_{12} \sin^2 \hat \theta_{23}\,.
\label{eq:th13the23thh23B1}
\ee
%

 The angle $\hat \theta_{23}$ 
results from the rearrangement of  
the product of matrices 
$R_{23}(\theta^e_{23})  \Psi R_{23}(\theta^{\nu}_{23})$ 
in the expression for $U$ given in eq.~(\ref{eq:Ue1223Unu2312}): 
\begin{equation}
R_{23}( \theta^e_{23})\, \Psi \, R_{23}(\theta^{\nu}_{23}) =
P_1\, \Phi\, R_{23}(\hat\theta_{23})\,Q_1\,.
\label{Phi}
\end{equation}
%
Here 
\begin{equation}
P_1={\rm diag}(1,1, \text{e}^{-\,i \alpha})\,, \quad
\Phi = {\rm diag}(1,\text{e}^{i \phi},1)\,, \quad
Q_1 = {\rm diag} \left(1,1, \text{e}^{i \beta} \right)\,,
\label{PPhitQ}
\end{equation}
%
where
\begin{equation}
\alpha = \gamma + \psi + \omega  \,, \quad
\beta = \gamma - \phi\,,
\label{alphabeta}
\end{equation}
%
and
\begin{align}
\gamma & = \arg \left (\,\text{e}^{ -i \psi}\cos\theta^e_{23}\sin\theta^\nu_{23}
+ \text{e}^{-i \omega}\sin\theta^e_{23}\cos\theta^\nu_{23}\right)\,,
\label{eq:gammaB1}
\\
\phi & = \arg \left (\text{e}^{ -i \psi}\cos \theta^e_{23}\cos\theta^\nu_{23}
- \text{e}^{-i \omega}\sin\theta^e_{23}\sin\theta^\nu_{23}\right)\,.
\label{eq:phiB1}
\end{align}
%
Equations~(\ref{eq:hattheta23genB1}), (\ref{eq:gammaB1}) and (\ref{eq:phiB1}) 
have been derived in \cite{Marzocca:2013cr}.

 The phase $\alpha$ in the matrix $P_1$ is unphysical.
The phase $\beta$ contributes
to the matrix of physical Majorana phases,
which now is equal to $\hat{Q} = Q_1\,Q_0$.
The phase $\phi$ serves as source for the Dirac phase $\delta$ 
and gives contributions also to the Majorana phases 
$\alpha_{21}$ and $\alpha_{31}$ \cite{Petcov:2014laa}. 
The PMNS matrix takes the form
\begin{equation}
U =
R_{12}(\theta^e_{12})\,\Phi(\phi)\, R_{23}(\hat\theta_{23})\,
R_{12}(\theta^{\nu}_{12})\,\hat{Q}\,,
\label{UPMNSthhat1}
\end{equation}
%
where $\theta^{\nu}_{12}$ has a fixed value 
which depends on the symmetry form of $\tilde{U}_\nu$ used.

 Before continuing further we note that we can consider 
both $\sin\theta^e_{12}$ and $\cos\theta^e_{12}$
to be positive without loss of generality.
Only their relative sign is physical.
If $\sin\theta^e_{12} > 0$ ( $\sin\theta^e_{12} < 0$) and 
$\cos\theta^e_{12} < 0$ ( $\cos\theta^e_{12} > 0$),
the negative sign can be absorbed in the phase $\phi$ 
by adding $\pm \pi$ to $\phi$. Similarly, 
we can consider both 
$\sin\theta^\nu_{12}$ and $\cos\theta^\nu_{12}$
to be positive: the negative signs of 
$\sin\theta^\nu_{12}$ and/or $\cos\theta^\nu_{12}$
can be absorbed in the phases
 $\xi_{21}/2$, $\xi_{31}/2$ and $\phi$~%
\footnote{If $\sin\theta^\nu_{12} < 0$ and $\cos\theta^\nu_{12} < 0$,
getting rid of the negative signs 
of $\sin\theta^\nu_{12}$ and $\cos\theta^\nu_{12}$
leads only to the change  $\xi_{31}/2 \rightarrow \xi_{31}/2 \pm \pi$.
If, however,  $\sin\theta^\nu_{12}\cos\theta^\nu_{12} < 0$,
the relevant negative signs can be absorbed 
in  $\xi_{21}/2$, $\xi_{31}/2$ and $\phi$, each of three phases being  
modified by $\pm \pi$.}. 
Nevertheless, for convenience of using  our 
results for making predictions in theoretical models in 
which the value of, e.g.,  $|\sin\theta^e_{12}|$
and the signs of $\sin\theta^e_{12}$ and $\cos\theta^e_{12}$ 
are specified, we will present the results for 
arbitrary signs of $\sin\theta^e_{12}$ and $\cos\theta^e_{12}$.

  The analytic results on the Majorana phases $\alpha_{21}$ and 
 $\alpha_{31}$, on the relation between the Dirac phase $\delta$ and 
the phase $\phi$, etc., derived in \cite{Petcov:2014laa}, 
do not depend explicitly on the value of the angle 
$\theta^{\nu}_{23}$ and are valid in the case under 
consideration. Thus, generalising 
eqs.~(88)~--~(91), (94) and (102) in 
\cite{Petcov:2014laa} for arbitrary sings of
$s^e_{12}$, $c^e_{12}$, $s^\nu_{12}$ and $c^\nu_{12}$, 
we have:
\begin{equation}
\frac{\alpha_{21}}{2} = \beta_{e2} - \beta_{e1} +\frac{\xi_{21}}{2}\,,~~  
\frac{\alpha_{31}}{2} = \beta_{e2} + \beta_{\mu3} - \phi + \beta + \frac{\xi_{31}}{2}\,,
\label{Majph2131B1}
\end{equation}
%
\begin{equation}
\delta = \beta_{e1} + \beta_{e2} + \beta_{\mu3} - \beta_{e3} -\phi\,,
\label{dphibetaB1}
\end{equation}
%
where
\begin{align}
\label{be1B1}
\beta_{e1} & = \arg\left(U_{e 1}\right) =  
\arg\left (c^e_{12} c^\nu_{12} - 
 s^e_{12}\hat{c}_{23} s^\nu_{12} e^{i \phi} \right )\,, \\[0.2cm]
\label{be2B1}
\beta_{e2} & =  \arg\left(U_{e 2}\,e^{-i\frac{\xi_{21}}{2}}\right) =  
\arg\left( c^e_{12} s^\nu_{12} 
+ s^e_{12}\hat{c}_{23} c^\nu_{12} e^{i \phi} \right )\,, \\[0.2cm]
\label{be3B1}
\beta_{e3} & = \arg\left(U_{e 3} e^{-i\left(\beta + \frac{\xi_{31}}{2}\right)}\right) = 
\arg\left(s^e_{12}\right) + \phi\,, \\[0.2cm]
\label{bmu3B1}
\beta_{\mu3} & = \arg\left(U_{\mu3} e^{-i\left(\beta + \frac{\xi_{31}}{2}\right)}\right) = 
\arg\left(c^e_{12}\right) + \phi\,,
\end{align}
%
with  $\hat{c}_{23} \equiv \cos\hat\theta_{23}$.
The preceding results can be obtained by 
casting $U$ in eq.~(\ref{UPMNSthhat1}) 
in the standard parametrisation form.
This leads, in particular, to additional 
contribution to the matrix $\hat{Q}$  of the 
Majorana phases, which takes the form
$\hat{Q} = Q_2\, Q_1\,Q_0$, where the generalisation 
of the corresponding expression for $Q_2$ 
in \cite{Petcov:2014laa} reads:
$Q_2 = {\rm diag} \left(1, e^{i (\beta_{e2} - \beta_{e1})}, 
e^{i (\beta_{e2} +\beta_{\mu3} - \phi)}\right)$. 
Note that we got rid of the common unphysical phase factor
$e^{-i (\beta_{e2} + \beta_{\mu3} - \phi)}$ in the matrix $Q_2$.

 The expressions for the phases 
$(\beta_{e2} + \beta_{\mu3} - \phi)$ and $(\beta_{e1} + \beta_{\mu3} - \phi)$ 
in terms of the angles $\theta_{12}$, $\theta_{13}$, $\theta_{23}$ 
and the Dirac phases $\delta$ of the standard parametrisation 
of the PMNS matrix have the form 
(cf. eqs.~(100) and (101) in ref. \cite{Petcov:2014laa}):
\begin{align}
\label{be2tau1phB1}
\beta_{e2} + \beta_{\mu3} - \phi & = 
\arg\left(U_{\tau 1}\right) - \beta_{\tau1} = 
\arg\left(s_{12} s_{23} - c_{12} c_{23} s_{13} e^{i \delta}\right) - \beta_{\tau1}\,,
\\[0.25cm]
\label{be1tau2phB1}
\beta_{e1} + \beta_{\mu3} - \phi & = 
\arg\left (U_{\tau 2} e^{-i\frac{\alpha_{21}}{2}}\right) - \beta_{\tau2}=   
\arg\left(- c_{12} s_{23} - s_{12} c_{23} s_{13} e^{i \delta}\right) - \beta_{\tau2}\,,
\end{align}
%
where 
\begin{equation}
\beta_{\tau1} = \arg(s^\nu_{12})\,, \quad 
\beta_{\tau2} = \arg(- c^\nu_{12})\,.
\end{equation}
%

We also have 
\begin{align}
\label{sindsinphiB1}
\sin\delta =
&\, - {\rm sgn} \left(\sin2\theta^e_{12}\right)
\frac{\sin2\theta^{\nu}_{12}}{\sin2\theta_{12}}\,\sin\phi\,.
\end{align}
%

A few comments are in order.
As like the cosine of the  Dirac phase $\delta$, 
$\cos\phi$ satisfies a sum rule by which it 
is expressed in terms of the three measured 
neutrino mixing angles $\theta_{12}$, $\theta_{13}$ 
and $\theta_{23}$, and is uniquely determined 
by the values of $\theta_{12}$, $\theta_{13}$ 
and $\theta_{23}$ \cite{Petcov:2014laa}.
The values of $\sin\delta$ and $\sin\phi$, however, 
are fixed up to a sign. 
Through eq.~(\ref{sindsinphiB1})
the signs $\sin\delta$ and $\sin\phi$ 
are correlated.
Thus, $\delta$ and $\phi$
are predicted with an ambiguity related to the 
ambiguity of the sign of 
$\sin\delta$ (or of $\sin\phi$).
Together with eqs.~(\ref{be2tau1phB1}) 
and  (\ref{be1tau2phB1}) this  implies that also 
the phases $\beta_{e 1}$ and $\beta_{e 2}$ 
are determined by the values of
$\theta_{12}$, $\theta_{13}$, $\theta_{23}$ and 
$\delta$ with a two-fold ambiguity.
The knowledge of the difference $(\beta_{e 2} - \beta_{e 1})$ 
allows to determine the Majorana phase 
$\alpha_{21}$ (up to the discussed two-fold ambiguity) 
if the value of the phase $\xi_{21}$ is known.   
In contrast, the knowledge of $\beta_{e 2}$ 
and $\xi_{31}$ is not enough to 
predict the value of the Majorana phase 
$\alpha_{31}$ since it receives a contribution 
also from the phase $\beta$ that cannot be fixed 
on general phenomenological grounds.
It is possible to determine the phase $\beta$
in certain specific cases (see \cite{Petcov:2014laa} for a 
detailed discussion of the cases when $\beta$ can be fixed).  
It should be noted, however, that the 
term involving the phase $\alpha_{31}$ 
in the $\betabeta$-decay effective Majorana mass 
$\mefff$ gives practically a negligible contribution 
in $\meff$ in the cases of neutrino mass spectrum with 
IO or of quasi-degenerate (QD) type 
\cite{BPP1,Petcov:2014laa}. 
In these cases we have \cite{PPSNO2bb} $\meff \gtap 0.014$~eV 
(see also, e.g., \cite{PDG2014,Petcov:2013poa}).
Values of  $\meff \gtap 0.014$~eV are in the range 
of planned sensitivity of the future large scale 
$\betabeta$-decay experiments 
(see, e.g., \cite{Dell'Oro:2016dbc}).

 Using eqs.~(\ref{Majph2131B1})~--~(\ref{bmu3B1}),
we can derive analytical formulae   
for $\sin (\alpha_{21}/2)$, $\sin (\alpha_{31}/2 - \delta)$ and
$\sin (\alpha_{31}/2)$ in terms of the parameters of 
$U$ given in eq.~(\ref{UPMNSthhat1}).
For arbitrary fixed values of the phases $\xi_{21}$ 
and $\xi_{31}$ we get:
\begin{align}
\sin (\alpha_{21}/2) & = \dfrac{1}{2|U_{e1}U_{e2}|}\,
\bigg[ \sin 2 \theta^e_{12} \cos \hat \theta_{23} 
\Big (\sin (\phi + \xi_{21}/2) - 
2\sin^2\theta^{\nu}_{12}\cos \phi  \sin (\xi_{21}/2) \Big )
\nonumber \\
& + \sin 2 \theta^{\nu}_{12} \sin (\xi_{21}/2)
\Big (\cos^2 \theta^e_{12} - \sin^2\theta^e_{12}\cos^2\hat\theta_{23}\Big ) 
\bigg] \,,
\label{eq:alpha21gen12e23eB1} \\
\sin(\alpha_{31}/2 - \delta) & = \frac{1}{|U_{e1}|}\,   
\bigg(\cos\theta^e_{12}\cos\theta^{\nu}_{12}\sin(\beta + \xi_{31}/2 + \phi) 
- \cos\hat\theta_{23}\sin\theta^e_{12}\sin\theta^{\nu}_{12} 
\sin(\beta + \xi_{31}/2)
\bigg)\,, 
\label{eq:alpha31dgen12e23eB1}\\
\sin (\alpha_{31}/2 ) & = \frac{1}{|U_{e2}|}\,
\bigg( \cos \theta^e_{12} \sin \theta^{\nu}_{12} \sin(\beta + \xi_{31}/2) 
+ \cos \hat \theta_{23} \sin \theta^e_{12} \cos \theta^{\nu}_{12} 
\sin( \phi + \beta + \xi_{31}/2) \bigg)\,,
\label{eq:alpha31gen12e23eB1}
\end{align}
%
where $U_{e1}$ and $U_{e2}$ are given in eqs.~(\ref{be1B1}) and 
(\ref{be2B1}). In the standard parametrisation 
of $U$ we have, as is well known,
$|U_{e1}| = \cos\theta_{12}\cos\theta_{13}$ 
and $|U_{e2}| = \sin\theta_{12}\cos\theta_{13}$.
The results for $\sin (\alpha_{21}/2)$, $\sin (\alpha_{31}/2 - \delta)$  
can also be obtained by comparing the expressions for the 
rephasing invariants $I_1$ and $I_2$ in the standard parametrisation 
of the PMNS matrix and in the parametrisation of 
$U$ in eq.~(\ref{UPMNSthhat1}). 

 In terms of the ``standard'' angles 
$\theta_{12}$, $\theta_{13}$, $\theta_{23}$ and the phase $\delta$, 
 $\sin (\alpha_{21}/2)$ and $\sin(\alpha_{31}/2)$ are given by
\begin{align}
\sin(\alpha_{21}/2) &= \frac{\cos(\beta_{\tau2} - \beta_{\tau1})}{2|U_{\tau1}U_{\tau 2}|}
\bigg[\sin2\theta_{23} \sin\theta_{13} 
\left(\sin( \delta + \xi_{21}/2) 
- 2 \sin^2\theta_{12} \cos\delta \sin(\xi_{21}/2)\right) 
\nonumber\\
&- \sin2\theta_{12} \left(\sin^2\theta_{23} 
- \cos^2\theta_{23} \sin^2\theta_{13}\right)
\sin(\xi_{21}/2)\bigg]\,,
\label{eq:alpha21genstandB1}
\\
\sin(\alpha_{31}/2) &= \frac{\cos\beta_{\tau1}}{|U_{\tau 1}|}
\bigg[\sin\theta_{12} \sin\theta_{23} \sin(\xi_{31}/2 + \beta) 
- \cos\theta_{12} \cos\theta_{23} \sin\theta_{13} 
\sin(\delta + \beta + \xi_{31}/2)\bigg]\,.
\label{eq:alpha31genstandB1}
\end{align}
%
In the parametrisation of $U$
defined in eq.~(\ref{UPMNSthhat1}) one has 
$|U_{\tau1}| = |\sin\theta^{\nu}_{12}| \sin\hat \theta_{23}$ and 
$|U_{\tau 2}| = |\cos\theta^{\nu}_{12}| \sin\hat \theta_{23}$.
The sign factors $\cos(\beta_{\tau2} - \beta_{\tau1})$ and $\cos\beta_{\tau1}$ 
are known once the angle $\theta^\nu_{12}$ is fixed: 
\begin{equation}
\cos(\beta_{\tau2} - \beta_{\tau1}) = -\,{\rm sgn} \left(s^\nu_{12} c^\nu_{12}\right)\,,
\quad
\cos\beta_{\tau1} = {\rm sgn} \left(s^\nu_{12}\right)\,.
\end{equation}
%

 The expressions for $\cos(\alpha_{21}/2)$ and $\cos (\alpha_{31}/2)$
can be obtained by replacing $\xi_{21}$ with $\xi_{21} + \pi$
and $\xi_{31}$ with $\xi_{31} + \pi$
in eq.~(\ref{eq:alpha21genstandB1})  and 
eq.~(\ref{eq:alpha31genstandB1}), respectively.

The expressions for  
$\sin(\alpha_{21}/2 -\xi_{21}/2)$, 
$\sin(\alpha_{31}/2 - \xi_{31}/2 -\beta)$ and 
$\sin(\alpha_{31}/2 - \delta - \xi_{31}/2 - \beta)$ 
have the following simple forms:
\begin{align}
\sin (\alpha_{21}/2 - \xi_{21}/2) 
& = \,\dfrac{\sin 2 \theta^e_{12}\,\cos \hat{\theta}_{23}}
{\cos^2 \theta_{13}\sin 2 \theta_{12}}\, \sin \phi \nonumber\\
&= \dfrac{\cos(\beta_{\tau2} - \beta_{\tau1})}
{2 |U_{\tau 1} U_{\tau 2}|}
\sin2\theta_{23} \sin\theta_{13} \sin\delta\,,
\label{eq:alpha21xi21B1}\\
\sin (\alpha_{31}/2 - \delta - \xi_{31}/2 - \beta) & = 
\,\dfrac{\sin 2\theta^e_{12}\cos \theta^{\nu}_{12}}
{\cos \theta_{12}\sin 2 \theta_{13}}\,  
\sin \hat{\theta}_{23}\, \sin \phi \nonumber\\
&= - \dfrac{\cos\beta_{\tau1}}{|U_{\tau 1}|}
\sin\theta_{12} \sin\theta_{23} \sin\delta\,, 
\label{eq:alpha31xi31betadB1}\\
\sin (\alpha_{31}/2 - \xi_{31}/2 - \beta) & = 
\, \dfrac{\sin \theta^e_{12} \cos \theta^\nu_{12}}
{\sin \theta_{12}  \cos \theta_{13}}\, 
\cos \hat{\theta}_{23}\,\sin\phi \nonumber\\
&= - \dfrac{\cos\beta_{\tau1}}{|U_{\tau 1}|}
\cos \theta_{12}\cos \theta_{23} \sin \theta_{13} \sin\delta\,.
\label{eq:alpha31xi31betaB1}
\end{align}
%

It follows from eqs.~(\ref{eq:alpha21xi21B1})~--~(\ref{eq:alpha31xi31betaB1}) 
that since 
$\sin\delta$ can be expressed in terms of the ``standard''
neutrino mixing angles $\theta_{12}$,  $\theta_{23}$ and 
$\theta_{13}$, $\sin(\alpha_{21}/2 -\xi_{21}/2)$, 
$\sin(\alpha_{31}/2 - \xi_{31}/2 -\beta)$ and 
$\sin(\alpha_{31}/2 - \delta - \xi_{31}/2 - \beta)$ 
are determined (up to an ambiguity related to 
the sign of $\sin\delta$) by the values of 
$\theta_{12}$,  $\theta_{23}$ and 
$\theta_{13}$.
Equations~(\ref{eq:alpha21xi21B1}) and 
(\ref{eq:alpha31xi31betaB1}) imply that also in the 
discussed case $|\sin(\alpha_{21}/2 - \xi_{21}/2 )|\propto \sin\theta_{13}$ 
and $|\sin(\alpha_{31}/2 - \xi_{31}/2 - \beta)|\propto \sin\theta_{13}$.
Predictions for the phases 
$(\alpha_{21}/2 - \xi_{21}/2)$ and 
$(\alpha_{31}/2 - \xi_{31}/2 - \beta)$ in the case 
considered in the present subsection  
will be given in Section~\ref{sec:predictions}.

%
\subsection{The Scheme with $\boldsymbol{(\theta^e_{13},\theta^e_{23}) - 
(\theta^\nu_{23}, \theta^\nu_{12})}$ Rotations (Case B2)}
\label{sec:13e23e23nu12nu}
%
%

In this subsection we consider the parametrisation of the PMNS
matrix as in eq.~(\ref{eq:Uijkl}) with $(ij) - (kl) = (13) - (23)$.
Analogously to the previous subsection, 
this parametrisation can be recast in the form
\begin{equation}
U =
R_{13}(\theta^e_{13})\,P_1(\alpha)\,R_{23}(\hat\theta_{23})\,
R_{12}(\theta^{\nu}_{12})\,\hat{Q}\,,
\label{UPMNSthhat1-13}
\end{equation}
%
where the angle $\hat\theta_{23}$ and the matrix $P_1$ 
are given by eqs.~(\ref{eq:hattheta23genB1}) and (\ref{PPhitQ}), respectively, 
and $\hat Q = Q_1\,Q_0$ with $Q_1$ as in eq.~(\ref{PPhitQ}). 
In explicit form eq.~(\ref{UPMNSthhat1-13}) reads: 
\begin{equation} 
\label{Uthenuhat-13}
U = \begin{pmatrix} 
c^e_{13} c^\nu_{12} + s^e_{13} \hat{s}_{23} s^\nu_{12} e^{- i \alpha}& 
c^e_{13} s^\nu_{12} - s^e_{13} \hat{s}_{23} c^\nu_{12} e^{- i \alpha}& 
s^e_{13} \hat{c}_{23} e^{- i \alpha}  \\[0.2cm] 
-\hat{c}_{23} s^\nu_{12}& 
\hat{c}_{23} c^\nu_{12}& 
\hat{s}_{23}& \\[0.2cm] 
- s^e_{13} c^\nu_{12} + c^e_{13} \hat{s}_{23} s^\nu_{12} e^{- i \alpha}& 
- s^e_{13} s^\nu_{12} - c^e_{13} \hat{s}_{23} c^\nu_{12} e^{- i \alpha}& 
c^e_{13} \hat{c}_{23} e^{- i \alpha}
\end{pmatrix}
\hat Q\,.
\end{equation}
%
To bring this matrix to the standard parametrisation form, 
we first rewrite it as follows: 
\begin{equation} 
\label{|U|phases-13}
U = \begin{pmatrix} 
|U_{e1}| e^{i \beta_{e1}} & |U_{e2}| e^{i \beta_{e2}} & |U_{e3}| e^{i \beta_{e3}} \\[0.2cm]  
|U_{\mu1}| e^{i \beta_{\mu1}} &  |U_{\mu2}| e^{i \beta_{\mu2}} & |U_{\mu3}| \\[0.2cm]
|U_{\tau1}| e^{i \beta_{\tau1}}& |U_{\tau2}| e^{i \beta_{\tau2}} & |U_{\tau3}| e^{i \beta_{\tau3}}
\end{pmatrix}
\hat Q\,,
\end{equation}
%
where
\begin{align}
\label{be1B2}
\beta_{e1} & = 
\arg\left(c^e_{13} c^\nu_{12} + s^e_{13} \hat{s}_{23} s^\nu_{12} e^{- i \alpha}\right )\,, 
\\[0.2cm]
\label{be2B2}
\beta_{e2} & = 
\arg\left(c^e_{13} s^\nu_{12} - s^e_{13} \hat{s}_{23} c^\nu_{12} e^{- i \alpha}\right )\,, 
\\[0.2cm]
\label{be3B2}
\beta_{e3} & = 
\arg\left(s^e_{13}\right ) - \alpha\,, 
\\[0.2cm]
\label{bmu1B2}
\beta_{\mu1} & = 
\arg\left(- s^\nu_{12}\right )\,, 
\\[0.2cm]
\label{bmu2B2}
\beta_{\mu2} & = 
\arg\left(c^\nu_{12}\right )\,, 
\\[0.2cm]
\label{btau1B2}
\beta_{\tau1} & = 
\arg\left(- s^e_{13} c^\nu_{12} + c^e_{13} \hat{s}_{23} s^\nu_{12} e^{- i \alpha}\right )\,, 
\\[0.2cm]
\label{btau2B2}
\beta_{\tau2} & = 
\arg\left(- s^e_{13} s^\nu_{12} - c^e_{13} \hat{s}_{23} c^\nu_{12} e^{- i \alpha}\right )\,,
\\[0.2cm]
\label{btau3B2}
\beta_{\tau3} & = 
\arg\left(c^e_{13}\right) - \alpha\,.
\end{align}
%
We recall that the angle $\hat\theta_{23}$ belongs to the first quadrant 
by construction (see eq.~(\ref{eq:hattheta23genB1})).

 Further, comparing the expressions for the $J_{\rm CP}$ invariant 
in the standard parametrisation and in the parametrisation given 
in eq.~(\ref{UPMNSthhat1-13}), we have~%
\footnote{This relation is the generalisation of eq.~(43) 
in ref.~\cite{Girardi:2015vha}, where we considered 
$\theta^e_{13}$ to be in the first quadrant.} 
\begin{equation}
\sin\delta = {\rm sgn} \left(\sin2\theta^e_{13}\right)
\frac{\sin2\theta^\nu_{12}}{\sin2\theta_{12}} \sin\alpha\,.
\label{sindsinalphaB2}
\end{equation}
%
It is not difficult to check that this relation holds if
\begin{equation}
\delta = \beta_{e1} + \beta_{e2} + \beta_{\tau3} - \beta_{e3} + \alpha\,, \quad
\beta_{\tau3} - \beta_{e3} = 0~{\rm or}~\pi\,,
\label{dB2}
\end{equation}
%
which, in turn, suggests what rearrangement of the phases 
in the PMNS matrix in eq.~(\ref{|U|phases-13}) one has to do to bring it 
to the standard parametrisation form. Namely, the required 
rearrangement should be made in the following way:
\begin{equation} 
\label{Uphases-13}
U = P_2\, 
\begin{pmatrix} 
|U_{e1}| & |U_{e2}|  & |U_{e3}| e^{- i \delta} \\[0.2cm]
|U_{\mu1}| e^{i (\beta_{\mu1} + \beta_{e2} + \beta_{\tau3} + \alpha)} & 
|U_{\mu2}| e^{ i (\beta_{\mu2} + \beta_{e1} + \beta_{\tau3} + \alpha)} & 
|U_{\mu3}| \\[0.2cm]
|U_{\tau1}| e^{ i (\beta_{\tau1} + \beta_{e2} + \alpha)}&
|U_{\tau2}| e^{ i (\beta_{\tau2} + \beta_{e1} + \alpha)}& 
|U_{\tau3}|
\end{pmatrix}
Q_{2}\,\hat Q\,,
\end{equation}
%
where 
\begin{align}
\label{P2}
P_2 &=\diag\left(e^{i (\beta_{e1} + \beta_{e2} + \beta_{\tau3} + \alpha)}, 1, 
e^{i \beta_{\tau3}}\right)\,, \\[0.2cm]
\label{Q2}
Q_2 &= \diag\left(e^{-i (\beta_{e2} + \beta_{\tau3} + \alpha)}, 
e^{-i (\beta_{e1} + \beta_{\tau3} + \alpha)}, 1\right) \nonumber\\
&= e^{-i (\beta_{e2} + \beta_{\tau3} + \alpha)}\,\diag\left(1, 
e^{i (\beta_{e2} - \beta_{e1})}, e^{i (\beta_{e2} + \beta_{\tau3} + \alpha)}\right)\,.
\end{align}
%
The phases in the matrix $P_2$ are unphysical. 
The phases  $(\beta_{e2} - \beta_{e1})$ and 
$(\beta_{e2} + \beta_{\tau3} + \alpha)$ 
in the matrix $Q_2$ contribute to the Majorana phases $\alpha_{21}$ and 
$\alpha_{31}$, respectively, while the common phase 
$(- \beta_{e2} - \beta_{\tau3} - \alpha)$ in 
this matrix is unphysical and we will not keep it further.
Thus, the Majorana phases in the PMNS matrix are determined by the phases 
in the product $Q_2\,\hat Q$:  
\begin{equation}
\frac{\alpha_{21}}{2} = \beta_{e2} - \beta_{e1} +\frac{\xi_{21}}{2}\,,
\quad 
\frac{\alpha_{31}}{2} = \beta_{e2} + \beta_{\tau3} + \alpha + \beta + \frac{\xi_{31}}{2}\,, \quad
\beta_{\tau3} + \alpha=0~{\rm or}~\pi\,.
\label{Majph2131B2}
\end{equation}
%

 In terms of the standard parametrisation mixing angles 
$\theta_{12}$, $\theta_{23}$, $\theta_{13}$ and the 
Dirac phase $\delta$ the phases $(\beta_{e1} + \beta_{\tau3} + \alpha)$ 
and $(\beta_{e2} + \beta_{\tau3} + \alpha)$ read: 
\begin{align}
\label{be1mu2phB2}
\beta_{e1} + \beta_{\tau3} + \alpha &= 
\arg\left(U_{\mu2} e^{-i \frac{\alpha_{21}}{2}}\right) - \beta_{\mu2}=   
\arg\left(c_{12} c_{23} - s_{12} s_{23} s_{13} e^{i \delta}\right) - \beta_{\mu2}\,,
\\[0.2cm]
\label{be2mu1phB2}
\beta_{e2} + \beta_{\tau3} + \alpha &= 
\arg\left(U_{\mu1}\right) - \beta_{\mu1} = 
\arg\left(- s_{12} c_{23} - c_{12} s_{23} s_{13} e^{i \delta}\right) - \beta_{\mu1}\,.
\end{align}
%
 
 The relevant expressions for the parameters 
$\sin^2 \theta^e_{13}$, $\sin^2 \hat \theta_{23}$ and $\cos \alpha$ 
in terms of the neutrino mixing angles $\theta_{12}$,
$\theta_{13}$, $\theta_{23}$ and the angles
contained in $\tilde U_{\nu}$ have been derived in \cite{Girardi:2015vha}:
\begin{align}
\sin^2 \theta_{13} &=  \sin^2 \theta^e_{13} \cos^2 \hat \theta_{23} \,, 
\\[0.2cm]
\sin^2 \hat \theta_{23} &= \sin^2 \theta_{23} \cos^2 \theta_{13} \,,
\end{align}
\begin{equation}
\cos \alpha = 2 \, \frac{\sin^2 \theta^{\nu}_{12} \cos^2 \theta_{23}
+ \cos^2 \theta^{\nu}_{12} \sin^2 \theta_{23} \sin^2 \theta_{13}
- \sin^2 \theta_{12} \left(1 - \sin^2 \theta_{23} \cos^2 \theta_{13}\right)}
{\sin 2 \theta^{\nu}_{12} \sin 2 \theta_{23}  \sin \theta_{13}} \,.
\label{eq:cosalpha13e23e}
\end{equation}
%

From eqs.~(\ref{be1B2})~--~(\ref{be3B2}), (\ref{btau3B2}), 
(\ref{dB2}) and (\ref{Majph2131B2})  
we find:
\begin{align}
\sin (\alpha_{21}/2) & = \frac{1}{2 |U_{e1} U_{e2}|}
\bigg[ \sin2\theta^e_{13} \sin\hat\theta_{23} 
\Big( \sin(\alpha - \xi_{21}/2) + 2 \sin^2\theta^\nu_{12} \cos\alpha \sin(\xi_{21}/2) \Big) \nonumber \\
& + \sin2\theta^{\nu}_{12} 
\left(\cos^2 \theta^e_{13} - \sin^2\theta^e_{13} \sin^2\hat\theta_{23} \right)
\sin(\xi_{21}/2) \bigg] \,,
\label{eq:alpha21gen13e23e} \\
\sin (\alpha_{31}/2 - \delta) & = 
\frac{1}{|U_{e1}|}
\bigg[ \sin\theta^e_{13} \sin\hat\theta_{23} \sin\theta^{\nu}_{12} \sin(\beta + \xi_{31}/2)
- \cos\theta^e_{13} \cos\theta^\nu_{12} \sin(\alpha - \beta -\xi_{31}/2) \bigg] \,,
\label{eq:alpha31gen13e23e}\\
\sin (\alpha_{31}/2) & = 
\frac{1}{|U_{e2}|} 
\bigg[ \cos\theta^e_{13} \sin\theta^{\nu}_{12} \sin(\beta + \xi_{31}/2) 
+ \sin\theta^e_{13}  \sin\hat\theta_{23} \cos\theta^{\nu}_{12} \sin(\alpha - \beta - \xi_{31}/2) \bigg] \,.
\label{eq:alpha31gen13e23e-2}
\end{align}
%
The results given in eqs.~(\ref{eq:alpha21gen13e23e}) and 
(\ref{eq:alpha31gen13e23e})  can be derived  
also by comparing the expressions for the rephasing invariants 
$I_1$ and $I_2$ in the standard parametrisation of the PMNS matrix 
and in the parametrisation given in eq.~(\ref{UPMNSthhat1-13}).
The formulae for $\cos (\alpha_{21}/2)$ and  
$\cos (\alpha_{31}/2 - \delta)$ ($\cos (\alpha_{31}/2)$)
can be obtained formally from eqs.~(\ref{eq:alpha21gen13e23e}) and 
(\ref{eq:alpha31gen13e23e}) ((\ref{eq:alpha31gen13e23e-2}))
by replacing $\xi_{21}$ with $\xi_{21} + \pi$ and $\xi_{31}$ 
with $\xi_{31} + \pi$, respectively.
Similarly to case B1, $\sin (\alpha_{21}/2)$ can be determined once
$\xi_{21}$ is fixed, while $\sin (\alpha_{31}/2 - \delta)$ and 
$\sin (\alpha_{31}/2)$ are functions of $\xi_{31}$ and 
of the free parameter $\beta$. 
The comment from the preceding subsection concerning 
the dependence of $\alpha_{31}$ on $\beta$ 
and its subdominant effect on the values of the 
absolute value of the 
$\betabeta$-decay effective Majorana mass $\meff$ 
in cases of IO and QD neutrino mass spectra 
is valid also in this case.

 In terms of the neutrino mixing angles 
$\theta_{12}$, $\theta_{13}$, $\theta_{23}$ and the phase $\delta$ we have: 
\begin{align}
\sin(\alpha_{21}/2) &= - \frac{\cos(\beta_{\mu2} - \beta_{\mu1})}{2|U_{\mu1}U_{\mu2}|}
\bigg[\sin2\theta_{23} \sin\theta_{13} 
\left(\sin( \delta + \xi_{21}/2) 
- 2 \sin^2\theta_{12} \cos\delta \sin(\xi_{21}/2)\right) 
\nonumber\\
&+ \sin2\theta_{12} \left(\cos^2\theta_{23} 
- \sin^2\theta_{23} \sin^2\theta_{13}\right)
\sin(\xi_{21}/2)\bigg]\,,
\label{eq:alpha21genstandB2}
\\
\sin(\alpha_{31}/2) &= - \frac{\cos\beta_{\mu1}}{|U_{\mu1}|}
\bigg[\sin\theta_{12} \cos\theta_{23} \sin(\beta + \xi_{31}/2) 
+ \cos\theta_{12} \sin\theta_{23} \sin\theta_{13} 
\sin(\delta + \beta + \xi_{31}/2)\bigg]\,.
\label{eq:alpha31genstandB2}
\end{align}
%
In the parametrisation defined in eq.~(\ref{UPMNSthhat1-13}) 
we have (as it follows from eq. (\ref{Uthenuhat-13})): 
$|U_{\mu1}| = \hat{c}_{23} |s^{\nu}_{12}|$ 
and   $|U_{\mu 2}| =\hat{c}_{23} |c^{\nu}_{12}|$. 
Given the angle  $\theta^\nu_{12}$, the sign factors 
$\cos(\beta_{\mu2} - \beta_{\mu1})$ and $\cos\beta_{\mu1}$ 
are fixed, since 
\begin{equation}
\cos(\beta_{\mu2} - \beta_{\mu1}) = -\,{\rm sgn} \left(s^\nu_{12} c^\nu_{12}\right)\,, \quad
\cos\beta_{\mu1} = -\,{\rm sgn} \left(s^\nu_{12}\right)\,. 
\end{equation}
%

 As in the previous subsections, the expressions for 
 $\sin (\alpha_{21}/2 - \xi_{21}/2)$, 
$\sin (\alpha_{31}/2 - \xi_{31}/2 - \beta)$ 
and $\sin (\alpha_{31}/2 - \delta - \xi_{31}/2 - \beta)$ 
are somewhat simpler:
\begin{align}
\sin (\alpha_{21}/2 - \xi_{21}/2) & = 
\frac{\sin2\theta^e_{13} \sin\hat\theta_{23}}
{\cos^2\theta_{13} \sin2\theta_{12}} \sin\alpha 
\nonumber\\
&= - \frac{\cos(\beta_{\mu2} - \beta_{\mu1})}{2|U_{\mu1} U_{\mu2}|} 
\sin2\theta_{23} \sin\theta_{13} \sin\delta \,,
\label{eq:13e23ealpha21B2} \\[0.2cm]
\sin (\alpha_{31}/2 - \xi_{31}/2 - \beta) & = 
\frac{\sin\theta^e_{13} \cos\theta^\nu_{12}}
{\sin\theta_{12} \cos\theta_{13}} \sin\hat\theta_{23} \sin\alpha \nonumber\\
&= - \frac{\cos\beta_{\mu1}}{|U_{\mu1}|} 
\cos\theta_{12} \sin\theta_{23}\sin\theta_{13} \sin\delta \,, 
\label{eq:13e23ealpha31B2} \\[0.2cm]
\sin (\alpha_{31}/2 - \delta - \xi_{31}/2 - \beta) & = 
-\frac{\sin2\theta^e_{13} \cos\theta^\nu_{12}}
{\cos\theta_{12} \sin2\theta_{13}} \cos\hat\theta_{23} \sin\alpha \nonumber\\
&= \frac{\cos\beta_{\mu1}}{|U_{\mu1}|} 
\sin\theta_{12} \cos\theta_{23} \sin\delta \,.
\label{eq:13e23ealpha31deltaB2}
\end{align}
%
Also in this case we have 
$|\sin (\alpha_{21}/2 - \xi_{21}/2)| \propto \sin\theta_{13}$ and 
$|\sin (\alpha_{31}/2 - \xi_{31}/2 - \beta)| \propto \sin\theta_{13}$.
As we have already mentioned earlier,
predictions for the phases $(\alpha_{21}/2 - \xi_{21}/2)$ and 
$(\alpha_{31}/2 - \xi_{31}/2 - \beta)$ in the case analysed 
in this subsection will be presented in Section~\ref{sec:predictions}.

 We would like to note finally that formulae in 
eqs.~(\ref{eq:alpha21gen13e23e})~--~(\ref{eq:alpha31genstandB2}) 
and eqs.~(\ref{eq:13e23ealpha21B2})~--~(\ref{eq:13e23ealpha31deltaB2})
can be obtained formally from 
the corresponding formulae in subsection~\ref{sec:12e23e23nu12nu}, 
eqs.~(\ref{eq:alpha21gen12e23eB1})~--~(\ref{eq:alpha31genstandB1}) 
and eqs.~(\ref{eq:alpha21xi21B1})~--~(\ref{eq:alpha31xi31betaB1}),
by making the following substitutions: 
\begin{equation}
\phi \rightarrow -\alpha\,, \quad
\theta^e_{12} \rightarrow \theta^e_{13}\,, \quad 
\hat\theta_{23} \rightarrow \hat\theta_{23} + \frac{\pi}{2}, \quad
\theta_{23} \rightarrow \theta_{23} - \frac{\pi}{2}\, \quad
\text{and} \quad
\tau \rightarrow \mu \,.
\end{equation}
%

%
\subsection{The Scheme with $\boldsymbol{(\theta^e_{12},\theta^e_{13}) - 
(\theta^\nu_{23}, \theta^\nu_{12})}$ Rotations (Case B3)}
\label{sec:12e13e23nu12nu}
%
%

 In this subsection we switch to the parametrisation of the PMNS
matrix $U$ given in eq.~(\ref{eq:Uijkl}) 
with $(ij) - (kl) = (12) - (13)$, i.e., 
\begin{equation}
U = R_{12}(\theta^e_{12}) \, R_{13}(\theta^e_{13})  \, \Psi \, 
R_{23}(\theta^{\nu}_{23}) \, R_{12}(\theta^{\nu}_{12}) \, Q_0 \,.
\label{eq:UB3a} 
\end{equation}
%
In explicit form this matrix reads: 
\begin{equation} 
U = \begin{pmatrix} 
|U_{e1}| e^{i \beta_{e1}} & |U_{e2}| e^{i \beta_{e2}} & |U_{e3}| e^{i \beta_{e3}} \\[0.2cm]  
|U_{\mu1}| e^{i \beta_{\mu1}} &  |U_{\mu2}| e^{i \beta_{\mu2}} & |U_{\mu3}| e^{i \beta_{\mu3}} \\[0.2cm]
|U_{\tau1}| e^{i \beta_{\tau1}}& |U_{\tau2}| e^{i \beta_{\tau2}} & |U_{\tau3}| e^{i \beta_{\tau3}}
\end{pmatrix}
Q_0\,,
\label{eq:UB3b}
\end{equation}
%
where
\begin{align}
|U_{e1}| e^{i \beta_{e1}} &= c^e_{12} c^e_{13} c^\nu_{12} - s^\nu_{12} 
\left(s^e_{12} c^\nu_{23} e^{- i \psi} - c^e_{12} s^e_{13} s^\nu_{23} e^{- i \omega}\right)\,,\\
|U_{e2}| e^{i \beta_{e2}} &= c^e_{12} c^e_{13} s^\nu_{12} + c^\nu_{12} 
\left(s^e_{12} c^\nu_{23} e^{- i \psi} - c^e_{12} s^e_{13} s^\nu_{23} e^{- i \omega}\right)\,,\\
|U_{e3}| e^{i \beta_{e3}} &= s^e_{12} s^\nu_{23} e^{- i \psi} 
+ c^e_{12} s^e_{13} c^\nu_{23} e^{- i \omega}\,,\\
|U_{\mu1}| e^{i \beta_{\mu1}} &= - s^e_{12} c^e_{13} c^\nu_{12} - s^\nu_{12} 
\left(c^e_{12} c^\nu_{23} e^{- i \psi} + s^e_{12} s^e_{13} s^\nu_{23} e^{- i \omega}\right)\,,\\
|U_{\mu2}| e^{i \beta_{\mu2}} &= - s^e_{12} c^e_{13} s^\nu_{12} + c^\nu_{12} 
\left(c^e_{12} c^\nu_{23} e^{- i \psi} + s^e_{12} s^e_{13} s^\nu_{23} e^{- i \omega}\right)\,,\\
|U_{\mu3}| e^{i \beta_{\mu3}} &= c^e_{12} s^\nu_{23} e^{- i \psi} 
- s^e_{12} s^e_{13} c^\nu_{23} e^{- i \omega}\,,\\
|U_{\tau1}| e^{i \beta_{\tau1}} &= - s^e_{13} c^\nu_{12} 
+ c^e_{13} s^\nu_{12} s^\nu_{23} e^{- i \omega}\,,
\label{eq:tau1B3}\\
|U_{\tau2}| e^{i \beta_{\tau2}} &= - s^e_{13} s^\nu_{12} 
- c^e_{13} c^\nu_{12} s^\nu_{23} e^{- i \omega}\,,
\label{eq:tau2B3}\\
|U_{\tau3}| e^{i \beta_{\tau3}} &= c^e_{13} c^\nu_{23} e^{- i \omega}\,.
\label{eq:tau3B3}
\end{align}
%

 Comparing the expressions for the absolute value 
of the element $U_{\tau3}$ in the standard parametrisation 
of the PMNS matrix and the parametrisation we are considering here, 
we have \cite{Girardi:2015vha}
\begin{equation}
\cos^2\theta^e_{13} = \frac{\cos^2\theta_{23} \cos^2\theta_{13}}
{\cos^2\theta^\nu_{23}}\,.
\label{eq:the13B3}
\end{equation}
%
Hence, the angle $\theta^e_{13}$ is expressed in terms of 
the known angles and can be determined up to a quadrant.
The phase $\omega$ is a free phase parameter, which 
enters, e.g., the sum rule for $\cos\delta$ 
(see eq.~(63) in ref.~\cite{Girardi:2015vha}), 
so its presence is expected as well in the sum rules for the 
Majorana phases we are going to derive.

 We aim as before to find an appropriate phase rearrangement 
in order to bring $U$ to the standard parametrisation form.
For that reason we compare first the expressions for the $J_{\rm CP}$ 
invariant in the standard parametrisation and in the parametrisation 
given in eq.~(\ref{eq:UB3a}) and find
\begin{equation}
\sin\delta = \frac{8\,\mathcal{J}}
{\sin2\theta_{12} \sin2\theta_{23} \sin2\theta_{13} \cos\theta_{13}}\,,
\label{eq:sindeltaB3}
\end{equation}
%
where $\mathcal{J}$ is the expression for $J_{\rm CP}$ in 
the parametrisation of $U$ given in eq.~(\ref{eq:UB3a}):
\begin{align}
\mathcal{J} &= \frac{1}{8} \cos\theta^e_{13} \bigg[
\sin2\theta^e_{12} \Big\{
2 \sin2\theta^\nu_{12} \cos\theta^\nu_{23} 
\left[
\left(\cos^2\theta^e_{13} - \cos^2\theta^\nu_{23}\right) \sin\psi 
- \sin^2\theta^e_{13} \sin^2\theta^\nu_{23} \sin(\psi - 2\omega)
\right] \nonumber\\
&- \sin2\theta^e_{13} \cos2\theta^\nu_{12} \sin2\theta^\nu_{23} \sin(\psi - \omega)
\Big\} 
+ 2 \cos2\theta^e_{12} \sin\theta^e_{13} \sin2\theta^\nu_{12} \sin2\theta^\nu_{23} 
\cos\theta^\nu_{23} \sin\omega
\bigg]\,.
\end{align}
%
This expression looks cumbersome, but one can verify 
that the relation in eq.~(\ref{eq:sindeltaB3}) holds if 
$\delta$ is given by
\begin{equation}
\delta = \beta_{e1} + \beta_{e2} + \beta_{\mu3} + \beta_{\tau3} - \beta_{e3} + \psi + \omega\,.
\label{eq:deltaB3}
\end{equation}
%

 Now we can cast $U$ in the following form: 
\begin{equation} 
U = P_2\, 
\begin{pmatrix} 
|U_{e1}| & |U_{e2}|  & |U_{e3}| e^{- i \delta} \\[0.2cm]
|U_{\mu1}| e^{i (\beta_{\mu1} + \beta_{e2} + \beta_{\tau3} + \psi + \omega)}
& |U_{\mu2}| e^{ i (\beta_{\mu2} +\beta_{e1} + \beta_{\tau3} + \psi + \omega)} 
& |U_{\mu3}| \\[0.2cm]
|U_{\tau1}| e^{ i (\beta_{\tau1} + \beta_{e2} + \beta_{\mu3} + \psi + \omega)}&
|U_{\tau2}| e^{ i (\beta_{\tau2} + \beta_{e1} + \beta_{\mu3} + \psi + \omega)}& 
|U_{\tau3}|
\end{pmatrix}
Q_2\,Q_0\,,
\label{eq:UB3c}
\end{equation}
%
where 
\begin{align}
\label{P2B3}
P_2 &= \diag\left(
e^{i (\beta_{e1} + \beta_{e2} + \beta_{\mu3} + \beta_{\tau3} + \psi + \omega)}, 
e^{i \beta_{\mu3}},
e^{i \beta_{\tau3}}
\right)\,, \\[0.2cm]
\label{Q2B3}
Q_2 &= \diag\left(
e^{-i (\beta_{e2} + \beta_{\mu3} + \beta_{\tau3} + \psi + \omega)}, 
e^{-i (\beta_{e1} + \beta_{\mu3} + \beta_{\tau3} + \psi + \omega)}, 
1
\right) \nonumber\\[0.2cm]
&= e^{-i (\beta_{e2} + \beta_{\mu3} + \beta_{\tau3} + \psi + \omega)}\,
\diag\left(
1, 
e^{i (\beta_{e2} - \beta_{e1})}, 
e^{i (\beta_{e2} + \beta_{\mu3} + \beta_{\tau3} + \psi + \omega)}\right)\,.
\end{align}
%
The phases in the matrix $P_2$ as well as the overall phase in the matrix $Q_2$ 
are unphysical. Thus, for the Majorana phases we get: 
\begin{equation}
\frac{\alpha_{21}}{2} = \beta_{e2} - \beta_{e1} +\frac{\xi_{21}}{2}\,,
\quad 
\frac{\alpha_{31}}{2} = \beta_{e2} + \beta_{\mu3} + \beta_{\tau3} 
+ \psi + \omega + \frac{\xi_{31}}{2}\,.
\label{Majph2131B3}
\end{equation}
%

 In terms of the standard parametrisation mixing angles 
$\theta_{12}$, $\theta_{23}$, $\theta_{13}$ and the 
Dirac phase $\delta$ we have:
\begin{align}
\label{be1tau2phB3}
\beta_{e1} + \beta_{\mu3} + \psi + \omega & = 
\arg\left(U_{\tau2} e^{-i \frac{\alpha_{21}}{2}}\right) - \beta_{\tau2} = 
\arg\left(- c_{12} s_{23} - s_{12} c_{23} s_{13} e^{i \delta}\right) - \beta_{\tau2}\,,
\\[0.2cm]
\label{be2tau1phB3}
\beta_{e2} + \beta_{\mu3} + \psi + \omega & = 
\arg\left(U_{\tau1}\right) - \beta_{\tau1} = 
\arg\left(s_{12} s_{23} - c_{12} c_{23} s_{13} e^{i \delta}\right) - \beta_{\tau1}\,,
\end{align}
%
where $\beta_{\tau1}$ and $\beta_{\tau2}$ are the arguments of  
the expressions given in 
eqs.~(\ref{eq:tau1B3}) and (\ref{eq:tau2B3}), respectively. 
They are fixed once the angles $\theta^\nu_{12}$ and $\theta^\nu_{23}$,  
the quadrant to which $\theta^e_{13}$ belongs and the phase $\omega$ 
are known. Finally, we find:
\begin{align}
\frac{\alpha_{21}}{2} &= 
\arg\left(U_{\tau1} U^*_{\tau2} e^{i \frac{\alpha_{21}}{2}}\right) 
+\beta_{\tau2} - \beta_{\tau1} + \frac{\xi_{21}}{2}\,, 
\label{eq:alpha21B3} \\
\frac{\alpha_{31}}{2} &= \arg\left(U_{\tau1}\right) 
+\beta_{\tau3} - \beta_{\tau1} + \frac{\xi_{31}}{2}\,,
\label{eq:alpha31B3}
\end{align}
%
where $\beta_{\tau3}$ is the argument of the 
expression in eq.~(\ref{eq:tau3B3}), which is 
fixed under the conditions specified above 
for $\beta_{\tau1}$ and $\beta_{\tau2}$.

 The mixing angles $\theta_{12}$, $\theta_{23}$ and $\theta_{13}$ 
of the standard parametrisation are related with the angles 
$\theta^e_{ij}$, $\theta^\nu_{kl}$ and the phases $\psi$ and $\omega$ 
present in the parametrisation of $U$ given in eq.~(\ref{eq:UB3a}) 
in the following way:
\begin{align}
\sin^2 \theta_{13} & = |U_{e3}|^2  = 
\sin^2\theta^e_{12} \sin^2\theta^\nu_{23} 
+ \cos^2\theta^e_{12} \sin^2\theta^e_{13} \cos^2\theta^\nu_{23} 
- X\,, 
\label{eq:th13B3}\\
\sin^2 \theta_{23} & = \frac{|U_{\mu3}|^2}{1-|U_{e3}|^2} = \frac{1}{1 - \sin^2\theta_{13}}
\bigg[\cos^2\theta^e_{12} \sin^2\theta^\nu_{23} 
+ \sin^2\theta^e_{12} \sin^2\theta^e_{13} \cos^2 \theta^\nu_{23} 
+ X\bigg]\,,
\label{eq:th23B3} \\
\sin^2 \theta_{12} & = \frac{|U_{e2}|^2}{1-|U_{e3}|^2} = \frac{1}{1 - \sin^2\theta_{13}}
\bigg[\cos^2\theta^e_{12} \cos^2\theta^e_{13} \sin^2\theta^\nu_{12} \nonumber\\
&+ \frac{1}{2} \sin2\theta^\nu_{12} 
\left(\sin2\theta^e_{12} \cos\theta^e_{13} \cos\theta^\nu_{23} \cos\psi 
- \cos^2\theta^e_{12} \sin2\theta^e_{13} \sin\theta^\nu_{23} \cos\omega\right) \nonumber\\
&+ \cos^2\theta^\nu_{12} 
\left(\sin^2\theta^e_{12} \cos^2\theta^\nu_{23} 
+ \cos^2\theta^e_{12} \sin^2\theta^e_{13} \sin^2\theta^\nu_{23} 
+ X\right)\bigg]\,,
\label{eq:th12B3}
\end{align}
%
where
\begin{equation}
X = - \frac{1}{2} \sin2\theta^e_{12} \sin\theta^e_{13} \sin2\theta^\nu_{23} \cos(\psi - \omega)\,.
\end{equation}
%
The sum of eqs.~(\ref{eq:th13B3}) and (\ref{eq:th23B3}) 
leads to the result given in eq.~(\ref{eq:the13B3}), i.e., 
the angle $\theta^e_{13}$ is known (up to a quadrant). 
Then, solving eq.~(\ref{eq:th13B3}) for $X$ and substituting 
the solution in eq.~(\ref{eq:th12B3}), we find $\cos\psi$ 
as a function of $\theta_{12}$, $\theta_{13}$, $\theta^e_{12}$, 
$\theta^e_{13}$ and $\omega$: 
\begin{align}
\cos\psi &= \frac{2}
{\sin2\theta^e_{12} \cos\theta^e_{13} \sin2\theta^\nu_{12} \cos\theta^\nu_{23}}
\bigg[\cos^2\theta_{13} \left(\sin^2\theta_{12}  - \cos^2\theta^\nu_{12}\right) 
\nonumber\\
&+ \cos^2\theta^e_{12} \cos^2\theta^e_{13} \cos2\theta^\nu_{12} 
+ \frac{1}{2} \cos^2\theta^e_{12} \sin2\theta^e_{13} 
\sin2\theta^\nu_{12} \sin\theta^\nu_{23} \cos\omega\bigg]\,.
\end{align}
%
Finally, substituting $\cos\psi$ and $\sin\psi = \pm \sqrt{1-\cos^2\psi}$ 
in eq.~(\ref{eq:th13B3}), one can express $\theta^e_{12}$ in terms 
of the known angles.

 As in the previous subsections, we give the formulae for 
$\sin (\alpha_{21}/2 - \xi_{21}/2)$, 
$\sin (\alpha_{31}/2 - \xi_{31}/2)$ 
and $\sin (\alpha_{31}/2 - \delta - \xi_{31}/2)$, 
which in the case under consideration read:
\begin{align}
\sin (\alpha_{21}/2 - \xi_{21}/2) & = \frac{1}{2 |U_{e1} U_{e2}|}
\Big[\cos^2\theta^e_{12} \sin2\theta^e_{13} \sin\theta^\nu_{23} \sin\omega
- \sin2\theta^e_{12} \cos\theta^e_{13} \cos\theta^\nu_{23} \sin\psi\Big] 
\nonumber\\
&= \frac{1}{2 |U_{\tau1} U_{\tau2}|}
\Big[\cos(\beta_{\tau2} - \beta_{\tau1}) \sin2\theta_{23} \sin\theta_{13} \sin\delta 
+ \sin(\beta_{\tau2} - \beta_{\tau1}) \nonumber\\
&\times \left(\sin2\theta_{12} 
\left(\cos^2\theta_{23} \sin^2\theta_{13} - \sin^2\theta_{23}\right)
+ \cos2\theta_{12} \cos\delta
\right)\Big]\,,
\label{eq:alpha21xi21B3} \\[0.2cm]
\sin (\alpha_{31}/2 - \xi_{31}/2) & = - \frac{1}{2 |U_{e2} U_{\mu3} U_{\tau3}|}
\Big[\sin2\theta^e_{12} \cos\theta^e_{13} \cos^2\theta^\nu_{23} 
\big[\cos\theta^\nu_{12} \sin\theta^\nu_{23} \nonumber\\
&\times \left(\sin\psi - \sin^2\theta^e_{13} \sin(\psi - 2\omega)\right) 
+ \frac{1}{2} \sin2\theta^e_{13} \sin\theta^\nu_{12} \sin(\psi - \omega)\big] \nonumber\\
&+ \frac{1}{2} \sin2\theta^e_{13} \cos\theta^\nu_{12} \cos\theta^\nu_{23} 
\left(\cos2\theta^e_{12} \cos2\theta^\nu_{23} - 1\right) \sin\omega\Big]
\nonumber\\
&= - \frac{1}{|U_{\tau1}|} 
\Big[\cos(\beta_{\tau3} - \beta_{\tau1}) \cos\theta_{12} \cos\theta_{23} 
\sin\theta_{13} \sin\delta \nonumber\\
&+ \sin(\beta_{\tau3} - \beta_{\tau1}) 
\left(\cos\theta_{12} \cos\theta_{23} \sin\theta_{13} \cos\delta 
-\sin\theta_{12} \sin\theta_{23}\right)\Big]\,,
\label{eq:alpha31xi31B3} \\[0.2cm]
\sin (\alpha_{31}/2 - \delta - \xi_{31}/2) & = - \frac{1}{2 |U_{e1} U_{e3}|}
\Big[\sin2\theta^e_{12} \sin\theta^e_{13} \sin\theta^\nu_{12} \sin(\psi - \omega) \nonumber\\
&+ \cos\theta^e_{13} \cos\theta^\nu_{12} 
\left(\sin2\theta^e_{12} \sin\theta^\nu_{23} \sin\psi 
+ 2\cos^2\theta^e_{12} \sin\theta^e_{13} \cos\theta^\nu_{23} \sin\omega\right)
\nonumber\\
&= - \frac{1}{|U_{\tau1}|}
\Big[\cos(\beta_{\tau3} - \beta_{\tau1}) \sin\theta_{12} \sin\theta_{23} \sin\delta \nonumber\\
&+ \sin(\beta_{\tau3} - \beta_{\tau1})
\left(\cos\theta_{12} \cos\theta_{23} \sin\theta_{13}  
- \sin\theta_{12} \sin\theta_{23} \cos\delta\right)\Big]\,.
\label{eq:alpha31deltaxi31B3}
\end{align}
%

Given the angles $\theta^\nu_{12}$ and $\theta^\nu_{23}$,  
the quadrant to which $\theta^e_{13}$ belongs, 
the phases $(\beta_{\tau2} - \beta_{\tau1})$ and $(\beta_{\tau3} - \beta_{\tau1})$
in eqs.~(\ref{eq:alpha21xi21B3}) and (\ref{eq:alpha31xi31B3}), 
$\sin (\alpha_{21}/2 - \xi_{21}/2)$ and  $\sin (\alpha_{31}/2 - \xi_{31}/2)$ 
depend on the free phase parameter $\omega$. 
The phases $(\alpha_{21}/2 - \xi_{21}/2 - (\beta_{\tau2} - \beta_{\tau1}))$ and 
$(\alpha_{31}/2 - \xi_{31}/2 -(\beta_{\tau3} - \beta_{\tau1}))$, 
as it follows from eqs.~(\ref{eq:alpha21B3}) and  (\ref{eq:alpha31B3}), 
are completely determined by the values of the standard parametrisation angles 
$\theta_{12}$, $\theta_{23}$ and $\theta_{13}$,   
and of the Dirac phase $\delta$. 
The expression for, e.g., 
 $\sin(\alpha_{21}/2 - \xi_{21}/2 - (\beta_{\tau2} - \beta_{\tau1}))$ 
($\sin (\alpha_{31}/2 - \xi_{31}/2 -(\beta_{\tau3} - \beta_{\tau1}))$) 
can formally be obtained from eq.~(\ref{eq:alpha21xi21B3}) 
(eq.~(\ref{eq:alpha31xi31B3})) by setting 
 $\sin(\beta_{\tau2} - \beta_{\tau1}) = 0$, 
$\cos(\beta_{\tau2} - \beta_{\tau1}) = 1$ 
($\sin(\beta_{\tau3} - \beta_{\tau1}) = 0$, 
$\cos(\beta_{\tau3} - \beta_{\tau1}) = 1$).
It follows from the results thus obtained that both 
$|\sin(\alpha_{21}/2 - \xi_{21}/2 - (\beta_{\tau2} - \beta_{\tau1}))|
\propto \sin\theta_{13}$ 
and  $|\sin (\alpha_{31}/2 - \xi_{31}/2 -(\beta_{\tau3} - \beta_{\tau1}))| 
\propto \sin\theta_{13}$. 
It should be noted, however, that in the considered scheme
the phase $\delta$ also depends on the phase $\omega$ 
and as long as  $\delta$ is not fixed (e.g., measured directly 
or determined in a global data analysis), 
the phases  $(\alpha_{21}/2 - \xi_{21}/2 - (\beta_{\tau2} - \beta_{\tau1}))$ and 
$(\alpha_{31}/2 - \xi_{31}/2 -(\beta_{\tau3} - \beta_{\tau1}))$ 
will depend on $\omega$ via $\delta$. 
Therefore in \cite{Girardi:2015vha} we have given predictions for 
$\delta$ for $\omega = 0$ and ${\rm sgn}(\sin2\theta^e_{13}) = 1$.
Correspondingly, in Section~\ref{sec:predictions} we will derive predictions for 
the values of the phases 
$(\alpha_{21}/2 - \xi_{21}/2)$ and
$(\alpha_{31}/2 - \xi_{31}/2)$ for the same values of 
$\omega = 0$ and ${\rm sgn}(\sin2\theta^e_{13}) = 1$,
for which the predicted value of $\delta$ lies in its 
$2\sigma$ allowed interval quoted in
eq.~(\ref{deltaexp}).  

 We note finally that $\sin^2\theta_{23}$ is constrained 
by the requirements that 
$\cos\psi$, $\sin^2\theta^e_{12}$ and  
$\sin^2\theta^e_{13}$ possess 
physically acceptable values, 
to lie for both the NO and IO 
spectra in the following narrow 
intervals \cite{Girardi:2015vha}: 
\begin{align*}
&(0.489,0.498)  ~ \text{for TBM}, \\
&(0.489,0.496)  ~ \text{ for GRA}, \\
&(0.489,0.499)  ~ \text{ for GRB}, \\
&(0.489,0.499)  ~ \text{ for HG}, \\
&(0.489,0.521)  ~ \text{ for BM}.
\end{align*}
%
Thus, we will present results for the phases of interest 
for the NO (IO) spectrum for
$\sin^2\theta_{23} = 0.48907$ ($\sin^2\theta_{23} = 0.48886$)~%
\footnote{For $\sin^2 \theta_{23} < 0.48907$ ($\sin^2\theta_{23} < 0.48886$), 
$\cos\delta$ acquires an unphysical (complex) value.}.

%
\section{The Cases of $\boldsymbol{\theta^e_{ij} - (\theta^\nu_{23}, \theta^\nu_{13}, \theta^\nu_{12})}$ 
 Rotations }
\label{sec:ije23nu13nu12nu}
%
%

We consider next a generalisation of the cases analysed
in Section \ref{sec:ije23nu12nu} with the presence of 
a third rotation matrix in 
$\tilde{U}_{\nu}$ arising from the neutrino
sector, i.e., we employ the parametrisation of 
$U$ given in eq.~(\ref{eq:Uija}). 
Non-zero values of $\theta^{\nu}_{13}$ are
inspired by certain types of flavour symmetries
\cite{Bazzocchi:2011ax,Rodejohann:2014xoa}.
In the numerical analysis of the predictions 
for $\alpha_{21}$, $\alpha_{31}$ and $\meff$ we will perform 
in Section~\ref{sec:predictions}, 
we will consider three representative values of 
$\theta^{\nu}_{13}$ discussed in the literature:
$\theta^{\nu}_{13} = \pi/20,~\pi/10$ and $\sin^{-1} (1 / 3)$. 
We are not going to consider the case in which
the $U$ matrix is parametrised as
in eq.~(\ref{eq:Uija}) with $(ij) = (23)$
for the reasons explained in \cite{Girardi:2015vha}, i.e.,
the absence of a correlation between the Dirac CPV phase
$\delta$ and the mixing angles. It should be noted that
for this and other cases for which it is not possible to derive 
such a correlation, different symmetry forms of $\tilde U_{\nu}$ can 
still be tested with an improvement of the precision in the measurement
of the neutrino mixing angles. For instance, in the case 
corresponding to eq.~(\ref{eq:Uija}) with $(ij) = (23)$,
one has, as was shown in \cite{Girardi:2015vha}, 
$\sin^2 \theta_{13} = \sin^2 \theta^{\nu}_{13}$
and $\sin^2 \theta_{12} = \sin^2 \theta^{\nu}_{12}$, 
i.e., the angles $\theta_{13}$ and $\theta_{12}$ are predicted 
to have particular values when the angles 
$\theta^{\nu}_{13}$ and $\theta^\nu_{23}$ are fixed 
by a symmetry.

%
\subsection{The Scheme with $\boldsymbol{\theta^e_{12} - 
(\theta^\nu_{23}, \theta^\nu_{13}, \theta^\nu_{12})}$ Rotations (Case C1)}
\label{sec:12e23nu13nu12nu} 
%
%

In this subsection we consider the parametrisation
of the PMNS matrix $U$ given in eq.~(\ref{eq:Uija}) with $(ij) = (12)$,
i.e.,
\begin{equation}
U = R_{12}(\theta^e_{12}) \, \Psi \, R_{23}(\theta^{\nu}_{23})\,
R_{13}(\theta^{\nu}_{13}) \, R_{12}(\theta^{\nu}_{12}) \, Q_0 \,.
\label{eq:UC1a} 
\end{equation}
%
In this case the the matrix $\Psi$ contains 
only one physical phase $\phi$,
$\Psi = \diag \, (1, e ^{i \phi}, 1)$ (we have denoted $\phi \equiv - \psi$),
since the phase $\omega$ in $\Psi$ is unphysical and we have dropped it.
The explicit form of the matrix $U$ reads:
\begin{equation} 
U = \begin{pmatrix} 
|U_{e1}| e^{i \beta_{e1}} & |U_{e2}| e^{i \beta_{e2}} & |U_{e3}| e^{i \beta_{e3}} \\[0.2cm]  
|U_{\mu1}| e^{i \beta_{\mu1}} &  |U_{\mu2}| e^{i \beta_{\mu2}} & |U_{\mu3}| e^{i \beta_{\mu3}} \\[0.2cm]
|U_{\tau1}| e^{i \beta_{\tau1}}& |U_{\tau2}| e^{i \beta_{\tau2}} & |U_{\tau3}| e^{i \beta_{\tau3}}
\end{pmatrix}
Q_0\,,
\label{eq:UC1b}
\end{equation}
%
where
\begin{align}
\label{Ue1C1}
|U_{e1}| e^{i \beta_{e1}} &= c^e_{12} c^\nu_{12} c^\nu_{13} - s^e_{12} 
\left(s^\nu_{12} c^\nu_{23} + c^\nu_{12} s^\nu_{23} s^\nu_{13}\right) e^{i \phi}\,,\\
\label{Ue2C1}
|U_{e2}| e^{i \beta_{e2}} &= c^e_{12} s^\nu_{12} c^\nu_{13} + s^e_{12} 
\left(c^\nu_{12} c^\nu_{23} - s^\nu_{12} s^\nu_{23} s^\nu_{13}\right) e^{i \phi}\,,\\
\label{Ue3C1}
|U_{e3}| e^{i \beta_{e3}} &= c^e_{12} s^\nu_{13} + s^e_{12} s^\nu_{23} c^\nu_{13} e^{i \phi}\,,\\
|U_{\mu1}| e^{i \beta_{\mu1}} &= - s^e_{12} c^\nu_{12} c^\nu_{13} - c^e_{12} 
\left(s^\nu_{12} c^\nu_{23} + c^\nu_{12} s^\nu_{23} s^\nu_{13}\right) e^{i \phi}\,,\\
|U_{\mu2}| e^{i \beta_{\mu2}} &= - s^e_{12} s^\nu_{12} c^\nu_{13} + c^e_{12} 
\left(c^\nu_{12} c^\nu_{23} - s^\nu_{12} s^\nu_{23} s^\nu_{13}\right) e^{i \phi}\,,\\
\label{Umu3C1}
|U_{\mu3}| e^{i \beta_{\mu3}} &= - s^e_{12} s^\nu_{13} + c^e_{12} s^\nu_{23} c^\nu_{13} e^{i \phi}\,,\\
|U_{\tau1}| e^{i \beta_{\tau1}} &= s^\nu_{12} s^\nu_{23} - c^\nu_{12} c^\nu_{23} s^\nu_{13}\,,
\label{eq:tau1C1}\\
|U_{\tau2}| e^{i \beta_{\tau2}} &= - c^\nu_{12} s^\nu_{23} - s^\nu_{12} c^\nu_{23} s^\nu_{13}\,,\label{eq:tau2C1}\\
|U_{\tau3}| e^{i \beta_{\tau3}} &= c^\nu_{23} c^\nu_{13}\,.
\label{eq:UC1elements}
\end{align}
%
Comparing the expressions for the $J_{\rm CP}$ invariant 
in the standard parametrisation and in the parametrisation given 
in eq.~(\ref{eq:UC1a}), one finds the following relation 
between $\sin\delta$ and $\sin\phi$~%
\footnote{For $\theta^\nu_{23} = -\pi/4$ this relation reduces to 
eq.~(75) in ref.~\cite{Girardi:2015vha}.}:
\begin{align}
\sin\delta = - \frac{\sin2\theta^e_{12}
\left[\left(\cos^2\theta^\nu_{13} + (\cos^2\theta^\nu_{13} - 2) \cos2\theta^\nu_{23}\right)
\sin2\theta^\nu_{12}
- 2 \cos2\theta^\nu_{12} \sin2\theta^\nu_{23} \sin\theta^\nu_{13}\right]}
{2\,{\rm sgn}(\cos\theta^\nu_{23} \cos\theta^\nu_{13}) \sin2\theta_{12} \sin2\theta_{13} \sin\theta_{23}} \sin\phi\,,
\label{eq:JCPC1}
\end{align}
%
where we have used that in this scheme 
$\cos^2\theta_{23} \cos^2\theta_{13} = \cos^2\theta^\nu_{23} \cos^2\theta^\nu_{13}$.
The relation in  eq.~(\ref{eq:JCPC1}) 
suggests the required rearrangement of the phases 
one has to perform to bring $U$ given in eq.~(\ref{eq:UC1b}) 
to the standard parametrisation form. 
Namely, it can be shown that eq.~(\ref{eq:JCPC1}) holds if 
\begin{equation}
\delta = \beta_{e1} + \beta_{e2} + \beta_{\mu3} - \beta_{e3} - \phi + \beta_{\tau3}\,,
\quad
\beta_{\tau3}=0~{\rm or}~\pi\,,
\label{eq:deltaC1}
\end{equation}
%
where $\beta_{\tau3} = \arg(c^\nu_{23} c^\nu_{13})$. 
The phase $\beta_{\tau3}$ provides the sign factor 
${\rm sgn(\cos\theta^\nu_{23} \cos\theta^\nu_{13})}$ 
in the relation between $\sin\delta$ and $\sin\phi$,
when one calculates $\sin\delta$ from eq.~(\ref{eq:deltaC1}). 
Now we can cast $U$ in the following form:
\begin{equation} 
U = P_2\, 
\begin{pmatrix} 
|U_{e1}| & |U_{e2}|  & |U_{e3}| e^{- i \delta} \\[0.2cm]
|U_{\mu1}| e^{i (\beta_{\mu1} + \beta_{e2} - \phi + \beta_{\tau3})}
& |U_{\mu2}| e^{ i (\beta_{\mu2} +\beta_{e1} -\phi + \beta_{\tau3})} 
& |U_{\mu3}| \\[0.2cm]
|U_{\tau1}| e^{ i (\beta_{\tau1} + \beta_{e2} + \beta_{\mu3} - \phi)}&
|U_{\tau2}| e^{ i (\beta_{\tau2} + \beta_{e1} + \beta_{\mu3} - \phi)}& 
|U_{\tau3}|
\end{pmatrix}
Q_2\,Q_0\,,
\label{eq:UC1c}
\end{equation}
%
where
\begin{align}
\label{P2C1}
P_2 &=\diag\left(e^{i (\beta_{e1} + \beta_{e2} + \beta_{\mu3} - \phi)}, 
e^{i (\beta_{\mu3} - \beta_{\tau3})}, 1\right)\,, \\[0.2cm]
\label{Q2C1}
Q_2 &= \diag\left(e^{-i (\beta_{e2} + \beta_{\mu3} - \phi)}, 
e^{-i (\beta_{e1} + \beta_{\mu3} - \phi)}, e^{i \beta_{\tau3}}\right) \nonumber\\[0.2cm]
&= e^{-i (\beta_{e2} + \beta_{\mu3} - \phi)}\,
\diag\left(1, e^{i (\beta_{e2} - \beta_{e1})}, 
e^{i (\beta_{e2} + \beta_{\mu3} - \phi + \beta_{\tau3})}\right)\,.
\end{align}
%
The phases in the matrix $P_2$ are unphysical. 
The Majorana phases get contribution from the matrix $Q_2Q_0$ and read:
\begin{equation}
\frac{\alpha_{21}}{2} = \beta_{e2} - \beta_{e1} +\frac{\xi_{21}}{2}\,,
\quad 
\frac{\alpha_{31}}{2} = \beta_{e2} + \beta_{\mu3} - \phi + \beta_{\tau3} + \frac{\xi_{31}}{2}\,,
\quad
\beta_{\tau3}=0~{\rm or}~\pi\,.
\label{Majph2131C1}
\end{equation}
%
 
 In terms of the standard parametrisation mixing angles 
$\theta_{12}$, $\theta_{23}$, $\theta_{13}$ and the 
Dirac phase $\delta$ we have:
\begin{align}
\label{be1tau2phC1}
\beta_{e1} + \beta_{\mu3} - \phi & = 
\arg\left(U_{\tau2} e^{-i \frac{\alpha_{21}}{2}}\right) - \beta_{\tau2} =   
\arg\left(- c_{12} s_{23} - s_{12} c_{23} s_{13} e^{i \delta}\right) - \beta_{\tau2}\,,
\\[0.2cm]
\label{be2tau1phC1}
\beta_{e2} + \beta_{\mu3} - \phi & = 
\arg\left(U_{\tau1}\right) - \beta_{\tau1} = 
\arg\left(s_{12} s_{23} - c_{12} c_{23} s_{13} e^{i \delta}\right) - \beta_{\tau1}\,,
\end{align}
%
where $\beta_{\tau1}$ and $\beta_{\tau2}$ can be $0$ or $\pi$ 
and are known when the angles $\theta^\nu_{12}$, $\theta^\nu_{23}$ 
and $\theta^\nu_{13}$ are fixed 
(see eqs.~(\ref{eq:tau1C1}) and (\ref{eq:tau2C1})). 

 The mixing angles $\theta_{12}$, $\theta_{23}$ and $\theta_{13}$ 
of the standard parametrisation are related with the angles 
$\theta^e_{12}$, $\theta^\nu_{ij}$ and the phase $\phi$ 
present in the parametrisation of $U$ given in eq.~(\ref{eq:UC1a}) 
as follows:
\begin{align}
\sin^2 \theta_{13} & = |U_{e3}|^2  = 
\sin^2 \theta^e_{12} \sin^2 \theta^{\nu}_{23} \cos^2 \theta^{\nu}_{13} 
+ \cos^2 \theta^e_{12} \sin^2 \theta^{\nu}_{13} - X_{12} \sin \theta^{\nu}_{13} \,, 
\label{eq:th13C1}\\
\sin^2 \theta_{23} & = \frac{|U_{\mu3}|^2}{1-|U_{e3}|^2} = 
1 - \frac{\cos^2 \theta^{\nu}_{23} \cos^2 \theta^{\nu}_{13}}
{1 - \sin^2 \theta_{13}} \,,
\label{eq:th23C1} \\
\sin^2 \theta_{12} & = \frac{|U_{e2}|^2}{1-|U_{e3}|^2} = \frac{1}{1-\sin^2 \theta_{13}} 
\bigg[\sin^2 \theta^e_{12} \left( \cos \theta^{\nu}_{12} \cos \theta^{\nu}_{23} 
- \sin \theta^{\nu}_{12} \sin \theta^{\nu}_{23} \sin \theta^{\nu}_{13} \right)^2 \nonumber \\
& +  \cos^2 \theta^e_{12} \sin^2 \theta^{\nu}_{12} \cos^2 \theta^{\nu}_{13} 
- X_{12} \sin \theta^{\nu}_{12} \left(\cos \theta^{\nu}_{12} \cot \theta^{\nu}_{23} 
- \sin \theta^{\nu}_{12} \sin \theta^{\nu}_{13}\right) \bigg] 
\label{eq:th12C1} \,,
\end{align}
%
where
\begin{equation}
X_{12} = - \sin 2 \theta^e_{12} \sin \theta^{\nu}_{23} \cos \theta^{\nu}_{13} \cos \phi \,.
\end{equation}
%
We notice that eqs.~(\ref{eq:th13C1})~--~(\ref{eq:th12C1}) are the
generalisation of eqs.~(66)~--~(68) in ref.~\cite{Girardi:2015vha}
for an arbitrary fixed value of $\theta^{\nu}_{23}$.
Solving eq.~(\ref{eq:th13C1}) for $X_{12}$ and inserting 
the solution in eq.~(\ref{eq:th12C1}), 
we find $\sin^2 \theta_{12}$
as a function of $\theta_{13}$, $\theta^{\nu}_{12}$, $\theta^{\nu}_{13}$, 
$\theta^\nu_{23}$ and $\theta^e_{12}$:
\be
\sin^2 \theta_{12} = 
\frac{\alpha \sin^2 \theta^e_{12} + \beta}{1-\sin^2 \theta_{13}} \,.
\ee
%
Here 
\begin{align}
\alpha & = \cos 2 \theta^{\nu}_{12} \cos^2 \theta^{\nu}_{23} +  \dfrac{1}{2} \sin 2 \theta^{\nu}_{12} \cos \theta^{\nu}_{23} \sin \theta^{\nu}_{13} \left ( \dfrac{\cos^2 \theta^{\nu}_{23}}{\sin \theta^{\nu}_{23}} - \dfrac{\sin \theta^{\nu}_{23}}{\sin^2 \theta^{\nu}_{13}} \right )\,, \\
\beta & = \sin \theta^{\nu}_{12} \left[ \cos^2 \theta_{13} \sin \theta^{\nu}_{12} 
- \cos \theta^{\nu}_{12} \cot \theta^{\nu}_{23} \left( \sin \theta^{\nu}_{13} 
- \frac{\sin^2 \theta_{13}}{\sin \theta^{\nu}_{13}} \right)\right] \,.
\end{align}
%
Inverting the formula for $\sin^2 \theta_{12}$ allows us to express
$\sin^2 \theta^e_{12}$ in terms of $\theta_{12}$, $\theta_{13}$, 
$\theta^{\nu}_{12}$, $\theta^{\nu}_{13}$ and $\theta^{\nu}_{23}$:
\begin{equation}
\sin^2\theta^e_{12} = \frac
{2\cos^2\theta_{13} \tan\theta^\nu_{23} \sin\theta^\nu_{13} 
\left(\sin^2\theta_{12} - \sin^2\theta^\nu_{12}\right)
- \sin2\theta^\nu_{12} \left(\sin^2\theta_{13} - \sin^2\theta^\nu_{13}\right)}
{\cos2\theta^\nu_{12} \sin2\theta^\nu_{23} \sin\theta^\nu_{13}
+ \sin2\theta^\nu_{12} \left(\cos2\theta^\nu_{23} - \cos^2\theta^\nu_{23} \cos^2\theta^\nu_{13}\right)}\,.
\label{eq:SSe12C1}
\end{equation}
%

 Using eq.~(\ref{eq:th13C1}), we can express
$\cos\phi$ in terms of the angle $\theta_{13}$, 
the angles $\theta^{\nu}_{12}$, $\theta^{\nu}_{13}$ and $\theta^{\nu}_{23}$ 
which are assumed to have known values 
and the angle $\theta^e_{12}$ whose value is fixed by eq.~(\ref{eq:SSe12C1}):
\begin{equation}
\label{eq:phiC1}
\cos \phi  = \frac
{\sin^2\theta_{13} - \cos^2\theta^e_{12} \sin^2\theta^\nu_{13} 
- \sin^2\theta^e_{12} \sin^2\theta^\nu_{23} \cos^2\theta^\nu_{13}}
{\sin2\theta^e_{12} \sin\theta^\nu_{23} \sin\theta^\nu_{13} \cos\theta^\nu_{13}}\,.
\end{equation}
%
 
 Using eqs.~(\ref{Ue1C1})~--~(\ref{Ue3C1}), (\ref{Umu3C1}), (\ref{eq:UC1elements}), 
(\ref{eq:deltaC1}) and (\ref{Majph2131C1}), we find:
\begin{align}
\sin(\alpha_{21}/2) &= \frac{1}{2 |U_{e1} U_{e2}|}
\bigg[\cos^2\theta^e_{12} \sin2\theta^{\nu}_{12} \cos^2\theta^{\nu}_{13} \sin(\xi_{21}/2) 
- 2 \sin^2\theta^e_{12} \sin (\xi_{21}/2) \nonumber \\
&\times \big(\sin\theta^{\nu}_{12} \cos\theta^{\nu}_{23} 
+ \cos\theta^{\nu}_{12} \sin\theta^{\nu}_{23} \sin\theta^{\nu}_{13}\big) 
\big(\cos\theta^{\nu}_{12} \cos\theta^{\nu}_{23} 
- \sin\theta^{\nu}_{12} \sin\theta^{\nu}_{23} \sin\theta^{\nu}_{13}\big) \nonumber \\
&+ \sin2\theta^e_{12} \cos\theta^{\nu}_{13} 
\Big((\cos2\theta^{\nu}_{12} \cos\theta^{\nu}_{23}
- \sin2\theta^{\nu}_{12} \sin\theta^{\nu}_{23} \sin\theta^{\nu}_{13}) 
\cos\phi \sin(\xi_{21}/2) \nonumber \\
&+ \cos\theta^{\nu}_{23} \sin\phi \cos(\xi_{21}/2)\Big) \bigg]\,,
\label{eq:alpha21C1a}\\
\sin(\alpha_{31}/2 - \delta) &= \frac{1}{2 |U_{e1} U_{e3}|}
\bigg[ \cos^2\theta^e_{12} \cos\theta^{\nu}_{12} \sin2\theta^{\nu}_{13} \sin(\xi_{31}/2) 
- 2 \sin^2\theta^e_{12}  \sin\theta^{\nu}_{23} \cos\theta^{\nu}_{13}  
\nonumber \\
&\times (\sin\theta^{\nu}_{12} \cos\theta^{\nu}_{23} 
+ \cos\theta^{\nu}_{12} \sin\theta^{\nu}_{23} \sin\theta^{\nu}_{13}) \sin(\xi_{31}/2) 
\nonumber \\
&+ \sin2\theta^e_{12} \Big( \cos \theta^{\nu}_{12} \sin \theta^{\nu}_{23} 
(\cos 2 \theta^{\nu}_{13} \cos\phi \sin(\xi_{31}/2) + \sin\phi \cos(\xi_{31}/2)) 
\nonumber \\
&- \sin\theta^{\nu}_{12} \cos\theta^{\nu}_{23} \sin\theta^{\nu}_{13} \sin(\xi_{31}/2 - \phi)\Big)\bigg]\,,
\label{eq:alpha31dC1a}\\
\sin(\alpha_{31}/2) &= \frac{\cos\theta^\nu_{23} \cos\theta^\nu_{13}}
{2 |U_{e2} U_{\mu3} U_{\tau3}|}
\bigg[\sin\theta^\nu_{12} \Big( \left(1 + \cos2\theta^e_{12} \cos2\theta^\nu_{13}\right) 
\sin\theta^\nu_{23} \sin(\xi_{31}/2) \nonumber\\
&- \frac{1}{2} \sin2\theta^e_{12} \sin2\theta^\nu_{13} 
\left(\sin(\xi_{31}/2 - \phi) + \sin^2\theta^\nu_{23} \sin(\xi_{31}/2 + \phi)\right) \Big) 
+ \cos\theta^\nu_{12} \nonumber\\
&\times \Big(\frac{1}{2} 
\sin2\theta^e_{12} \sin2\theta^\nu_{23} \cos\theta^\nu_{13} \sin(\xi_{31}/2 + \phi) 
- 2 \sin^2\theta^e_{12} \cos\theta^\nu_{23} \sin\theta^\nu_{13} \sin(\xi_{31}/2)\Big)\bigg]\,.
\label{eq:alpha31C1a}
\end{align}
%
It is straightforward to check that in the limit of zero 
$\theta^\nu_{13}$ eqs.~(\ref{eq:alpha21C1a}), (\ref{eq:alpha31dC1a}) 
and (\ref{eq:alpha31C1a}) reduce to 
eqs.~(\ref{eq:alpha21A0gen}), (\ref{eq:alpha31dA0gen}) and 
(\ref{eq:alpha31A0gen}), respectively.
 The formulae for $\cos (\alpha_{21}/2)$, $\cos (\alpha_{31}/2 - \delta)$ and 
$\cos (\alpha_{31}/2)$
can be obtained from eqs.~(\ref{eq:alpha21C1a})~--~(\ref{eq:alpha31C1a}) by 
changing $\xi_{21}$ to $\xi_{21} + \pi$ and $\xi_{31}$ to $\xi_{31} + \pi$, 
respectively.
The results for  $\sin(\alpha_{21}/2)$ and $\sin (\alpha_{31}/2 - \delta)$
can also be obtained by comparing the expressions for the rephasing invariants 
$I_1$ and $I_2$ in the standard parametrisation of the PMNS matrix 
and in the parametrisation given in eq.~(\ref{eq:UC1a}).

 Using eqs.~(\ref{be1tau2phC1}) and (\ref{be2tau1phC1}) we get 
in terms of the standard parametrisation mixing angles 
$\theta_{12}$, $\theta_{13}$, $\theta_{23}$ and 
the Dirac phase $\delta$:
\begin{align}
\sin (\alpha_{21}/2) & = \frac{\cos(\beta_{\tau2} - \beta_{\tau1})}{2 |U_{\tau1} U_{\tau2}|}
\bigg[ \sin2\theta_{23} \sin\theta_{13}
\big(\sin(\delta + \xi_{21}/2)
- 2\sin^2\theta_{12} \cos\delta \sin(\xi_{21}/2)\big) 
\nonumber 
\label{eq:alpha21C1b}\\
& - \sin 2 \theta_{12}\big(\sin^2 \theta_{23} -  
\cos^2 \theta_{23} \sin^2\theta_{13}
  \big) \sin(\xi_{21}/2) \bigg] \,, 
\quad
\cos(\beta_{\tau2} - \beta_{\tau1})=+1~{\rm or}~(-1)\,, \\
\sin (\alpha_{31}/2) & = \frac{\cos(\beta_{\tau3} - \beta_{\tau1})}{|U_{\tau1}|} 
\bigg[
\sin\theta_{12}\sin\theta_{23} \sin(\xi_{31}/2) 
\nonumber
\\
& - \cos\theta_{12}\cos\theta_{23}\sin\theta_{13} 
\sin (\delta + \xi_{31}/2) \bigg] \,, 
\quad
\cos(\beta_{\tau3} - \beta_{\tau1}) = +1~{\rm or}~(-1)\,,
\label{eq:alpha31C1b}
\end{align}
%
where, according to eq.~(\ref{eq:th23C1}),
$\cos^2\theta_{23} \cos^2\theta_{13} = \cos^2\theta^\nu_{23} \cos^2\theta^\nu_{13}$.
Note that, as it follows from eqs.~(\ref{eq:tau1C1})~--~(\ref{eq:UC1elements}), 
the sign factors $\cos(\beta_{\tau2} - \beta_{\tau1})$ 
and $\cos(\beta_{\tau3} - \beta_{\tau1})$ are known 
when the angles $\theta^\nu_{ij}$ are fixed.

 Finally, we give the expressions for 
$\sin (\alpha_{21}/2 - \xi_{21}/2)$, 
$\sin (\alpha_{31}/2 - \xi_{31}/2)$ 
and $\sin (\alpha_{31}/2 - \delta - \xi_{31}/2)$, 
which have a simpler form:
\begin{align}
\sin(\alpha_{21}/2 - \xi_{21}/2) &= \frac{\sin2\theta^e_{12}}{2 |U_{e1} U_{e2}|}
\cos\theta^\nu_{23} \cos\theta^\nu_{13} \sin\phi \nonumber\\ 
& = \frac{\cos(\beta_{\tau2} - \beta_{\tau1})}{2 |U_{\tau1} U_{\tau2}|}
\sin2\theta_{23} \sin\theta_{13} \sin\delta\,, 
\label{a21x21C1}
\\[0.2cm]
\sin(\alpha_{31}/2 - \xi_{31}/2) &= \frac{\sin2\theta^e_{12} \cos^2\theta^\nu_{23}}
{2 |U_{e2} U_{\mu3} U_{\tau3}|}
\cos^2\theta^\nu_{13}
\left(\cos\theta^\nu_{12} \sin\theta^\nu_{23} 
+ \sin\theta^\nu_{12} \cos\theta^\nu_{23} \sin\theta^\nu_{13}\right) \sin\phi 
\nonumber\\
\label{a31x31C1}
&= - \frac{\cos(\beta_{\tau3} - \beta_{\tau1})}{|U_{\tau1}|}
\cos\theta_{12} \cos\theta_{23} \sin\theta_{13} \sin\delta\,, \\[0.2cm]
\sin(\alpha_{31}/2 - \delta - \xi_{31}/2) &= \frac{\sin2\theta^e_{12}}{2 |U_{e1} U_{e3}|}
\left(\cos\theta^\nu_{12} \sin\theta^\nu_{23} 
+ \sin\theta^\nu_{12} \cos\theta^\nu_{23} \sin\theta^\nu_{13}\right) \sin\phi \nonumber\\
&= - \frac{\cos(\beta_{\tau3} - \beta_{\tau1})}{|U_{\tau1}|}
\sin\theta_{12} \sin\theta_{23} \sin\delta\,.
\end{align}
%
Equations (\ref{a21x21C1}) and (\ref{a31x31C1}) imply, in particular, that 
$|\sin(\alpha_{21(31)}/2 - \xi_{21(31)}/2)| \propto \sin\theta_{13}$.

%
\subsection{The Scheme with $\boldsymbol{\theta^e_{13} - 
(\theta^\nu_{23}, \theta^\nu_{13}, \theta^\nu_{12})}$ Rotations (Case C2)}
\label{sec:13e23nu13nu12nu}
%

 In this subsection we derive the formulae for the Majorana phases 
in the case when the PMNS matrix $U$ is parametrised 
as in eq.~(\ref{eq:Uija}) with $(ij) = (13)$,
i.e.,
\begin{equation}
U = R_{13}(\theta^e_{13}) \, \Psi \, R_{23}(\theta^{\nu}_{23})\,
R_{13}(\theta^{\nu}_{13}) \, R_{12}(\theta^{\nu}_{12}) \, Q_0 \,.
\label{eq:UC2a} 
\end{equation}
%
In this case the phase $\psi$ in the matrix $\Psi$ is unphysical, 
and $\Psi = \diag \, (1, 1, e ^{-i \omega})$. 
We will proceed in analogy with the previous subsection.
We start by writing the matrix $U$ in explicit form:
\begin{equation} 
U = \begin{pmatrix} 
|U_{e1}| e^{i \beta_{e1}} & |U_{e2}| e^{i \beta_{e2}} & |U_{e3}| e^{i \beta_{e3}} \\[0.2cm]  
|U_{\mu1}| e^{i \beta_{\mu1}} &  |U_{\mu2}| e^{i \beta_{\mu2}} & |U_{\mu3}| e^{i \beta_{\mu3}} \\[0.2cm]
|U_{\tau1}| e^{i \beta_{\tau1}}& |U_{\tau2}| e^{i \beta_{\tau2}} & |U_{\tau3}| e^{i \beta_{\tau3}}
\end{pmatrix}
Q_0\,,
\label{eq:UC2b}
\end{equation}
%
where
\begin{align}
\label{be1C2}
|U_{e1}| e^{i \beta_{e1}} &= c^e_{13} c^\nu_{12} c^\nu_{13} + s^e_{13} 
\left(s^\nu_{12} s^\nu_{23} - c^\nu_{12} c^\nu_{23} s^\nu_{13}\right) e^{- i \omega}\,,\\
\label{be2C2}
|U_{e2}| e^{i \beta_{e2}} &= c^e_{13} s^\nu_{12} c^\nu_{13} - s^e_{13} 
\left(c^\nu_{12} s^\nu_{23} + s^\nu_{12} c^\nu_{23} s^\nu_{13}\right) e^{-i \omega}\,,\\
\label{be3C2}
|U_{e3}| e^{i \beta_{e3}} &= c^e_{13} s^\nu_{13} + s^e_{13} c^\nu_{23} c^\nu_{13} e^{- i \omega}\,,\\
\label{Umu1C2}
|U_{\mu1}| e^{i \beta_{\mu1}} &= - s^\nu_{12} c^\nu_{23} - c^\nu_{12} s^\nu_{23} s^\nu_{13}\,, \\
\label{Umu2C2}
|U_{\mu2}| e^{i \beta_{\mu2}} &= c^\nu_{12} c^\nu_{23} - s^\nu_{12} s^\nu_{23} s^\nu_{13}\,, \\
\label{Umu3C2}
|U_{\mu3}| e^{i \beta_{\mu3}} &= s^\nu_{23} c^\nu_{13}\,, \\
|U_{\tau1}| e^{i \beta_{\tau1}} &= - s^e_{13} c^\nu_{12} c^\nu_{13} + c^e_{13} 
\left(s^\nu_{12} s^\nu_{23} - c^\nu_{12} c^\nu_{23} s^\nu_{13}\right) e^{- i \omega}\,,\\
|U_{\tau2}| e^{i \beta_{\tau2}} &= - s^e_{13} s^\nu_{12} c^\nu_{13} - c^e_{13} 
\left(c^\nu_{12} s^\nu_{23} + s^\nu_{12} c^\nu_{23} s^\nu_{13}\right) e^{- i \omega}\,,\\
\label{btau3C2}
|U_{\tau3}| e^{i \beta_{\tau3}} &= - s^e_{13} s^\nu_{13} + c^e_{13} c^\nu_{23} c^\nu_{13} e^{- i \omega}\,.
\end{align}
%
From the comparison of the expressions for $J_{\rm CP}$ 
in the standard parametrisation and in the parametrisation given 
in eq.~(\ref{eq:UC2a}), it follows that~%
\footnote{For $\theta^\nu_{23} = -\pi/4$ this relation reduces to 
eq.~(91) in ref.~\cite{Girardi:2015vha}.} 
\begin{align}
\sin\delta = \frac{\sin2\theta^e_{13}
\left[\left(\cos^2\theta^\nu_{13} - (\cos^2\theta^\nu_{13} - 2) \cos2\theta^\nu_{23}\right)
\sin2\theta^\nu_{12}
+ 2 \cos2\theta^\nu_{12} \sin2\theta^\nu_{23} \sin\theta^\nu_{13}\right]}
{2\,{\rm sgn}(\sin\theta^\nu_{23} \cos\theta^\nu_{13}) \sin2\theta_{12} \sin2\theta_{13} \cos\theta_{23}} \sin\omega\,,
\label{eq:JCPC2}
\end{align}
%
where we have used the equality 
$\sin^2\theta_{23} \cos^2\theta_{13} = \sin^2\theta^\nu_{23} \cos^2\theta^\nu_{13}$ valid 
in this scheme. 
As can be shown, the relation between $\sin\delta$ and $\sin\omega$ 
in eq.~(\ref{eq:JCPC2}) takes place if 
\begin{equation}
\delta = \beta_{e1} + \beta_{e2} + \beta_{\tau3} - \beta_{e3} 
+ \omega + \beta_{\mu3}\,, 
\quad
\beta_{\mu3}=0~{\rm or}~\pi\,,
\label{eq:deltaC2}
\end{equation}
%
where $\beta_{\mu3} = \arg(s^\nu_{23} c^\nu_{13})$.
Knowing the expression for $\delta$ allows us to rearrange 
the phases in eq. (\ref{eq:UC2b}) in such a way as to render $U$ 
in the standard parametrisation form:
\begin{equation} 
U = P_2\, 
\begin{pmatrix} 
|U_{e1}| & |U_{e2}|  & |U_{e3}| e^{- i \delta} \\[0.2cm]
|U_{\mu1}| e^{i (\beta_{\mu1} + \beta_{e2} + \beta_{\tau3}  + \omega)}
& |U_{\mu2}| e^{ i (\beta_{\mu2} +\beta_{e1} + \beta_{\tau3} + \omega)} 
& |U_{\mu3}| \\[0.2cm]
|U_{\tau1}| e^{ i (\beta_{\tau1} + \beta_{e2} + \omega + \beta_{\mu3})}&
|U_{\tau2}| e^{ i (\beta_{\tau2} + \beta_{e1} + \omega + \beta_{\mu3})}& 
|U_{\tau3}|
\end{pmatrix}
Q_2\,Q_0\,,
\label{eq:UC2c}
\end{equation}
%
with
\begin{align}
\label{P2C2}
P_2 &=\diag\left(e^{i (\beta_{e1} + \beta_{e2} + \beta_{\tau3} + \omega)}, 1,
e^{i (\beta_{\tau3} - \beta_{\mu3})}\right)\,, \\[0.2cm]
\label{Q2C2}
Q_2 &= \diag\left(e^{-i (\beta_{e2} + \beta_{\tau3} + \omega)}, 
e^{-i (\beta_{e1} + \beta_{\tau3} + \omega)}, e^{i \beta_{\mu3}}\right) \nonumber\\[0.2cm]
&= e^{-i (\beta_{e2} + \beta_{\tau3} + \omega)}\,
\diag\left(1, e^{i (\beta_{e2} - \beta_{e1})}, 
e^{i (\beta_{e2} + \beta_{\tau3} + \omega + \beta_{\mu3})}\right)\,.
\end{align}
%
The matrix $P_2$ contains unphysical phases which can be removed. 
The Majorana phases are determined by the phases in the product $Q_2\,Q_0$:
\begin{equation}
\frac{\alpha_{21}}{2} = \beta_{e2} - \beta_{e1} +\frac{\xi_{21}}{2}\,,
\quad 
\frac{\alpha_{31}}{2} = \beta_{e2} + \beta_{\tau3} + \omega + \beta_{\mu3} + \frac{\xi_{31}}{2}\,, 
\quad
\beta_{\mu3}=0~{\rm or}~\pi\,. 
\label{Majph2131C2}
\end{equation}
%

 In terms of the ``standard'' mixing angles $\theta_{12}$, $\theta_{23}$, 
$\theta_{13}$ and the Dirac phase $\delta$ one has: 
\begin{align}
\label{be1mu2phC2}
\beta_{e1} + \beta_{\tau3} + \omega & = 
\arg\left(U_{\mu2} e^{-i \frac{\alpha_{21}}{2}}\right) - \beta_{\mu2} =   
\arg\left(c_{12} c_{23} - s_{12} s_{23} s_{13} e^{i \delta}\right) - \beta_{\mu2}\,,
\\[0.2cm]
\label{be2mu1phC2}
\beta_{e2} + \beta_{\tau3} + \omega & = 
\arg\left(U_{\mu1}\right) - \beta_{\mu1} = 
\arg\left(- s_{12} c_{23} - c_{12} s_{23} s_{13} e^{i \delta}\right) - \beta_{\mu1}\,,
\end{align}
%
where $\beta_{\mu1} = 
{\rm arg}( - s^\nu_{12} c^\nu_{23} - c^\nu_{12} s^\nu_{23} s^\nu_{13})$ and 
$\beta_{\mu2}  = 
{\rm arg}(c^\nu_{12} c^\nu_{23} - s^\nu_{12} s^\nu_{23} s^\nu_{13})$ 
can take values of $0$ or $\pi$ 
and are known when the angles $\theta^\nu_{12}$, $\theta^\nu_{23}$ 
and $\theta^\nu_{13}$ are fixed.

 The mixing angles $\theta_{12}$, $\theta_{23}$ and $\theta_{13}$ 
of the standard parametrisation are related with the angles 
$\theta^e_{13}$, $\theta^\nu_{ij}$ and the phase $\omega$ 
present in the parametrisation of $U$ given in eq.~(\ref{eq:UC2a}) 
in the following way:
\begin{align}
\sin^2 \theta_{13} &= |U_{e3}|^2 = 
\sin^2\theta^e_{13} \cos^2\theta^{\nu}_{23} \cos^2\theta^{\nu}_{13} 
+ \cos^2\theta^e_{13} \sin^2\theta^{\nu}_{13} + X_{13} \sin\theta^{\nu}_{13}\,, 
\label{eq:th13C2}\\
\sin^2 \theta_{23} & = \frac{|U_{\mu3}|^2}{1-|U_{e3}|^2} = 
\frac{\sin^2\theta^{\nu}_{23} \cos^2\theta^{\nu}_{13}}{1 - \sin^2\theta_{13}}\,, 
\label{eq:th23C2}\\
\sin^2 \theta_{12} & = \frac{|U_{e2}|^2}{1-|U_{e3}|^2} = \frac{1}{1-\sin^2 \theta_{13}} 
\bigg[ \sin^2\theta^e_{13} \left(\cos\theta^{\nu}_{12} \sin\theta^{\nu}_{23} 
+ \sin\theta^{\nu}_{12} \cos\theta^{\nu}_{23} \sin\theta^{\nu}_{13}\right)^2 \nonumber \\
&+ \cos^2\theta^e_{13} \sin^2\theta^{\nu}_{12} \cos^2\theta^{\nu}_{13} 
- X_{13} \sin\theta^{\nu}_{12} \left(\cos \theta^{\nu}_{12} \tan\theta^{\nu}_{23} 
+ \sin\theta^{\nu}_{12} \sin\theta^{\nu}_{13}\right) \bigg]\,,
\label{eq:th12C2}
\end{align}
%
where
\begin{equation}
X_{13} = \sin2\theta^e_{13} \cos\theta^{\nu}_{23} \cos\theta^{\nu}_{13} \cos\omega\,.
\end{equation}
%
Equations~(\ref{eq:th13C2})~--~(\ref{eq:th12C2}) are the
generalisation of eqs.~(82)~--~(84) in ref.~\cite{Girardi:2015vha}
for an arbitrary fixed value of $\theta^{\nu}_{23}$.
Solving eq.~(\ref{eq:th13C2}) for $X_{13}$ and inserting 
the solution in eq.~(\ref{eq:th12C2}), one finds $\sin^2\theta_{12}$
as a function of $\theta_{13}$, $\theta^{\nu}_{12}$, $\theta^{\nu}_{13}$, 
$\theta^\nu_{23}$ and $\theta^e_{13}$:
\begin{equation}
\sin^2 \theta_{12}  = \frac{\rho \sin^2\theta^e_{13} + \eta}{1 - \sin^2 \theta_{13}}\,,
\label{s2th12rhoeta}
\end{equation}
%
where $\rho$ and $\eta$ are given by
%
\begin{align}
\rho &= \cos2\theta^{\nu}_{12} \sin^2\theta^{\nu}_{23} 
- \frac{1}{2} \sin2\theta^\nu_{12} \sin\theta^\nu_{23} \sin\theta^\nu_{13} 
\left(\frac{\sin^2\theta^\nu_{23}}{\cos\theta^\nu_{23}} 
- \frac{\cos\theta^\nu_{23}}{\sin^2\theta^\nu_{13}}\right)\,, 
\label{eq:rhoC2} \\
\eta & = \sin \theta^{\nu}_{12} \left[ \cos^2 \theta_{13} \sin \theta^{\nu}_{12} + \cos \theta^{\nu}_{12} \tan \theta^{\nu}_{23} \left( \sin \theta^{\nu}_{13} - \frac{\sin^2 \theta_{13}}{\sin \theta^{\nu}_{13}} \right)\right] \,.
\label{eq:eta_rho_13b}
\end{align}
%
From eq.~(\ref{s2th12rhoeta}) we can express $\sin^2\theta^e_{13}$ 
as a function of $\theta_{12}$, $\theta_{13}$, 
$\theta^{\nu}_{12}$, $\theta^{\nu}_{13}$ and $\theta^{\nu}_{23}$:
\begin{equation}
\sin^2\theta^e_{13} = \frac
{2\cos^2\theta_{13} \cot\theta^\nu_{23} \sin\theta^\nu_{13} 
\left(\sin^2\theta_{12} - \sin^2\theta^\nu_{12}\right)
+ \sin2\theta^\nu_{12} \left(\sin^2\theta_{13} - \sin^2\theta^\nu_{13}\right)}
{\cos2\theta^\nu_{12} \sin2\theta^\nu_{23} \sin\theta^\nu_{13}
+ \sin2\theta^\nu_{12} \left(\cos2\theta^\nu_{23} + \sin^2\theta^\nu_{23} \cos^2\theta^\nu_{13}\right)}\,.
\label{eq:SSe13C2}
\end{equation}
%

 Using eq.~(\ref{eq:th13C2}), we can write
$\cos\omega$ in terms of the angle $\theta_{13}$, 
the angles $\theta^{\nu}_{12}$, $\theta^{\nu}_{13}$ and $\theta^{\nu}_{23}$  
which are assumed to have known values  
and the angle $\theta^e_{13}$ whose 
value is fixed by eq.~(\ref{eq:SSe13C2}):
\begin{equation}
\label{eq:omegaC2}
\cos \omega  = \frac
{\sin^2\theta_{13} - \cos^2\theta^e_{13} \sin^2\theta^\nu_{13} 
- \sin^2\theta^e_{13} \cos^2\theta^\nu_{23} \cos^2\theta^\nu_{13}}
{\sin2\theta^e_{13} \cos\theta^\nu_{23} \sin\theta^\nu_{13} \cos\theta^\nu_{13}}\,.
\end{equation}
%
Thus, we have at our disposal expressions for 
$\sin^2\theta^e_{13}$ and $\cos\omega$ in terms of the 
known angles.

 Further, using eqs.~(\ref{be1C2})~--~(\ref{be3C2}), (\ref{Umu3C2}), (\ref{btau3C2}), 
(\ref{eq:deltaC2}) and (\ref{Majph2131C2}), we get:
\begin{align}
\sin(\alpha_{21}/2) &= \frac{1}{2 |U_{e1} U_{e2}|}
\bigg[\cos^2\theta^e_{13} \sin2\theta^{\nu}_{12} \cos^2\theta^{\nu}_{13} \sin(\xi_{21}/2) 
- 2 \sin^2\theta^e_{13} \sin (\xi_{21}/2) \nonumber \\
&\times \big(\sin\theta^{\nu}_{12} \sin\theta^{\nu}_{23} 
- \cos\theta^{\nu}_{12} \cos\theta^{\nu}_{23} \sin\theta^{\nu}_{13}\big) 
\big(\cos\theta^{\nu}_{12} \sin\theta^{\nu}_{23} 
+ \sin\theta^{\nu}_{12} \cos\theta^{\nu}_{23} \sin\theta^{\nu}_{13}\big) \nonumber \\
&- \sin2\theta^e_{13} \cos\theta^{\nu}_{13} 
\Big((\cos2\theta^{\nu}_{12} \sin\theta^{\nu}_{23}
+ \sin2\theta^{\nu}_{12} \cos\theta^{\nu}_{23} \sin\theta^{\nu}_{13}) 
\cos\omega \sin(\xi_{21}/2) \nonumber \\
&- \sin\theta^{\nu}_{23} \sin\omega \cos(\xi_{21}/2)\Big) \bigg]\,,
\label{eq:alpha21C2a}\\
\sin(\alpha_{31}/2 - \delta) &= \frac{1}{2 |U_{e1} U_{e3}|}
\bigg[ \cos^2\theta^e_{13} \cos\theta^{\nu}_{12} \sin2\theta^{\nu}_{13} \sin(\xi_{31}/2) 
+ 2 \sin^2\theta^e_{13} \cos\theta^{\nu}_{23} \cos\theta^{\nu}_{13}  
\nonumber \\
&\times \left(\sin\theta^{\nu}_{12} \sin\theta^{\nu}_{23} 
- \cos\theta^{\nu}_{12} \cos\theta^{\nu}_{23} \sin\theta^{\nu}_{13}\right) \sin(\xi_{31}/2) 
\nonumber \\
&+ \sin2\theta^e_{13} \Big( \cos \theta^{\nu}_{12} \cos\theta^{\nu}_{23} 
\left(\cos2\theta^{\nu}_{13} \cos\omega \sin(\xi_{31}/2) - \sin\omega \cos(\xi_{31}/2)\right) 
\nonumber \\
&+ \sin\theta^{\nu}_{12} \sin\theta^{\nu}_{23} \sin\theta^{\nu}_{13} \sin(\xi_{31}/2 + \omega)\Big)\bigg]\,,
\label{eq:alpha31dC2a}\\
\sin(\alpha_{31}/2) &= \frac{\sin\theta^\nu_{23} \cos\theta^\nu_{13}}
{2 |U_{e2} U_{\mu3} U_{\tau3}|}
\bigg[\sin\theta^\nu_{12} \Big( \left(1 + \cos2\theta^e_{13} \cos2\theta^\nu_{13}\right) 
\cos\theta^\nu_{23} \sin(\xi_{31}/2) \nonumber\\
&- \frac{1}{2} \sin2\theta^e_{13} \sin2\theta^\nu_{13} 
\left(\sin(\xi_{31}/2 + \omega) + \cos^2\theta^\nu_{23} \sin(\xi_{31}/2 - \omega)\right) \Big) 
- \cos\theta^\nu_{12} \nonumber\\
&\times \Big(\frac{1}{2} 
\sin2\theta^e_{13} \sin2\theta^\nu_{23} \cos\theta^\nu_{13} \sin(\xi_{31}/2 - \omega) 
- 2 \sin^2\theta^e_{13} \sin\theta^\nu_{23} \sin\theta^\nu_{13} \sin(\xi_{31}/2)\Big)\bigg]\,.
\label{eq:alpha31C2a}
\end{align}
%
The results given in eqs. (\ref{eq:alpha21C2a}) and  (\ref{eq:alpha31dC2a})
can also be derived by comparing the expressions for the rephasing invariants 
$I_1$ and $I_2$ in the standard parametrisation of the PMNS matrix 
and in the parametrisation given in eq.~(\ref{eq:UC2a}).
The formulae for $\cos (\alpha_{21}/2)$,  $\cos (\alpha_{31}/2 - \delta)$ and 
$\cos (\alpha_{31}/2)$
can be obtained from eqs.~(\ref{eq:alpha21C2a}), (\ref{eq:alpha31dC2a}) and 
(\ref{eq:alpha31C2a}) by changing $\xi_{21}$ to $\xi_{21} + \pi$ and $\xi_{31}$ to $\xi_{31} + \pi$, 
respectively. In the limit of zero $\theta^\nu_{13}$, 
eqs.~(\ref{eq:alpha21C2a}), (\ref{eq:alpha31dC2a}) 
and (\ref{eq:alpha31C2a}) reduce to 
eqs.~(\ref{eq:alpha21B0gen}), (\ref{eq:alpha31dB0gen}) and 
(\ref{eq:alpha31B0gen}), respectively, as could be expected.

 Using relations in eqs.~(\ref{be1mu2phC2}) and (\ref{be2mu1phC2}) 
we get in terms of the standard parametrisation mixing angles 
$\theta_{12}$, $\theta_{13}$, $\theta_{23}$ and the Dirac phase $\delta$:
\begin{align}
\sin (\alpha_{21}/2) & = - \frac{\cos(\beta_{\mu2} - \beta_{\mu1})}{2 |U_{\mu1} U_{\mu2}|}
\bigg[ \sin2\theta_{23} \sin\theta_{13}
\big(\sin(\delta + \xi_{21}/2)
- 2\sin^2\theta_{12} \cos\delta \sin(\xi_{21}/2)\big) 
\nonumber 
\label{eq:alpha21C1b}\\
&+ \sin 2 \theta_{12}\big(\cos^2 \theta_{23} -  
\sin^2 \theta_{23} \sin^2\theta_{13}
\big) \sin(\xi_{21}/2) \bigg] \,,
\quad
\cos(\beta_{\mu2} - \beta_{\mu1}) = +1~{\rm or}~(-1)\,, 
\\
\sin (\alpha_{31}/2) & = - \frac{\cos(\beta_{\mu3} - \beta_{\mu1})}{|U_{\mu1}|} 
\bigg[
\sin\theta_{12} \cos\theta_{23} \sin(\xi_{31}/2) 
\nonumber
\\
&+ \cos\theta_{12} \sin\theta_{23} \sin\theta_{13} 
\sin (\delta + \xi_{31}/2) \bigg] \,,
\quad
\cos(\beta_{\mu3} - \beta_{\mu1}) = +1~{\rm or}~(-1)\,, 
\label{eq:alpha31C1b}
\end{align}
%
where, according to eq.~(\ref{eq:th23C2}),
$\sin^2\theta_{23} \cos^2\theta_{13} = \sin^2\theta^\nu_{23} \cos^2\theta^\nu_{13}$.
As it follows from eqs.~(\ref{Umu1C2})~--~(\ref{Umu3C2}), the sign factors 
$\cos(\beta_{\mu2} - \beta_{\mu1})$ and $\cos(\beta_{\mu3} - \beta_{\mu1})$ 
are known once the angles $\theta^\nu_{ij}$ are fixed.

 Finally, we provide the expressions for 
$\sin (\alpha_{21}/2 - \xi_{21}/2)$, 
$\sin (\alpha_{31}/2 - \xi_{31}/2)$ 
and $\sin (\alpha_{31}/2 - \delta - \xi_{31}/2)$:
\begin{align}
\sin(\alpha_{21}/2 - \xi_{21}/2) &= \frac{\sin2\theta^e_{13}}{2 |U_{e1} U_{e2}|}
\sin\theta^\nu_{23} \cos\theta^\nu_{13} \sin\omega \nonumber\\ 
& = - \frac{\cos(\beta_{\mu2} - \beta_{\mu1})}{2 |U_{\mu1} U_{\mu2}|}
\sin2\theta_{23} \sin\theta_{13} \sin\delta\,, \\[0.2cm]
\sin(\alpha_{31}/2 - \xi_{31}/2) &= \frac{\sin2\theta^e_{13} \sin^2\theta^\nu_{23}}
{2 |U_{e2} U_{\mu3} U_{\tau3}|}
\cos^2\theta^\nu_{13}
\left(\cos\theta^\nu_{12} \cos\theta^\nu_{23} 
- \sin\theta^\nu_{12} \sin\theta^\nu_{23} \sin\theta^\nu_{13}\right) \sin\omega \nonumber\\
&= - \frac{\cos(\beta_{\mu3} - \beta_{\mu1})}{|U_{\mu1}|}
\cos\theta_{12} \sin\theta_{23} \sin\theta_{13} \sin\delta\,, \\[0.2cm]
\sin(\alpha_{31}/2 - \delta - \xi_{31}/2) &= - \frac{\sin2\theta^e_{13}}{2 |U_{e1} U_{e3}|}
\left(\cos\theta^\nu_{12} \cos\theta^\nu_{23} 
- \sin\theta^\nu_{12} \sin\theta^\nu_{23} \sin\theta^\nu_{13}\right) \sin\omega \nonumber\\
&= \frac{\cos(\beta_{\mu3} - \beta_{\mu1})}{|U_{\mu1}|}
\sin\theta_{12} \cos\theta_{23} \sin\delta\,.
\end{align}
%
As in the cases analysed in the preceding subsections we have 
$|\sin(\alpha_{21}/2 - \xi_{21}/2)| \propto \sin\theta_{13}$ and  
$|\sin(\alpha_{31}/2 - \xi_{31}/2)| \propto \sin\theta_{13}$.

%
\section{Summary of the Sum Rules for the Majorana Phases}
\label{sec:sumrules} 
%
%

 In the present Section we summarise the sum rules for the Majorana phases 
obtained in the previous Sections. Throughout this Section the neutrino 
mixing matrix $U$ is assumed to be in the standard parametrisation.

 In schemes A1, B1, B3 and C1 the sum rules for $\alpha_{21}/2$ 
and $\alpha_{31}/2$ 
can be cast in the form:
\begin{align}
\frac{\alpha_{21}}{2} 
&= \arg\left(U_{\tau1} U^*_{\tau2} e^{i \frac{\alpha_{21}}{2}}\right) 
+\varkappa_{21} + \frac{\xi_{21}}{2}\,, 
\label{eq:alpha21ABC1} \\
\frac{\alpha_{31}}{2} &= \arg\left(U_{\tau1}\right) 
+\varkappa_{31} + \frac{\xi_{31}}{2}\,,
\label{eq:alpha31ABC1}
\end{align}
%
where the expressions for the phases $\varkappa_{21}$ and $\varkappa_{31}$, 
which should be used in these sum rules in each particular case, 
are given in Table~\ref{tab:summary}. 
In schemes A1 and C1 the phases  
$\varkappa_{21}$ and $\varkappa_{31}$ take values 
$0$ or $\pi$ and are known once the angles 
$\theta^\nu_{ij}$ are fixed. In scheme B1 (B3), $\varkappa_{31}$
($\varkappa_{21}$ and $\varkappa_{31}$) 
depends (depend) on the free phase parameter 
$\beta$ ($\omega$).

 In schemes A2, B2 and C2 we similarly have:
\begin{align}
\frac{\alpha_{21}}{2} &= \arg\left(U_{\mu1} U^*_{\mu2} e^{i \frac{\alpha_{21}}{2}}\right) 
+\varkappa_{21} + \frac{\xi_{21}}{2}\,, 
\label{eq:alpha21ABC2} \\
\frac{\alpha_{31}}{2} &= \arg\left(U_{\mu1}\right) 
+\varkappa_{31} + \frac{\xi_{31}}{2}\,,
\label{eq:alpha31ABC2}
\end{align}
%
where the corresponding expressions for 
$\varkappa_{21}$ and $\varkappa_{31}$ 
are given again in Table~\ref{tab:summary}. 
In cases A2 and C2 the phases $\varkappa_{21}$ and 
$\varkappa_{31}$ can take values 0 or $\pi$. 
They are fixed when the angles $\theta^\nu_{ij}$ are 
given. The phase $\beta$, which is a free parameter 
as long as it is not fixed by 
additional arguments, enters the sum rule 
for $\alpha_{31}/2$ in scheme B2. 
 
 In all schemes considered, A1, A2, B1, B2, B3, C1 and C2, 
the phases $(\alpha_{21}/2 - \xi_{21}/2 -\varkappa_{21})$ and 
 $(\alpha_{31}/2 - \xi_{31}/2 -\varkappa_{31})$ are determined  
by the values of the neutrino mixing angles $\theta_{12}$, 
$\theta_{23}$ and $\theta_{13}$, and of the Dirac phase $\delta$.  
The Dirac phase is determined in each scheme by a corresponding sum rule.
In schemes A1, A2, C1 and C2 there is a correlation between 
the values of $\sin^2\theta_{23}$ and $\sin^2\theta_{13}$.   
The sum rules for $\cos\delta$ and 
the relevant expressions for $\sin^2\theta_{23}$ 
in the cases of interest,   
which should be used in eqs.~(\ref{eq:alpha21ABC1})~--~(\ref{eq:alpha31ABC2}), 
are given, e.g., in  Tables~1 and 2 of ref.~\cite{Girardi:2015vha}.
\begin{table}[t]
\hspace{-0.7cm}
\begin{tabular}{lll}
\toprule
Case & $\varkappa_{21}$ & $\varkappa_{31}$ \\
\midrule
A1 & 
$\arg\left(- s^\nu_{12} c^\nu_{12}\right)$ &
$\arg\left(s^\nu_{12} s^\nu_{23} c^\nu_{23}\right)$ \\[0.2cm]
A2 & 
$\arg\left(- s^\nu_{12} c^\nu_{12}\right)$ &
$\arg\left(- s^\nu_{12} s^\nu_{23} c^\nu_{23}\right)$ \\[0.2cm]
B1 & 
$\arg\left(- s^\nu_{12} c^\nu_{12}\right)$ &
$\arg\left(s^\nu_{12}\right) + \beta$ \\[0.2cm]
B2 & 
$\arg\left(- s^\nu_{12} c^\nu_{12}\right)$ &
$\arg\left(- s^\nu_{12}\right) + \beta$ \\[0.2cm]
B3 & 
$\arg\left[ 
\left(s^e_{13} s^\nu_{12} + c^e_{13} c^\nu_{12} s^\nu_{23} e^{- i \omega}\right)
\left(s^e_{13} c^\nu_{12} - c^e_{13} s^\nu_{12} s^\nu_{23} e^{i \omega} \right)
\right]$ &
$\arg\left[
c^e_{13} c^\nu_{23}
\left(c^e_{13} s^\nu_{12} s^\nu_{23} - s^e_{13} c^\nu_{12} e^{-i \omega}\right)
\right]$ \\[0.2cm]
C1 & 
$\arg\left[
- \left(c^\nu_{12} s^\nu_{23} + s^\nu_{12} c^\nu_{23} s^\nu_{13}\right)
\left(s^\nu_{12} s^\nu_{23} - c^\nu_{12} c^\nu_{23} s^\nu_{13}\right)
\right]$ &
$\arg\left[
c^\nu_{23} c^\nu_{13}
\left(s^\nu_{12} s^\nu_{23} - c^\nu_{12} c^\nu_{23} s^\nu_{13}\right)
\right]$ \\[0.2cm]
C2 & 
$\arg\left[
- \left(c^\nu_{12} c^\nu_{23} - s^\nu_{12} s^\nu_{23} s^\nu_{13}\right)
\left(s^\nu_{12} c^\nu_{23} + c^\nu_{12} s^\nu_{23} s^\nu_{13}\right)
\right]$ &
$\arg\left[
- s^\nu_{23} c^\nu_{13}
\left(s^\nu_{12} c^\nu_{23} + c^\nu_{12} s^\nu_{23} s^\nu_{13}\right)
\right]$ \\
\bottomrule
\end{tabular}
\caption{The phases $\varkappa_{21}$ and $\varkappa_{31}$ 
entering the sum rules for the Majorana phases given in 
eqs.~(\ref{eq:alpha21ABC1})~--~(\ref{eq:alpha31ABC2}) 
for all the cases considered.}
\label{tab:summary}
\end{table}
%
In the following Section we use the sum rules 
given in eqs.~(\ref{eq:alpha21ABC1})~--~(\ref{eq:alpha31ABC2}) 
to obtain the numerical predictions for the 
Majorana phases in the PMNS matrix.

%
\section{Predictions}
\label{sec:predictions}
%
%
%
\subsection{Dirac Phase}
\label{sec:predictionsDiracPh}
%
%

In Table~\ref{tab:deltaNO}~%
\footnote{This table is an updated version of Table~4 in \cite{Girardi:2015vha}.}
we show predictions 
for the Dirac phase $\delta$, obtained from the sum rules, 
derived in refs. \cite{Petcov:2014laa,Girardi:2015vha}
and summarised in Table~1 in ref.~\cite{Girardi:2015vha}.
The numerical values are obtained 
using the best fit values 
of the neutrino mixing parameters given in 
eqs.~(\ref{th12values})~--~(\ref{th13values}) 
for both the NO and IO spectra. 
In the BM (LC) case, the sum rules for $\cos\delta$ 
lead to unphysical values of $|\cos\delta| > 1$ 
if one uses as input the current best fit values of 
$\sin^2\theta_{12}$, $\sin^2\theta_{23}$ and $\sin^2\theta_{13}$ 
\cite{Marzocca:2013cr,Petcov:2014laa,Girardi:2014faa,Girardi:2015vha}.
This is an indication of the fact that the current data disfavours 
the BM (LC) form of $\tilde{U}_{\nu}$. In the case of the B1 scheme and 
the NO spectrum, for example, the   BM (LC) form is disfavoured  
at approximately $2\sigma$ confidence level. 
Physical values of $\cos\delta$ are found for larger (smaller) 
values of $\sin^2\theta_{12}$ ($\sin^2\theta_{23}$) 
\cite{Petcov:2014laa,Girardi:2014faa,Girardi:2015vha}. 
For, e.g., $\sin^2\theta_{12} = 0.354$,  
which is the $3\sigma$ upper bound of 
$\sin^2\theta_{12}$, and the best fit values of 
$\sin^2\theta_{23}$ and $\sin^2\theta_{13}$,
we get  $|\cos\delta| \leq 1$ in most of the  
schemes considered in the present article, the exceptions being 
the schemes B1 with the IO spectrum, B2 with the NO spectrum and B3. 
The values of the Dirac phase
corresponding to the BM (LC) form
quoted in Table~\ref{tab:deltaNO}   
are obtained for 
$\sin^2\theta_{12}=0.354$ and the best fit values of 
$\sin^2\theta_{23}$ and $\sin^2\theta_{13}$.
\begin{table}[t]
\centering
\begin{tabular}{llllll}
\toprule
Case (O) & TBM & GRA & GRB & HG &  BM (LC) \\
\midrule
A1 (NO) & $101.9 \lor 258.1$ & $77.3 \lor 282.7$ & $107.2 \lor 252.8$ & $65.3 \lor 294.7$ & $176.5 \lor 183.5$ \\[0.2cm]
A1 (IO) & $101.7 \lor 258.3$ & $77.3 \lor 282.7$ & $107.0 \lor 253.0$ & $65.5 \lor 294.5$ & $171.1 \lor 188.9$ \\
\midrule
A2 (NO) & $78.1 \lor 281.9$ & $102.7 \lor 257.3$ & $72.8 \lor 287.2$ & $114.7 \lor 245.3$ & $3.5 \lor 356.5$ \\[0.2cm]
A2 (IO) & $78.3 \lor 281.7$ & $102.7 \lor 257.3$ & $73.0 \lor 287.0$ & $114.6 \lor 245.4$ & $8.9 \lor 351.1$ \\
\midrule
B1 (NO) & $99.9 \lor 260.1$ & $77.7 \lor 282.3$ & $104.8 \lor 255.2$ & $66.9 \lor 293.1$ & $153.4 \lor 206.6$ \\[0.2cm]
B1 (IO) & $104.9 \lor 255.1$ & $76.4 \lor 283.6$ & $111.3 \lor 248.7$ & $62.4 \lor 297.6$ & | \\
\midrule
B2 (NO) & $75.1 \lor 284.9$ & $103.6 \lor 256.4$ & $68.8 \lor 291.2$ & $117.6 \lor 242.4$ & | \\[0.2cm]
B2 (IO) & $80.5 \lor 279.5$ & $102.2 \lor 257.8$ & $75.7 \lor 284.3$ & $112.8 \lor 247.2$ & $29.1 \lor 330.9$ \\
\midrule
B3 (NO) & $103.5 \lor 256.5$ & $78.8 \lor 281.2$ & $108.9 \lor 251.1$ & $66.9 \lor 293.1$ & | \\[0.2cm]
B3 (IO) & $103.1 \lor 256.9$ & $78.6 \lor 281.4$ & $108.4 \lor 251.6$ & $66.8 \lor 293.2$ & | \\
\toprule
Case & $[\pi/20,-\pi/4]$ &  $[\pi/10,-\pi/4]$ & $[a,-\pi/4]$ & $[\pi/20,b]$ & $[\pi/20,\pi/6]$ \\
\midrule
C1 (NO) & $108.7 \lor 251.3$ & $44.8 \lor 315.2$ & $29.7 \lor 330.3$ & $154.9 \lor 205.1$ & $132.8 \lor 227.2$ \\[0.2cm]
C1 (IO) & $108.5 \lor 251.5$ & $45.2 \lor 314.8$ & $30.5 \lor 329.5$ & $153.7 \lor 206.3$ & $132.3 \lor 227.7$ \\
\toprule
Case & $[\pi/20,c]$ & $[\pi/20,\pi/4]$ & $[\pi/10,\pi/4]$ & $[a,\pi/4]$ & $[\pi/20,d]$ \\
\midrule
C2 (NO) & $146.0 \lor 214.0$ & $71.3 \lor 288.7$ & $135.2 \lor 224.8$ & $150.3 \lor 209.7$ & $138.5 \lor 221.5$ \\[0.2cm]
C2 (IO) & $145.3 \lor 214.7$ & $71.5 \lor 288.5$ & $134.8 \lor 225.2$ & $149.5 \lor 210.5$ & $138.1 \lor 221.9$ \\
\bottomrule
\end{tabular}
\caption{The Dirac CPV phase $\delta$ in degrees 
calculated from the sum rules derived in 
refs.  \cite{Petcov:2014laa,Girardi:2015vha} 
using the best fit values of the neutrino mixing angles 
quoted in eqs.~(\ref{th12values})~--~(\ref{th13values}), 
except for the B3 scheme and the BM (LC) form of $\tilde{U}_{\nu}$.
The results shown for the B3 scheme are obtained for $\omega = 0$,  
${\rm sgn}\,(\sin2\theta^e_{13}) = 1$, and for 
$\sin^2\theta_{23} = 0.48907$ (0.48886) 
for the NO (IO) spectrum. 
The numbers quoted for the BM (LC) form of $\tilde{U}_{\nu}$ 
are for $\sin^2\theta_{12}=0.354$, 
which is the $3\sigma$ upper bound.
For each cell the first number corresponds 
to $\delta = \cos^{-1}(\cos\delta)$, while 
the second number corresponds to $\delta = 2\pi - \cos^{-1}(\cos\delta)$. 
In cases C1 and C2, $\theta^\nu_{23} = -\,\pi/4$ 
and the values in square brackets are those of 
$[\theta^\nu_{13}, \theta^\nu_{12}]$ used.
The letters $a$, $b$, $c$ and $d$ stand for 
$\sin^{-1}(1/3)$, $\sin^{-1}(1/\sqrt{2 + r})$, 
$\sin^{-1}(1/\sqrt{3})$ and $\sin^{-1}(\sqrt{3 - r}/2)$, 
respectively. 
See text for further details.}
\label{tab:deltaNO}
\end{table}

 In each of cases C1 and C2 we report results 
for $\theta^\nu_{23} = -\,\pi/4$ and
five sets of values of $[\theta^\nu_{13},\theta^\nu_{12}]$, 
associated with, or inspired by, models of neutrino mixing.
These sets include the three values of 
$\theta^\nu_{13} = \pi/20$, $\pi/10$ and $a \equiv \sin^{-1}(1/3)$
and selected values of $\theta^\nu_{12}$ from the set: $\pm \pi/4$, 
$b \equiv \sin^{-1}(1/\sqrt{2 + r})$, $c \equiv \sin^{-1}(1/\sqrt{3})$ and 
$d \equiv \sin^{-1}(\sqrt{3 - r}/2)$.   
The values in square brackets in Table~\ref{tab:deltaNO}
are those of $[\theta^\nu_{13}, \theta^\nu_{12}]$ used.
In scheme C1 we define cases I, II, III, IV and V 
as the cases with $[\theta^\nu_{13}, \theta^\nu_{12}]$ being equal to 
$[\pi/20,-\pi/4]$, $[\pi/10,-\pi/4]$, $[a,-\pi/4]$, $[\pi/20,b]$ and $[\pi/20,\pi/6]$, 
respectively. In scheme C2 cases I, II, III, IV and V correspond to the following 
pairs: $[\pi/20,c]$, $[\pi/20,\pi/4]$, $[\pi/10,\pi/4]$, $[a,\pi/4]$ and $[\pi/20,d]$, 
respectively.

   As can be seen from Table~\ref{tab:deltaNO}, 
the values of $\delta$ for the IO spectrum 
differ insignificantly from the values 
obtained for the NO one in all the schemes considered, except 
for the B1 and B2 ones. 
The difference between the NO and IO values of $\delta$ in the B1 and B2 
schemes is a consequence of the difference between the best 
fit values of $\sin^2\theta_{23}$ corresponding to the NO and IO spectra~%
\footnote{We recall that $\sin^2\theta_{23}$ is a free parameter 
in schemes B1 and B2.}.
We use the values of $\delta$ from Table~\ref{tab:deltaNO} 
to obtain predictions for the Majorana phases in the next subsection.

%
\subsection{Majorana Phases}
\label{sec:predictionsMajPh}
%
%

 In this subsection we present results of the numerical analysis 
of the predictions for the Majorana phases,
performed using the best fit values of the 
neutrino mixing parameters given in 
eqs.~(\ref{th12values})~--~(\ref{th13values}). 
These predictions are obtained from the sum rules 
in eqs.~(\ref{eq:alpha21ABC1})~--~(\ref{eq:alpha31ABC2}), 
in which we have used the proper expressions 
for $\sin^2\theta_{23}$ and $\cos\delta$ from 
\cite{Petcov:2014laa,Girardi:2015vha}. 
We summarise the predictions for all the cases considered in 
the present study 
in Tables~\ref{tab:alpha21NO} and \ref{tab:alpha31NO}, 
in which we give, respectively, 
the values of the phase differences 
$(\alpha_{21}/2 - \xi_{21}/2)$ 
and $(\alpha_{31}/2 - \xi_{31}/2)$ found in schemes 
A1, A2, B3, C1 and C2. In the cases of  
schemes B1 and B2 we present 
in Table \ref{tab:alpha31NO}  
results for the difference $(\alpha_{31}/2 - \xi_{31}/2 - \beta)$, 
since the phase $\beta$, in general, 
is not fixed, unless some additional arguments 
are used that fix it.  
In the case of the B3 scheme the results are obtained for $\omega = 0$,  
${\rm sgn}\,(\sin2\theta^e_{13}) = 1$, and for 
$\sin^2\theta_{23} = 0.48907$ (0.48886) 
for the NO (IO) spectrum
(see subsection \ref{sec:12e13e23nu12nu} 
and ref. \cite{Girardi:2015vha} for details). 
%
%
%
\begin{table}[h]
\centering
\begin{tabular}{llllll}
\toprule
Case (O) & TBM & GRA & GRB & HG &  BM (LC) \\
\midrule
A1 (NO) & $342.3 \lor 17.7$ & $341.4 \lor 18.6$ & $342.9 \lor 17.1$ & $342.1 \lor 17.9$ & $359.0 \lor 1.0$ \\[0.2cm]
A1 (IO) & $342.1 \lor 17.9$ & $341.2 \lor 18.8$ & $342.7 \lor 17.3$ & $341.9 \lor 18.1$ & $357.4 \lor 2.6$ \\
\midrule
A2 (NO) & $17.7 \lor 342.3$ & $18.6 \lor 341.4$ & $17.1 \lor 342.9$ & $17.9 \lor 342.1$ & $1.0 \lor 359.0$ \\[0.2cm]
A2 (IO) & $17.9 \lor 342.1$ & $18.8 \lor 341.2$ & $17.3 \lor 342.7$ & $18.1 \lor 341.9$ & $2.6 \lor 357.4$ \\
\midrule
B1 (NO) & $340.3 \lor 19.7$ & $339.3 \lor 20.7$ & $340.8 \lor 19.2$ & $339.9 \lor 20.1$ & $351.7 \lor 8.3$ \\[0.2cm]
B1 (IO) & $345.0 \lor 15.0$ & $344.1 \lor 15.9$ & $345.7 \lor 14.3$ & $345.0 \lor 15.0$ & | \\
\midrule
B2 (NO) & $15.1 \lor 344.9$ & $16.0 \lor 344.0$ & $14.4 \lor 345.6$ & $15.0 \lor 345.0$ & | \\[0.2cm]
B2 (IO) & $20.2 \lor 339.8$ & $21.1 \lor 338.9$ & $19.6 \lor 340.4$ & $20.6 \lor 339.4$ & $9.2 \lor 350.8$ \\
\midrule
B3 (NO) & $342.5 \lor 17.5$ & $341.4 \lor 18.6$ & $343.1 \lor 16.9$ & $342.0 \lor 18.0$ & | \\[0.2cm]
B3 (IO) & $342.3 \lor 17.7$ & $341.2 \lor 18.8$ & $342.9 \lor 17.1$ & $341.8 \lor 18.2$ & | \\
\toprule
Case & $[\pi/20,-\pi/4]$ &  $[\pi/10,-\pi/4]$ & $[a,-\pi/4]$ & $[\pi/20,b]$ & $[\pi/20,\pi/6]$ \\
\midrule
C1 (NO) & $163.5 \lor 196.5$  &  $166.9 \lor 193.1$ & $170.7 \lor 189.3$ & $353.0 \lor 7.0$ & $347.6 \lor 12.4$ \\[0.2cm]
C1 (IO) & $163.3 \lor 196.7$  &  $166.6 \lor 193.4$ & $170.3 \lor 189.7$ & $352.6 \lor 7.4$ & $347.4 \lor 12.6$ \\
\toprule
Case & $[\pi/20,c]$ & $[\pi/20,\pi/4]$ & $[\pi/10,\pi/4]$ & $[a,\pi/4]$ & $[\pi/20,d]$ \\
\midrule
C2 (NO) & $11.6 \lor 348.4$ & $16.5 \lor 343.5$ & $13.1 \lor 346.9$ & $9.3 \lor 350.7$ & $13.5 \lor 346.5$ \\[0.2cm]
C2 (IO) & $11.9 \lor 348.1$ & $16.7 \lor 343.3$ & $13.4 \lor 346.6$ & $9.7 \lor 350.3$ & $13.7 \lor 346.3$ \\
\bottomrule
\end{tabular}
\caption{The phase difference $(\alpha_{21}/2 - \xi_{21}/2)$ in degrees 
calculated using the best fit values of the neutrino mixing angles 
quoted in eqs.~(\ref{th12values})~--~(\ref{th13values}), 
except for scheme B3 and the BM (LC) form of $\tilde{U}_{\nu}$.
For scheme B3 the results shown 
are obtained for $\omega = 0$,  
${\rm sgn}\,(\sin2\theta^e_{13}) = 1$ and 
$\sin^2\theta_{23} = 0.48907$ (0.48886) 
in the case of the NO (IO) spectrum. 
The numbers quoted for the BM (LC) form of $\tilde{U}_{\nu}$ 
are for the $3\sigma$ upper bound of
 $\sin^2\theta_{12}=0.354$.
For each cell the first number corresponds 
to $\delta = \cos^{-1}(\cos\delta)$, while 
the second number is obtained for $\delta = 2\pi - \cos^{-1}(\cos\delta)$. 
In cases C1 and C2, $\theta^\nu_{23} = -\,\pi/4$ 
and the values in square brackets are those of 
$[\theta^\nu_{13}, \theta^\nu_{12}]$ used. 
The letters $a$, $b$, $c$ and $d$ stand for 
$\sin^{-1}(1/3)$, $\sin^{-1}(1/\sqrt{2 + r})$, 
$\sin^{-1}(1/\sqrt{3})$ and $\sin^{-1}(\sqrt{3 - r}/2)$, respectively. 
See text for further details.}
\label{tab:alpha21NO}
\end{table}
%
%
%
%
\begin{table}[t]
\centering
\begin{tabular}{llllll}
\toprule
Case (O) & TBM & GRA & GRB & HG &  BM (LC) \\
\midrule
A1 (NO) & $167.9 \lor 192.1$ & $166.7 \lor 193.3$ & $168.4 \lor 191.6$ & $167.0 \lor 193.0$ & $179.4 \lor 180.6$ \\[0.2cm]
A1 (IO) & $167.7 \lor 192.3$ & $166.6 \lor 193.4$ & $168.3 \lor 191.7$ & $166.8 \lor 193.2$ & $178.5 \lor 181.5$ \\
\midrule
A2 (NO) & $192.1 \lor 167.9$ & $193.3 \lor 166.7$ & $191.6 \lor 168.4$ & $193.0 \lor 167.0$ & $180.6 \lor 179.4$ \\[0.2cm]
A2 (IO) & $192.3 \lor 167.7$ & $193.4 \lor 166.6$ & $191.7 \lor 168.3$ & $193.2 \lor 166.8$ & $181.5 \lor 178.5$ \\
\midrule
B1 (NO) & $346.4 \lor 13.6$ & $345.2 \lor 14.8$ & $346.9 \lor 13.1$ & $345.4 \lor 14.6$ & $355.2 \lor 4.8$ \\[0.2cm]
B1 (IO) & $349.7 \lor 10.3$ & $348.6 \lor 11.4$ & $350.2 \lor 9.8$ & $349.1 \lor 10.9$ & | \\
\midrule
B2 (NO) & $10.3 \lor 349.7$ & $11.4 \lor 348.6$ & $9.8 \lor 350.2$ & $11.0 \lor 349.0$ & | \\[0.2cm]
B2 (IO) & $13.9 \lor 346.1$ & $15.1 \lor 344.9$ & $13.4 \lor 346.6$ & $15.0 \lor 345.0$ & $5.3 \lor 354.7$ \\
\midrule
B3 (NO) & $168.0 \lor 192.0$ & $166.7 \lor 193.3$ & $168.6 \lor 191.4$ & $166.9 \lor 193.1$ & | \\[0.2cm]
B3 (IO) & $167.9 \lor 192.1$ & $166.6 \lor 193.4$ & $168.4 \lor 191.6$ & $166.8 \lor 193.2$ & | \\
\toprule
Case & $[\pi/20,-\pi/4]$ &  $[\pi/10,-\pi/4]$ & $[a,-\pi/4]$ & $[\pi/20,b]$ & $[\pi/20,\pi/6]$ \\
\midrule
C1 (NO) & $348.8 \lor 11.2$  &  $350.2 \lor 9.8$ & $352.9 \lor 7.1$ & $175.5 \lor 184.5$ & $171.9 \lor 188.1$ \\[0.2cm]
C1 (IO) & $348.7 \lor 11.3$ & $350.0 \lor 10.0$ & $352.7 \lor 7.3$ & $175.2 \lor 184.8$ & $171.7 \lor 188.3$ \\
\toprule
Case & $[\pi/20,c]$ & $[\pi/20,\pi/4]$ & $[\pi/10,\pi/4]$ & $[a,\pi/4]$ & $[\pi/20,d]$ \\
\midrule
C2 (NO) & $188.8 \lor 171.2$ & $191.2 \lor 168.8$ & $189.8 \lor 170.2$ & $187.1 \lor 172.9$ & $190.1 \lor 169.9$ \\[0.2cm]
C2 (IO) & $189.0 \lor 171.0$ & $191.3 \lor 168.7$ & $190.0 \lor 170.0$ & $187.3 \lor 172.7$ & $190.3 \lor 169.7$ \\
\bottomrule
\end{tabular}
\caption{The same as in Table~\ref{tab:alpha21NO}, but for 
the phase difference $(\alpha_{31}/2 - \xi_{31}/2)$ given in degrees. 
In cases B1 and B2 the presented numbers correspond to 
$(\alpha_{31}/2 - \xi_{31}/2 - \beta)$, where $\beta$ 
is a free phase parameter. 
See text for further details.}
\label{tab:alpha31NO}
\end{table}
%

 All the quoted phases are 
determined with a two-fold ambiguity owing to the  fact that 
the Dirac phase $\delta$, which enters into the expressions for
all the phases under discussion, is determined  with a two-fold ambiguity 
from the sum rules it satisfies in the schemes of interest 
(see \cite{Petcov:2014laa,Girardi:2015vha}). The absolute 
values of the sines of the phases quoted in Tables~\ref{tab:alpha21NO} 
and \ref{tab:alpha31NO} are all proportional to $\sin\theta_{13}$, and thus 
are relatively small. The results in cases A1 and B2 
for the TBM, BM (LC), GRA, GRB and HG symmetry 
forms of $\tilde{U}_{\nu}$ considered were first 
obtained in \cite{Petcov:2014laa} using the best fit values of 
$\sin^2\theta_{12}$, $\sin^2\theta_{23}$ and $\sin^2\theta_{13}$
from (the first e-archive version of) 
ref. \cite{Capozzi:2013csa}. 
Here, in particular, 
we update the results derived in \cite{Petcov:2014laa}.

As we have already noticed, in the BM (LC) case, 
the sum rules for $\cos\delta$ 
lead to unphysical values of $|\cos\delta| > 1$ 
if one uses as input the current best fit values of 
$\sin^2\theta_{12}$, $\sin^2\theta_{23}$ and $\sin^2\theta_{13}$ 
\cite{Marzocca:2013cr,Petcov:2014laa,Girardi:2015vha}.
Physical values of $\cos\delta$ are found for larger (smaller) 
values of $\sin^2\theta_{12}$ ($\sin^2\theta_{23}$) 
\cite{Petcov:2014laa,Girardi:2014faa,Girardi:2015vha}.  
The values of the phases given in  
Tables~\ref{tab:alpha21NO} and \ref{tab:alpha31NO} 
and corresponding to the BM (LC) mixing 
are obtained for the $3\sigma$ upper bound 
of  $\sin^2\theta_{12} = 0.354$
and the best fit values of 
$\sin^2\theta_{23}$ and $\sin^2\theta_{13}$.
For these values of the three mixing parameters 
$|\cos\delta|$ has an unphysical value 
greater than one only for schemes B1 with the IO spectrum, 
B2 with the NO spectrum and B3.
 
 A few comments on the results presented in  Tables 
\ref{tab:alpha21NO} and \ref{tab:alpha31NO}
are in order. 
These results show that 
for a given scheme and fixed form of 
the matrix $\tilde{U}_{\nu}$, the difference between the 
predictions of the phases 
$(\alpha_{21}/2 - \xi_{21}/2)$ and $(\alpha_{31}/2 - \xi_{31}/2)$
or $(\alpha_{31}/2 - \xi_{31}/2 - \beta)$ for the NO and IO 
neutrino mass spectra are relatively small. The largest 
difference is approximately of $5^\circ$ 
between the NO and IO values of $(\alpha_{21}/2 - \xi_{21}/2)$ 
in the B1 and B2 schemes. The same observation is valid for 
the variation of the phases with the 
variation of the form of  $\tilde{U}_{\nu}$ 
within a given scheme, the only exceptions being
i) the BM (LC) form, for which the phases differ from those 
for the TBM, GRA, GRB and HG forms of $\tilde{U}_{\nu}$ 
of schemes A1, A2, B1 (NO spectrum) and B2 (IO spectrum)
by approximately $10^\circ$ to $16^\circ$, and 
ii) the C1 scheme, in which the values of the phases  
 $(\alpha_{21}/2 - \xi_{21}/2)$ and $(\alpha_{31}/2 - \xi_{31}/2)$ 
differ relatively little within the group of the first three 
cases in Tables \ref{tab:alpha21NO} and \ref{tab:alpha31NO} 
and within the group of the last two ones, but change significantly~---~%
approximately by $\pi$~---~%
when switching from a case of one of the groups 
to a case in the second group.
\begin{figure}[t]
\centering
\includegraphics[width=\textwidth]{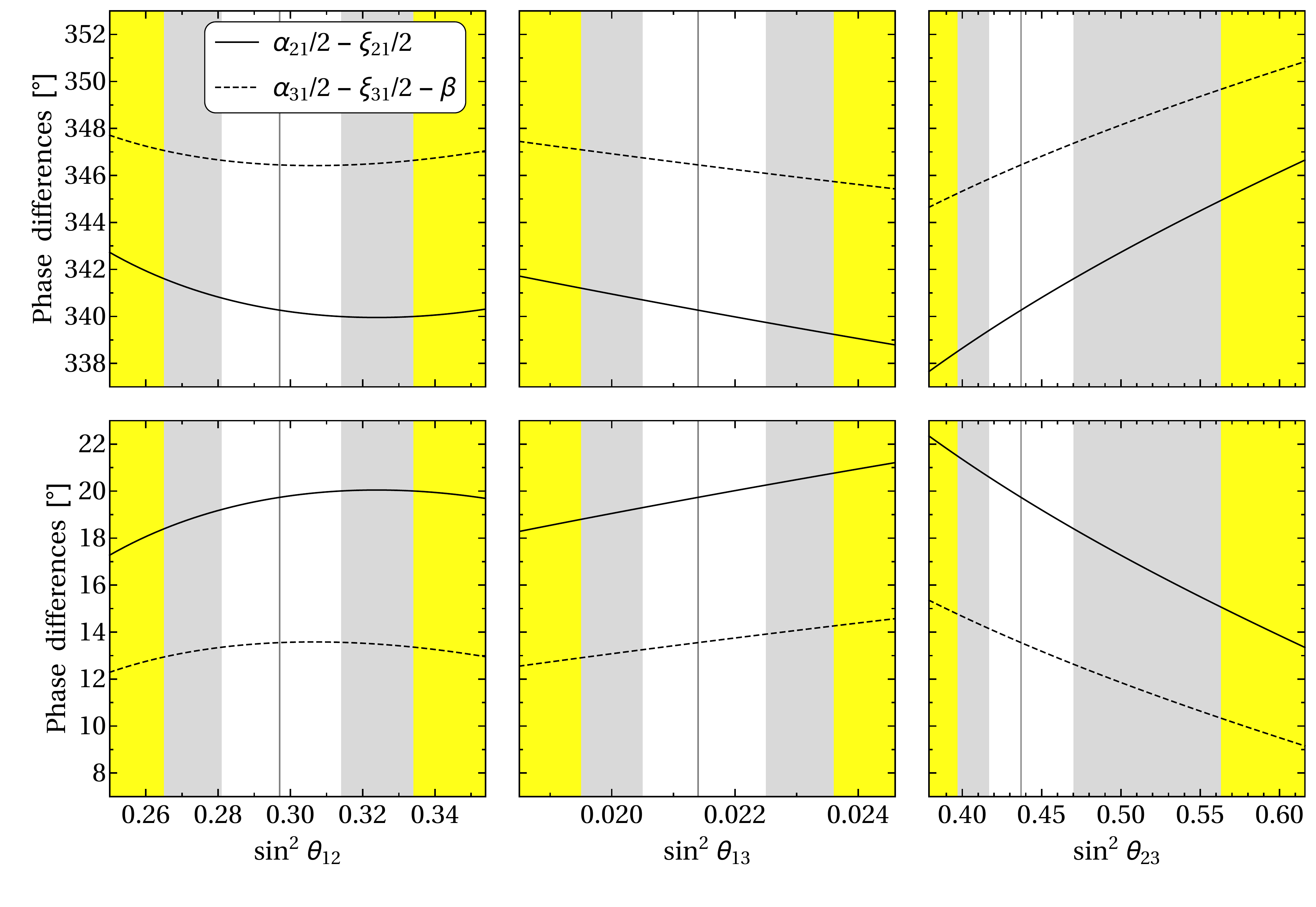}
\caption{\label{fig:MPs3sigmaB1}The phase differences 
$(\alpha_{21}/2 - \xi_{21}/2)$ (solid line) and 
$(\alpha_{31}/2 - \xi_{31}/2 - \beta$) (dashed line) 
as functions of $\sin^2\theta_{ij}$ 
in case B1 and for the TBM symmetry form of the matrix $\tilde U_\nu$.
The two other parameters, $\sin^2\theta_{kl}$ and $\sin^2\theta_{mn}$,  
$ij \neq kl \neq mn$, have been fixed to their best fit values for the NO spectrum.
The upper panels correspond to $\delta = \cos^{-1}(\cos\delta)$, 
while the lower panels correspond to $\delta = 2\pi - \cos^{-1}(\cos\delta)$.
The vertical line and the three coloured vertical bands 
indicate the best fit value and the $1\sigma$, $2\sigma$ and $3\sigma$ 
allowed ranges of $\sin^2\theta_{ij}$.}
\end{figure}
%
%
%
%
%
\begin{figure}[t]
\centering
\includegraphics[width=\textwidth]{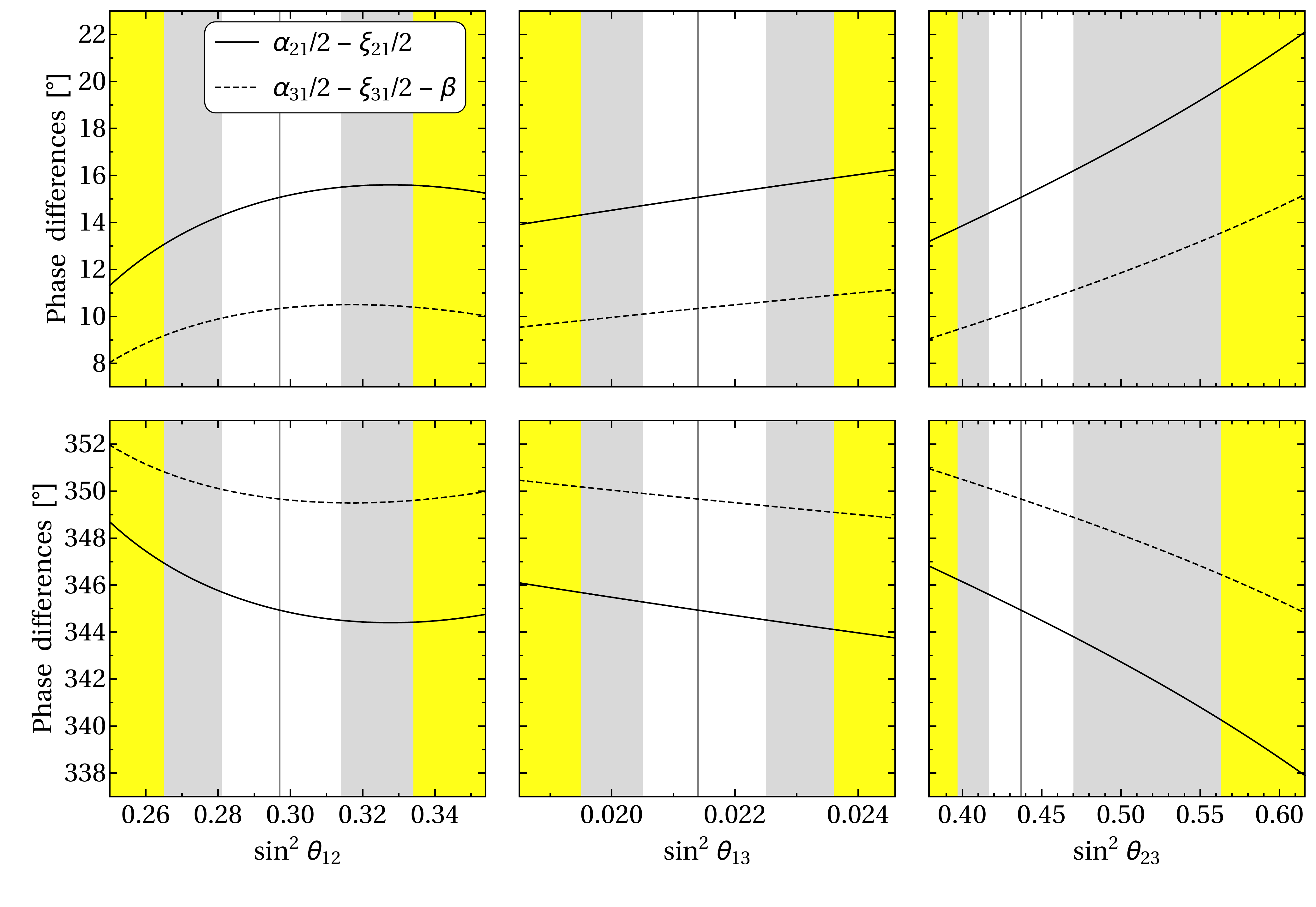}
\caption{\label{fig:MPs3sigmaB2}The same as in Fig.~\ref{fig:MPs3sigmaB1}, but for case B2.}
\end{figure}
%

  For a given symmetry form of  $\tilde{U}_{\nu}$~---~TBM, GRA, GRB and 
HG~---~the phase difference $(\alpha_{21}/2 - \xi_{21}/2)$ has 
very similar values for the A1, B1 and B3 schemes, they differ 
approximately by at most $2^\circ$, and for the A2 and B2 schemes, 
for which the difference does not exceed  $3^\circ$.
However, the predictions for $(\alpha_{21}/2 - \xi_{21}/2)$
for schemes A1, B1, B3 and 
A2, B2 differ significantly~---~the sum of the 
values of  $(\alpha_{21}/2 - \xi_{21}/2)$
for any of the  A1, B1, B3 schemes and for any of the A2, B2 
schemes being roughly equal to $2\pi$. In contrast, 
for a given symmetry form of  
$\tilde{U}_{\nu}$~---~TBM, GRA, GRB and HG~---~%
i) the values of the phase difference $(\alpha_{31}/2 - \xi_{31}/2)$ 
($(\alpha_{31}/2 - \xi_{31}/2 - \beta)$) for the schemes
A1 and A2 (B1 and B2) differ significantly~---~by up to $26^\circ$ 
($337^\circ$), and 
ii) the values of $(\alpha_{31}/2 - \xi_{31}/2)$ 
and $(\alpha_{31}/2 - \xi_{31}/2 - \beta)$ 
are drastically different.
At the same time, the values of $(\alpha_{31}/2 - \xi_{31}/2)$ for the A1 
and B3 schemes practically coincide. 
 
 Finally, for any given of the five cases of schemes C1 and C2, 
the values of the phase difference $(\alpha_{21}/2 - \xi_{21}/2)$
for schemes C1 and C2 differ drastically. 
The same conclusion is valid for the C1 and C2 values of 
the phase difference $(\alpha_{31}/2 - \xi_{31}/2)$ 
for any of the first three cases of these schemes listed in 
Table \ref{tab:alpha31NO}.  For the last two cases 
in  Table \ref{tab:alpha31NO} the difference between 
the C1 and C2 values of  $(\alpha_{31}/2 - \xi_{31}/2)$ 
is approximately $12^\circ$ and $18^\circ$.
 
 Further, we show how the predictions 
for the phase differences presented in 
Tables~\ref{tab:alpha21NO} and \ref{tab:alpha31NO} change when 
the uncertainties in  determination of the neutrino mixing parameters 
are taken into account. As an example, we consider the cases B1 and B2 
with the TBM form of the matrix $\tilde U_\nu$. 
We fix two of $\sin^2\theta_{ij}$ to their best fit values 
for the NO neutrino mass spectrum and 
vary the third one in its $3\sigma$ allowed range given in 
eqs.~(\ref{th12values})~--~(\ref{th13values}).
We show the results for cases B1 and B2 
in Figs.~\ref{fig:MPs3sigmaB1} and \ref{fig:MPs3sigmaB2}, respectively. 
As can be seen, the phase differences of interest depend weakly 
on $\sin^2\theta_{12}$ and $\sin^2\theta_{13}$. 
When these parameters are varied in their $3\sigma$ ranges, 
the variation of the phase differences is within a few degrees.
The dependence on $\sin^2\theta_{23}$ is stronger: 
the maximal variations of  $(\alpha_{21}/2 - \xi_{21}/2)$ 
and $(\alpha_{31}/2 - \xi_{31}/2 - \beta)$
are approximately of $9^\circ$ and $6^\circ$ in both cases. 
Another example, corresponding to the cases A1 and A2 with the TBM form 
of the matrix $\tilde U_\nu$, is considered in Appendix~\ref{app:A1A2}. 

 Performing a full statistical analysis of the predictions for 
$(\alpha_{21}/2 - \xi_{21}/2)$ and $(\alpha_{31}/2 - \xi_{31}/2)$ 
($(\alpha_{31}/2 - \xi_{31}/2 - \beta)$) is however outside the 
scope of the present study. Such an analysis will be presented 
elsewhere.

%
\subsection{Neutrinoless Double Beta Decay}
\label{sec:predictionsbb0nu}
%
%
 
 If the light neutrinos with definite mass $\nu_j$ 
are Majorana fermions, their exchange 
can trigger processes in which the 
total lepton charge changes by two units, 
$|\Delta L|= 2$: 
$K^+ \rightarrow \pi^- + \mu^+ + \mu^+$,
$e^- +(A,Z) \rightarrow e^+ + (A,Z-2)$, etc.
The experimental searches for $\betabeta$-decay,  
$(A,Z) \rightarrow (A,Z+2) + e^- + e^-$,
of even-even nuclei $^{48}Ca$, $^{76}Ge$, 
$^{82}Se$, $^{100}Mo$, $^{116}Cd$, $^{130}Te$, 
$^{136}Xe$, $^{150}Nd$, etc., 
are unique in reaching the sensitivity
that might allow to observe this process 
if it is triggered by the exchange of the light 
neutrinos  $\nu_j$  
(see, e.g., refs.~\cite{Petcov:2013poa,bb0nuth}). 
In  $\betabeta$-decay, two neutrons of the initial nucleus 
$(A,Z)$ transform by exchanging 
virtual $\nu_{1,2,3}$ into two protons of the final state 
nucleus $(A,Z+2)$ and two free electrons. 
The corresponding $\betabeta$-decay amplitude has the 
form (see, e.g., refs.~\cite{BiPet87,bb0nuth}):
$A(\betabeta) = G^2_{\rm F}\, \mefff\,M(A,Z)$, 
where  $G_{\rm F}$ is the Fermi constant, $\mefff$ is 
the  $\betabeta$-decay effective Majorana mass and 
$M(A,Z)$ is the nuclear matrix element (NME) of the process.
The $\betabeta$-decay effective Majorana mass $\mefff$  
contains all the dependence of $A(\betabeta)$ 
on the neutrino mixing parameters. 
The current experimental limits on $\meff$ are in the range of 
$(0.1 - 0.7)$~eV.  Most importantly, a large number of 
experiments of a new generation aim at
sensitivity to $\meff \sim (0.01 - 0.05)$~eV 
(for a detailed discussion of the current limits on 
$\meff$ and of the currently running and future planned 
$\betabeta$-decay experiments and their prospective
sensitivities see, e.g., the recent review article
\cite{Dell'Oro:2016dbc}).

The predictions for $\meff$ (see, e.g., \cite{BiPet87,BPP1,bb0nuth}), 
\begin{align}
\meff &= \left | \sum_{i=1}^3 m_i U_{ei}^2 \right| \nonumber \\
&= \left| m_1 \cos^2\theta_{12} \cos^2\theta_{13} 
+ m_2 \sin^2\theta_{12} \cos^2\theta_{13} e^{i \alpha_{21}}
+ m_3 \sin^2 \theta_{13} e^{i \left(\alpha_{31} - 2 \delta\right)} \right|\,,
\label{meff}
\end{align}
%
$m_{1,2,3}$ being the light Majorana neutrino masses,
depend on the values of the Majorana phase $\alpha_{21}$ 
and on the Majorana-Dirac phase difference 
$(\alpha_{31} - 2\delta)$. 
In what follows we will derive 
predictions for $\meff$ as a function of the lightest neutrino 
mass $m_{\rm min} \equiv {\rm min}(m_j)$, $j=1,2,3$,
for both the NO and IO neutrino mass spectra 
\footnote{For a discussion of the physics implications 
of a measurement of $\meff$, i.e., of the physics potential 
of the $\betabeta$-decay experiments see, e.g., \cite{bb0nuth,PPW}.}
and for two values of 
each of the phases $\xi_{21}$ and $\xi_{31}$:   $\xi_{21}=0$ or $\pi$,
$\xi_{31} = 0$ or $\pi$. 
 The choice of the two values of the phases $\xi_{21}$ and $\xi_{31}$ 
will be justified in the next Section where we show that the requirement of 
generalised CP invariance of the neutrino Majorana mass term 
in the cases of the $S_4$, $A_4$, $T^\prime$ and $A_5$ 
lepton flavour symmetries leads to the constraints 
$\xi_{21}=0$ or $\pi$, $\xi_{31}=0$ or $\pi$.

 We use the standard convention for 
numbering the neutrinos with definite masses in the 
cases of the NO and IO spectra (see, e.g., \cite{PDG2014}):
$m_1 < m_2 < m_3$ for the NO spectrum and 
$m_3 < m_1 < m_2$ for the IO one. We recall that 
the two heavier neutrino masses are expressed in terms of 
the lightest neutrino mass and the two independent 
neutrino mass squared differences measured in neutrino 
oscillation experiments as follows:
\begin{align}
\label{m23NO}
&m_2 = \sqrt{\Delta m^2_{21} + m^2_1}\,, \quad 
m_3 = \sqrt{\Delta m^2_{31} + m^2_1} \quad 
{\rm for~the~NO~spectrum}\,,\\
\label{m12IO}
&m_1 = \sqrt{\Delta m^2_{23} - \Delta m^2_{21} + m^2_3}\,, \quad
m_2 = \sqrt{\Delta m^2_{23} + m^2_3} \quad 
{\rm for~the~IO~spectrum}\,,
\end{align}
%
where $\Delta m^2_{ij} \equiv m^2_i - m^2_j$.
The best fit values and the $3\sigma$ allowed ranges of 
$\Delta m^2_{21}$ and $\Delta m^2_{31(23)}$
obtained in the global analysis of the neutrino oscillation data 
performed in \cite{Capozzi:2016rtj} we are going to 
use in our numerical study read:
\begin{align}
&(\Delta m^2_{21})_{\rm BF} = 7.37 \times 10^{-5}~\text{eV}^2\,, \quad 
6.93 \times 10^{-5}~\text{eV}^2 \leq \Delta m^2_{21} 
\leq 7.97 \times 10^{-5}~\text{eV}^2\,, 
\label{eq:dm21values} 
\\[0.30cm]
&(\Delta m^2_{31(23)})_{\rm BF} = 2.54~(2.50) \times 10^{-3}~\text{eV}^2\,, 
\nonumber
\\[0.30cm]
& 2.40~(2.36) \times 10^{-3}~\text{eV}^2 \leq \Delta m^2_{31(23)} 
\leq 2.67~(2.64) \times 10^{-3}~\text{eV}^2\,, 
\label{eq:dm3123values}
\end{align}
%
where the quoted values of $\Delta m^2_{31}$ and $\Delta m^2_{23}$
correspond to the NO and IO spectra, respectively.

As can be seen from Tables~\ref{tab:deltaNO}~--~\ref{tab:alpha31NO}, 
the values of all three phases, $\delta$, $\alpha_{21}$ and $\alpha_{31}$,
for scheme B3 with $\omega = 0$ and ${\rm sgn}\,(\sin2\theta^e_{13}) = 1$
are very close to the values for scheme A1. 
Thus, the predictions for $\meff$ in scheme B3 
are practically the same as those for scheme A1 
and we present predictions only for the latter.

\begin{figure}[t]
\vspace*{-1.5cm}
\centering
\includegraphics[width=14cm]{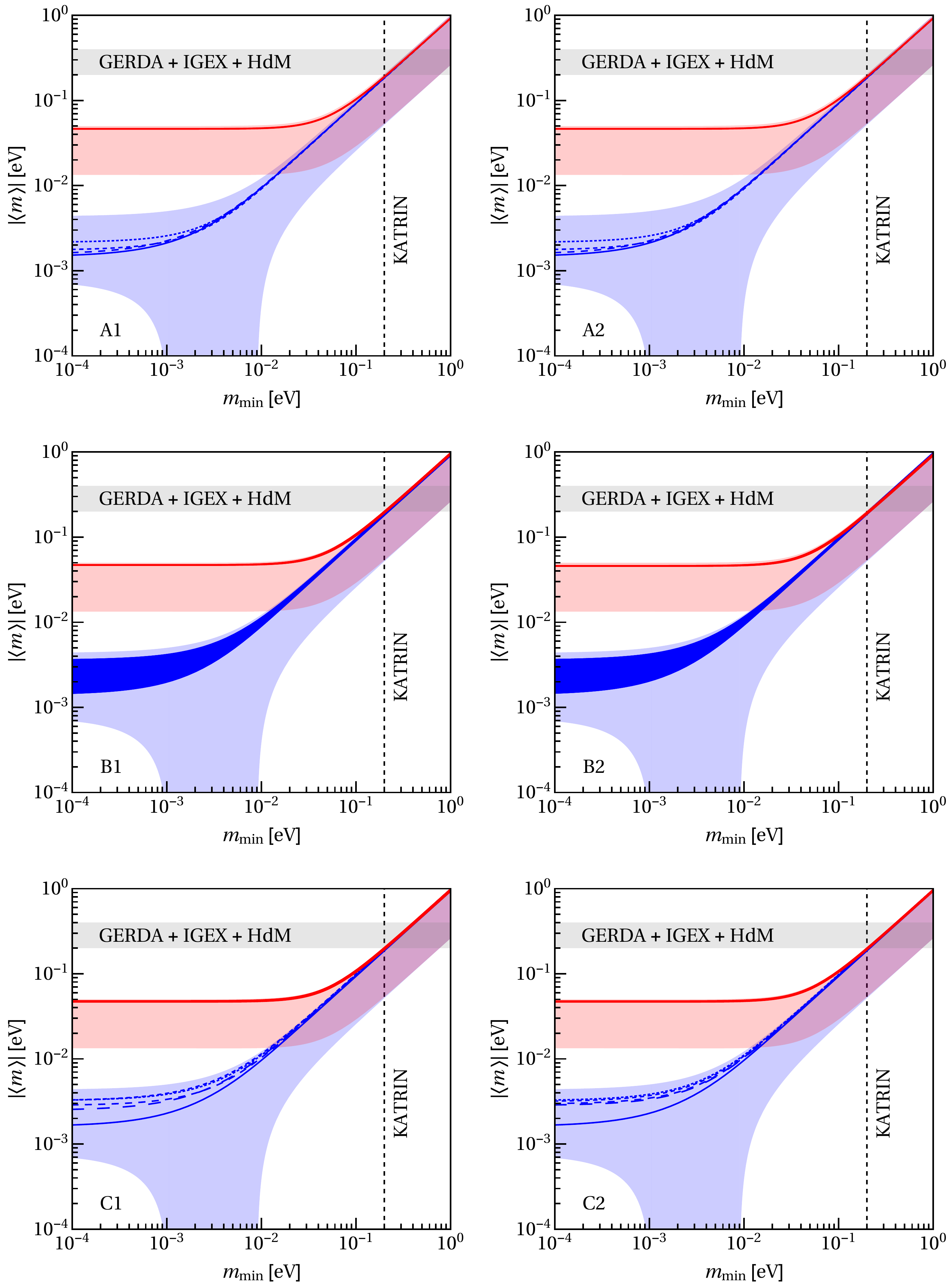}
\caption{\label{fig:1}%
The absolute value of the effective Majorana mass $\meff$ versus 
the lightest neutrino mass $m_{\rm min}$. 
The {\it blue (dark-red) lines} and {\it bands} correspond to $\meff$ 
computed using the best fit values of $\theta_{12}$ 
and $\theta_{13}$ for the NO (IO) spectrum 
and the values of $\delta$, $\alpha_{21}$ and $\alpha_{31}$ 
obtained using the corresponding sum rules and assuming 
$(\xi_{21},\xi_{31}) = (0,0)$.
In cases A1 and A2 the 
{\it solid blue line} corresponds to the TBM symmetry form, 
while the {\it medium, small} and {\it tiny dashed blue lines} are for the 
GRB, GRA and HG symmetry forms, respectively. 
In cases B1 and B2 the predicted values of $\meff$ for all 
the symmetry forms considered are 
within the blue and dark-red bands obtained varying the phase $\beta$ 
in the interval $[0,\pi]$.
In case C1 (C2) the {\it solid blue line} stands for case I (II), 
while the {\it large, medium, small} and {\it tiny dashed blue lines} 
are for cases V (III), II (V), IV (I) and III (IV), respectively. 
The {\it light-blue} and {\it light-red areas} are obtained varying 
the neutrino oscillation parameters 
$\theta_{12}$, $\theta_{13}$, $\Delta m^2_{21}$ and $\Delta m^2_{31(23)}$ 
in their respective $3\sigma$ ranges quoted in 
eqs.~(\ref{th12values}), (\ref{th13values}), 
(\ref{eq:dm21values}) and (\ref{eq:dm3123values})
and the phases $\alpha_{21}$ and $(\alpha_{31} - 2\delta)$ in the interval $[0,2\pi]$. 
The {\it horizontal grey band} indicates the upper bound 
$\meff \sim 0.2 - 0.4$~eV obtained in \cite{Agostini:2013mzu}. 
The {\it vertical dashed line} represents the prospective 
upper limit on $m_{\rm min}$ of $0.2$~eV from the KATRIN 
experiment \cite{Eitel:2005hg}.}
\end{figure}

\begin{figure}[t]
\centering
\includegraphics[width=14cm]{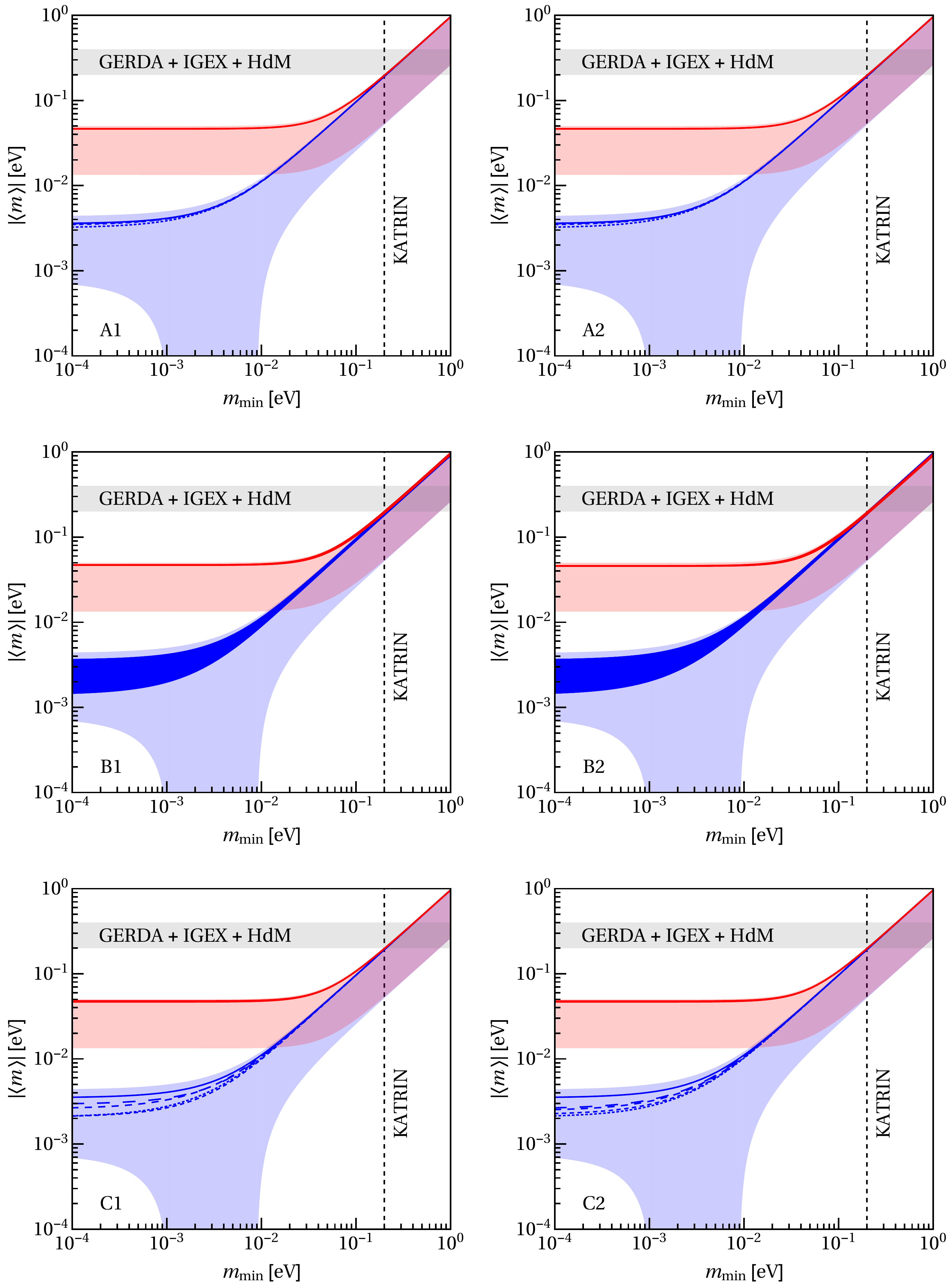}
\caption{\label{fig:2}%
The same as in Fig.~{\ref{fig:1}}, but for $(\xi_{21},\xi_{31}) = (0,\pi)$.}
\end{figure}

\begin{figure}[t]
\centering
\includegraphics[width=14cm]{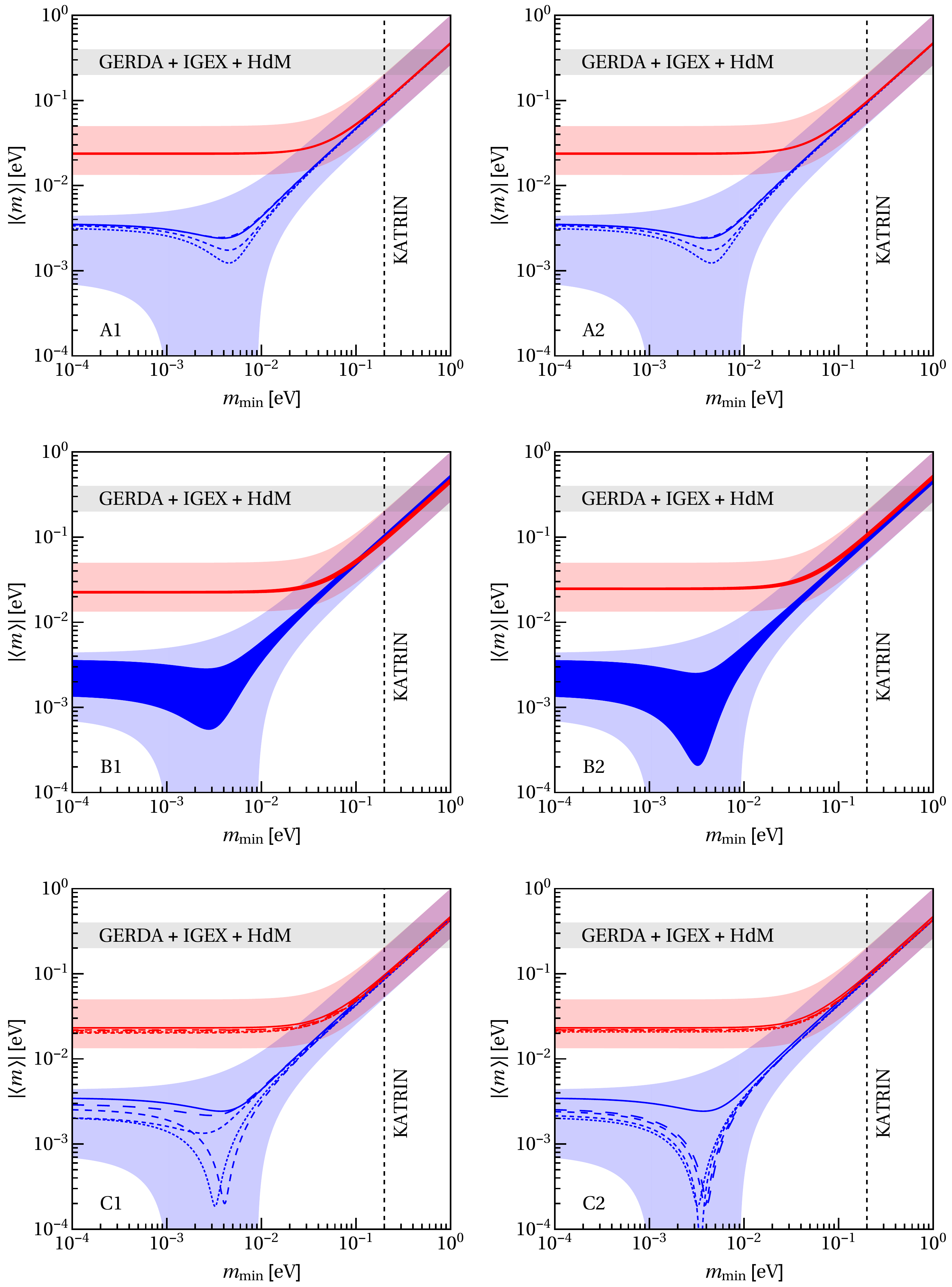}
\caption{\label{fig:3}%
The same as in Fig.~{\ref{fig:1}}, but for $(\xi_{21},\xi_{31}) = (\pi,0)$.}
\end{figure}

\begin{figure}[t]
\centering
\includegraphics[width=14cm]{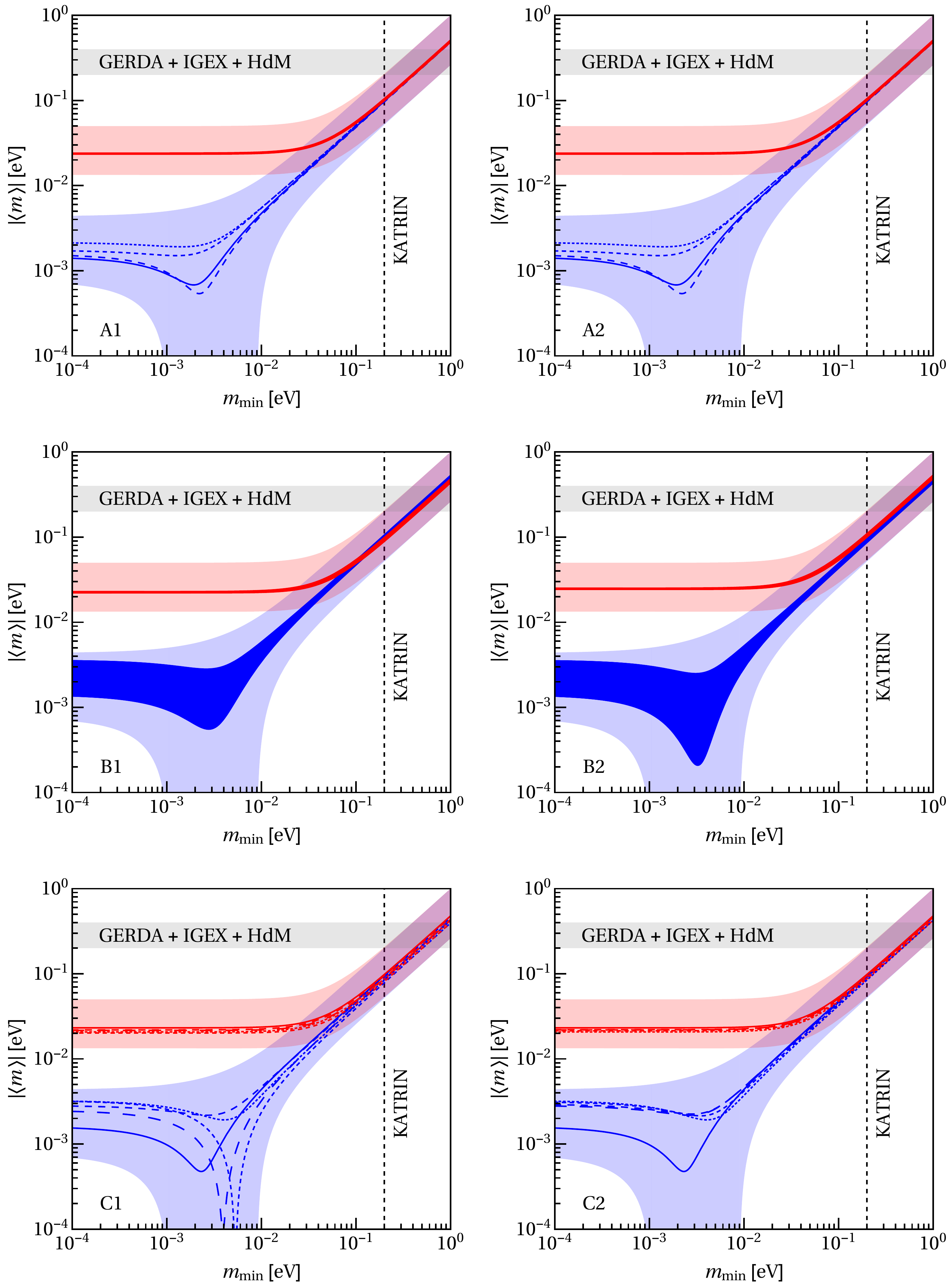}
\caption{\label{fig:4}%
The same as in Fig.~{\ref{fig:1}}, but for $(\xi_{21},\xi_{31}) = (\pi,\pi)$.}
\end{figure}

  In Fig.~\ref{fig:1} we show the absolute value of the effective 
Majorana mass $\meff$ versus the lightest neutrino mass 
$m_{\rm min}$ in the cases of schemes A1, A2, B1, B2, C1 and C2 for 
the NO (blue lines and bands) and IO (dark-red lines and bands) 
neutrino mass spectra, using the best fit values of the mixing 
angles $\theta_{12}$ and $\theta_{13}$ 
quoted in eqs.~(\ref{th12values}) and (\ref{th13values}), 
the best fit values of the two neutrino mass 
squared differences $\Delta m^2_{21}$ and 
$\Delta m^2_{31(23)}$ given 
in eqs.~(\ref{eq:dm21values}) and (\ref{eq:dm3123values}), 
the values of the Dirac phase $\delta$ from Table~\ref{tab:deltaNO} 
and the values of the  Majorana phases $\alpha_{21}$ and $\alpha_{31}$ 
extracted from Tables~\ref{tab:alpha21NO} and \ref{tab:alpha31NO}
setting $(\xi_{21},\xi_{31}) = (0,0)$.
In Figs.~\ref{fig:2}, \ref{fig:3} and \ref{fig:4} 
the values of $(\xi_{21},\xi_{31})$ 
are fixed to $(0,\pi)$, $(\pi,0)$ and $(\pi,\pi)$, respectively. 

 In cases A1 and A2 the 
solid blue line corresponds to the TBM symmetry form of 
the matrix $\tilde U_\nu$, 
while the medium, small and tiny dashed blue lines are for the 
GRB, GRA and HG symmetry forms, respectively. 
In cases B1 and B2 the predicted values of $\meff$ for all 
the symmetry forms considered are 
within the blue and dark-red bands obtained varying the phase $\beta$ 
within the interval $[0,\pi]$. 
In case C1 (C2) the solid blue line stands for case I (II) 
characterised by $[\theta^\nu_{13},\theta^\nu_{12}] = [\pi/20,-\pi/4]$ 
($[\pi/20,\pi/4]$), 
while the large, medium, small and tiny dashed blue lines 
are for cases V (III), II (V), IV (I) and III (IV), respectively, 
where the values of $[\theta^\nu_{13},\theta^\nu_{12}]$ 
in each of these cases are given in the text below Table~\ref{tab:deltaNO}.

 The light-blue and light-red areas are obtained varying 
the neutrino oscillation parameters 
$\theta_{12}$, $\theta_{13}$, $\Delta m^2_{21}$ and 
$\Delta m^2_{31(23)}$ within their respective $3\sigma$ ranges 
quoted in eqs.~(\ref{th12values}), (\ref{th13values}), 
(\ref{eq:dm21values}) and (\ref{eq:dm3123values}), 
and the phases  $\alpha_{21}$ and $(\alpha_{31} - 2\delta)$
within the interval~%
\footnote{The absolute value of the effective Majorana mass as a function 
of $\alpha_{21}$ and $(\alpha_{31} - 2\delta)$, 
$\meff = f(\alpha_{21},\alpha_{31} - 2\delta)$, possesses 
the following symmetry: 
$$f(\alpha_{21},\alpha_{31} - 2\delta) = 
f(2\pi - \alpha_{21}, 2\pi - (\alpha_{31} - 2\delta))\,.$$ 
Thus, it is enough to vary one phase (e.g., $\alpha_{21}$) 
in the interval $[0,\pi]$ 
and the second phase (e.g., $(\alpha_{31} - 2\delta)$) 
in the interval $[0,2\pi]$.}
$[0,2\pi]$.
The horizontal grey band indicates the upper bound on
$\meff$ of $(0.2 - 0.4)$~eV obtained in \cite{Agostini:2013mzu}. 
The vertical dashed line represents the prospective 
upper limit on $m_{\rm min}$ of $0.2$~eV from the KATRIN 
experiment \cite{Eitel:2005hg}. 

As Figs.~\ref{fig:1} and \ref{fig:2} show, for  
$(\xi_{21},\xi_{31})= (0,0)$ and $(0,\pi)$,
the absolute value of the effective Majorana mass $\meff$ for the IO spectrum 
has practically the maximal possible values  
for all schemes considered. In the case of the
NO spectrum and $(\xi_{21},\xi_{31})= (0,0)$, 
$\meff$ is always bigger than $(1.5 - 2.0)\times 10^{-3}$ eV. 
For $(\xi_{21},\xi_{31})= (0,\pi)$,  
$\meff$ has the maximal possible values 
in the A1 and A2 schemes as well in case 
I (II) of the C1 (C2) scheme; in the other cases of the 
C1 (C2) scheme, $\meff$ is always bigger than 
$2.0\times 10^{-3}$~eV. In the B1 and B2 schemes 
and the NO spectrum, $\meff$ can have the maximal 
possible values for both sets of values of 
$(\xi_{21},\xi_{31})= (0,0)$ and $(0,\pi)$.

For $(\xi_{21},\xi_{31})= (\pi,0)$ and $(\pi,\pi)$ 
(Figs.  \ref{fig:3} and \ref{fig:4}) and the IO spectrum, 
a partial  compensation between the three terms 
in $\meff$ takes place for all schemes considered. 
However, $\meff \gtap 2\times 10^{-2}$ eV for all 
cases analysed by us. The mutual compensation 
between the different terms in $\meff$ 
can be stronger in the case of the NO spectrum, 
when $\meff \ltap 10^{-3}$~eV in certain cases 
in specific intervals of values of 
$m_1$, typically between approximately 
$10^{-3}$~eV and $7\times 10^{-3}$~eV.

\clearpage
\thispagestyle{empty}
%
%
%
%
%
%
%
%
%
%
%
%
%
\clearpage

%
\section{Implications of Generalised CP Symmetry 
Combined with Flavour Symmetry}
\label{sec:GCP}
%
%

In the present Section we derive constraints 
on the phases $\xi_{21}$ and $\xi_{31}$
in the matrix $U_{\nu}$, which 
diagonalises the neutrino Majorana mass matrix $M_{\nu}$,
within the approach in which a lepton flavour symmetry $G_f$ 
is combined with a generalised CP symmetry $H_{\rm CP}$. 
We examine successively the cases of $G_f = A_4~(T^\prime)$, 
$S_4$ and $A_5$ with the three LH charged leptons and three 
LH flavour neutrinos transforming under a 3-dimensional 
representation $\rho$ of $G_f$. At low energies the 
flavour symmetry $G_f$ has necessarily to be broken down to residual 
symmetries $G_e$ and $G_\nu$ in the charged lepton 
and neutrino sectors, respectively. 
All the cases considered in the present study 
fall into the class of residual symmetries with 
trivial $G_e$ ($G_f$ being fully broken in the 
charged lepton sector) and $G_\nu = Z_2 \times Z_2$~%
\footnote{Note there are two possibilities for 
$G_\nu = Z_2 \times Z_2$ to be realised.
The first possibility is $G_\nu = Z_2 \times Z_2$ 
being an actual subgroup of $G_f$. Other possibility is 
that only one $Z_2$ subgroup of $G_f$ is preserved, 
while the second $Z_2$ arises accidentally.}.
 
 The residual symmetry $G_\nu$ alone does not provide 
any information on the phases $\xi_{21}$ and $\xi_{31}$ 
of interest. Indeed, let $\bar U_\nu$ be a unitary matrix which 
diagonalises the complex symmetric neutrino Majorana mass matrix:
\begin{equation}
\bar U_\nu^T\, M_\nu\, \bar U_\nu = 
\diag\left(m_1 e^{-i \xi_1}, m_2 e^{-i \xi_2}, m_3 e^{-i \xi_3} \right)\,,
\end{equation}
%
where $m_i$ are non-negative non-degenerate masses~%
\footnote{It follows from the neutrino oscillation data 
that $m_1 \neq m_2 \neq m_3$, and that at least 
two of the three neutrino masses,  $m_{2,3}$ ($m_{1,2}$) 
in the case of the NO (IO) spectrum, are non-zero.
However, even if $m_1 = 0$ ($m_3 = 0$) at tree level 
and the zero value is not protected by a symmetry, 
$m_1$ ($m_3$) will get a non-zero contribution at 
least at two loop level \cite{CNLSPST83} 
and in the framework of a self-consistent (renormalisable) 
theory of neutrino mass generation this higher contribution 
will be finite.}
and $\xi_i$ are phases contributing to the Majorana phases in the 
PMNS matrix. Let us introduce the matrices 
\begin{equation}
\bar Q_0 = \diag\left(e^{i \frac{\xi_1}{2}}, e^{i \frac{\xi_2}{2}}, e^{i \frac{\xi_3}{2}}\right)\,,
\end{equation}
%
and $U_\nu \equiv \bar U_\nu \bar Q_0$, such that 
\begin{equation}
U_\nu^T\, M_\nu\, U_\nu = M_\nu^{\rm d} \equiv
\diag\left(m_1, m_2 , m_3\right)\,.
\label{eq:Mnudiag}
\end{equation}
%
Thus, 
\begin{equation}
U_\nu = \bar U_\nu\, \bar Q_0 = 
e^{i\frac{\xi_1}{2}}\Psi_{\nu}\tilde{U}_\nu \, Q_0\,,  
\label{eq:factorisation}
\end{equation}
%
where $\Psi_{\nu}$ is a diagonal phase matrix containing, 
in general, two phases, $\xi_1/2$ is a common unphysical 
phase, and 
\begin{equation}
 Q_0 = {\rm diag}\left(1,e^{i\frac{\xi_2 - \xi_1}{2}},e^{i\frac{\xi_3 - \xi_1}{2}}\right) 
= {\rm diag}\left(1,e^{i\frac{\xi_{21}}{2}},e^{i\frac{\xi_{31}}{2}}\right)\,.
\end{equation}
%
Clearly, the phases of interest are 
$\xi_{21} = \xi_2 - \xi_1$ and $\xi_{31} = \xi_3 - \xi_1$.
It is clear from eq.~(\ref{eq:factorisation}) that the common phases of the 
columns of $U_\nu$ have been factorised in the matrix $\bar Q_0$.

 The $G_\nu$ invariance of the neutrino mass matrix implies 
\begin{equation}
\rho(g_\nu)^T\, M_\nu\, \rho(g_\nu) = M_\nu
\quad
\forall\, g_\nu \in G_\nu\,.
\end{equation}
%
Further, using eq.~(\ref{eq:Mnudiag}), we find 
\begin{equation}
(\rho^{({\rm d})}(g_\nu))^T \, M_\nu^{\rm d}\, \rho^{({\rm d})}(g_\nu)= M_\nu^{\rm d}\,,
\quad
\text{with}
\quad
\rho^{({\rm d})}(g_\nu)= U_\nu^\dagger \, \rho(g_\nu) \, U_\nu\,.
\end{equation}
%
For $m_1 \neq m_2 \neq m_3$ and 
${\rm min}(m_j)\neq 0$, $j=1,2,3$,
as it is not difficult to show,
the matrix $\rho^{({\rm d})}(g_\nu)$
can have only the following form:
\begin{equation}
\rho^{({\rm d})}(g_\nu) = \diag(\pm1, \pm1, \pm1)\,,
\end{equation}
%
where the signs of the three non-zero entries in 
$\rho^{({\rm d})}(g_\nu)$ are not correlated. 
Finally, from the preceding two equations we get 
\begin{equation}
\rho^{({\rm d})}(g_\nu)= \bar Q_0 \, \rho^{({\rm d})}(g_\nu) \, \bar Q_0^* = 
\bar U_\nu^\dagger \, \rho(g_\nu) \, \bar U_\nu\,,
\end{equation}
%
i.e., the phases $\xi_i$ cancel out.
Therefore a lepton flavour symmetry alone does not lead 
to any constraints on the phases $\xi_i$, $i=1,2,3$, 
and thus on the phases $\xi_{21}$ and $\xi_{31}$. 

 Let us consider next the implications of 
a residual generalised CP symmetry $H_{\rm CP}^\nu \subset H_{\rm CP}$, 
which is preserved in the neutrino sector.
In this case the neutrino Majorana mass matrix satisfies the 
following condition: 
\begin{equation}
X_i^T \, M_\nu \, X_i = M_\nu^*\,,
\end{equation}
%
where $X_i \in H_{\rm CP}^\nu$ are the generalised CP transformations. 
Substituting $M_\nu$ from eq.~(\ref{eq:Mnudiag}), we find 
\begin{equation}
(X_i^{\rm d})^T\, M_\nu^{\rm d}\, X_i^{\rm d} = M_\nu^{\rm d}\,,
\quad
\text{with}
\quad
X_i^{\rm d} = U_\nu^\dagger \, X_i \, U_\nu^*\,.
\end{equation}
%
Again, since the three neutrino masses in $M_\nu^{\rm d}$ 
have to be, as it follows from the data,
non-degenerate, we have 
\begin{equation}
X_i^{\rm d} = \diag(\pm1, \pm1, \pm1)\,.
\end{equation}
%
Finally, using that $U_\nu \equiv \bar U_\nu \bar Q_0$, 
we obtain \cite{Everett:2015oka} 
\begin{equation}
\diag\left(\pm e^{i \xi_1}, \pm e^{i \xi_2}, \pm e^{i \xi_3}\right) 
= \bar Q_0 \, X_i^{\rm d} \, \bar Q_0 = 
\bar U_\nu^\dagger \, X_i \, \bar U_\nu^*\,.
\label{eq:xiphases}
\end{equation}
%
Thus, we come to the conclusion that the phases $\xi_i$ 
will be known once i) the matrix $\bar U_\nu$ is fixed by the residual 
flavour symmetry $G_\nu$, and 
ii) the generalised CP transformations 
$X_i \in H_{\rm CP}^\nu$, which are consistent with $G_\nu$, 
are identified.

 Now we turn to concrete examples. 
For $G_f = A_4$ we choose to work in 
the Altarelli-Feruglio basis \cite{Altarelli:2005yx}. 
Preserving the $S$ generator leads to 
$\bar U_\nu = U_{\rm TBM}$, provided there is 
an additional accidental $\mu$\,--\,$\tau$ symmetry \cite{Altarelli:2010gt}.
Then, twelve generalised CP transformations consistent with 
the $A_4$ flavour symmetry for the triplet representation 
in the chosen basis have been found in \cite{Ding:2013bpa}, 
solving the consistency condition 
\begin{equation}
X \, \rho^*(g) \, X^{-1} = \rho (g^\prime)\,, 
\quad g, g^\prime \in A_4\,.
\label{eq:consistency}
\end{equation}
%
These transformations can be summarised in a compact way as follows:
\begin{equation}
X = \rho (g)\,, \quad g \in A_4\,,
\end{equation}
%
i.e., the generalised CP transformations consistent with the 
$A_4$ flavour symmetry are of the same form as the flavour 
symmetry group transformations \cite{Ding:2013bpa}. 
They are given in Table~1 in \cite{Ding:2013bpa} 
together with the elements $\hat S$ and $\hat T$ 
to which the generators $S$ and $T$ 
of $A_4$ are mapped by the consistency condition 
in eq.~(\ref{eq:consistency}).
Further, since in our case the residual flavour symmetry 
$G_\nu = Z_2 \times Z_2$, 
where one $Z_2$ factor corresponds to the preserved 
$S$ generator, only those $X$ are acceptable, 
for which $\hat S = S$. From Table~1 in \cite{Ding:2013bpa} 
it follows that there are four such generalised CP transformations, 
namely, $\rho(E)$, $\rho(S)$, $\rho(T^2ST)$ and $\rho(TST^2)$, 
where $E$ is the identity element of the group.
The last two transformations are not symmetric in the chosen basis, 
and, as shown in \cite{Ding:2013bpa}, 
lead to partially degenerate neutrino mass spectrum 
with two equal masses 
 (see also \cite{Feruglio:2012cw}), which is ruled out 
by the existing neutrino oscillation data.
Thus, we are left with two allowed generalised CP transformations, 
$\rho(E)$ and $\rho(S)$, for which we have:
\begin{align}
& U_{\rm TBM}^{\dagger} \, \rho(E) \, U^*_{\rm TBM} = \rho(E) = \diag(1, 1, 1)\,, \\
& U_{\rm TBM}^{\dagger} \, \rho(S) \, U^*_{\rm TBM} = \diag(-1, 1, -1)\,.
\end{align}
%
Finally, according to eq.~(\ref{eq:xiphases}), this implies 
that the phases $\xi_i$ can be either $0$ or $\pi$.
The same conclusion holds for a $T^{\prime}$ 
flavour symmetry, because restricting ourselves 
to the triplet representation
for the LH charged lepton and neutrino
fields, there is no way to distinguish $T^{\prime}$ from $A_4$
\cite{Feruglio:2007uu}.

 In the case of $G_f = S_4$ we choose to 
work in the basis given in \cite{Altarelli:2009gn}.
The residual symmetry $G_{\nu} =  Z_2 \times Z_2$, 
where one $Z_2$ factor corresponds to preserved $S$ generator 
in the chosen basis and the second one arises accidentally 
(a $\mu$\,--\,$\tau$ symmetry), leads to 
$\bar U_\nu = U_{\rm BM}$ \cite{Altarelli:2009gn}.
As in the previous example, the generalised CP transformations 
consistent with the $S_4$ flavour symmetry are 
of the same form as the flavour symmetry group transformations 
\cite{Holthausen:2012dk}.
Solving the consistency condition in eq.~(\ref{eq:consistency}),
we find ten symmetric generalised CP transformations consistent with 
the $S_4$ flavour symmetry for the triplet representation 
in the chosen basis. We summarise them in Table~\ref{tab:GCPS4} 
together with elements $\hat T$ and $\hat S$ to which 
the consistency condition maps the group generators $T$ and $S$.
\begin{table}[h]
\centering
\begin{tabular}{lcc}
\toprule
 $g$, $X = \rho(g)$  & $T \rightarrow \hat T$ & $S \rightarrow \hat S$ \\
 \midrule
$(ST^2)^2$ 
& $T$ & $S$ \\
$T^3$
& $T^3$ & $T^3ST$ \\
$E$ 
& $T^3$ & $S$ \\ 
$T$ 
& $T^3$ & $TST^3$ \\
$T^2ST^2$ 
& $STS$ & $S$ \\ 
$ST^2S$ 
& $T$ & $T^2ST^2$ \\
$S$ 
& $TST$ & $S$ \\ 
$T^2$ 
& $T^3$ & $T^2ST^2$ \\
$STS$ 
& $ST^2$ & $ST^2ST$ \\ 
$TST$ 
& $T^2S$ & $TST^2S$ \\
\bottomrule
\end{tabular}
\caption{The ten symmetric generalised CP transformations $X = \rho(g)$ 
consistent with the $S_4$ flavour symmetry for the triplet representation $\rho$
in the chosen basis \cite{Altarelli:2009gn} determined by the consistency condition in eq.~(\ref{eq:consistency}). 
The mapping $(T,S) \rightarrow (\hat T, \hat S)$ is realised 
via the consistency condition applied to the group generators 
$T$ and $S$, i.e., $X \rho^*(T) X^{-1} = \rho(\hat T)$ and 
$X \rho^*(S) X^{-1} = \rho(\hat S)$.
\label{tab:GCPS4}}
\end{table}
 
 From this table we see that there are four 
symmetric generalised CP transformations consistent with 
the preserved $S$ generator, namely,
$\rho(E)$, $\rho(S)$, $\rho(T^2ST^2)$ and $\rho(ST^2ST^2)$. 
Substituting them and $\bar U_\nu = U_{\rm BM}$ in eq.~(\ref{eq:xiphases}), 
we find: 
\begin{align}
& U_{\rm BM}^{\dagger} \, \rho(E) \, U^*_{\rm BM}  = \rho(E) = \diag(1,1,1)\,, \\
& U_{\rm BM}^{\dagger} \, \rho(S) \, U^*_{\rm BM}  = \diag(1,-1,1)\,, \\
& U_{\rm BM}^{\dagger} \, \rho(T^2ST^2) \, U^*_{\rm BM}  = \diag(-1,1,1)\,, \\
& U_{\rm BM}^{\dagger} \, \rho(ST^2ST^2) \, U^*_{\rm BM}  = \diag(-1,-1,1)\,.
\end{align}
Therefore also in this case the phases  $\xi_i$ are fixed by 
residual generalised CP symmetry to be either
$0$ or $\pi$.

 As a third example, we consider $G_f = A_5$. 
We employ the basis for the triplet representation 
of the generators $S$ and $T$ of this group 
given in \cite{Ding:2011cm}.
The residual symmetry $G_{\nu} =  Z_2 \times Z_2$
generated by $S$ and $T^3ST^2ST^3$ leads to GRA mixing, 
i.e., $\bar U_\nu = U_{\rm GRA}$, as is shown in \cite{Ding:2011cm}. 
It is stated in \cite{Li:2015jxa} that the generalised CP transformations 
consistent with $A_5$ are of the same form as the group transformations. 
Solving the consistency condition in eq.~(\ref{eq:consistency}),
we find 16 symmetric generalised CP transformations consistent with 
$A_5$ for the triplet representation in the working basis.
We summarise them in Table~\ref{tab:GCPA5}, where 
we present also the elements $\hat T$ and $\hat S$.
\begin{table}[h]
\centering
\begin{tabular}{lcc}
\toprule
 $g$, $X = \rho(g)$  & $T \rightarrow \hat T$ & $S \rightarrow \hat S$ \\
 \midrule
$T^3ST^2ST^3$ 
& $STS$ & $S$ \\
$S$ 
& $TST$ & $S$ \\
$(ST^2)^2S$ 
& $ST^3$ & $(T^2S)^2T^4$ \\
$TST$ 
& $T^2S$ & $TST^2S$ \\
$ST^3S$ 
& $T^2ST$ & $ST^3ST^2S$ \\
$T^3ST^3$ 
& $T^4ST^3$ & $T^2ST^2ST^3S$ \\
$T^3ST^2ST^3S$ 
& $T$ & $S$ \\
$T$ 
& $T^4$ & $TST^4$ \\
$T^2$ 
& $T^4$ & $T^2ST^3$ \\
$E$ 
& $T^4$ & $S$ \\
$T^3$ 
& $T^4$ & $T^3ST^2$ \\
$T^4$ 
& $T^4$ & $T^4ST$ \\
$ST^2S$ 
& $TST^2$ & $ST^2ST^3S$ \\
$T^2ST^2$ 
& $T^3ST^4$ & $T^4ST^2ST^3S$ \\
$STS$ 
& $ST^2$ & $ST^2ST$ \\
$(T^2S)^2T^2$ 
& $T^3S$ & $T^4(ST^2)^2$ \\
\bottomrule
\end{tabular}
\caption{The 16 symmetric generalised CP transformations $X = \rho(g)$ 
consistent with the $A_5$ flavour symmetry for the triplet representation $\rho$
in the chosen basis \cite{Ding:2011cm} determined by the consistency condition in eq.~(\ref{eq:consistency}). 
The mapping $(T,S) \rightarrow (\hat T, \hat S)$ is realised 
via the consistency condition applied to the group generators 
$T$ and $S$, i.e., $X \rho^*(T) X^{-1} = \rho(\hat T)$ and 
$X \rho^*(S) X^{-1} = \rho(\hat S)$.
\label{tab:GCPA5}}
\end{table}

It follows from this table that the generalised CP transformations consistent 
with $G_{\nu} =  Z_2 \times Z_2$ of interest are of the same 
form of $G_\nu$. Namely, they are 
$\rho(E)$, $\rho(S)$, $\rho(T^3ST^2ST^3)$ and $\rho(T^3ST^2ST^3S)$,
and we have: 
\begin{align}
& U_{\rm GRA}^{\dagger} \, \rho(E) \, U^*_{\rm GRA} = \rho(E) = \diag(1,1,1)\,, \\
& U_{\rm GRA}^{\dagger} \, \rho(S) \, U^*_{\rm GRA} = \diag(1,-1,-1)\,, \\
& U_{\rm GRA}^{\dagger} \, \rho(T^3ST^2ST^3) \, U^*_{\rm GRA} = \diag(-1,1,-1)\,, \\
& U_{\rm GRA}^{\dagger} \, \rho(T^3ST^2ST^3S) \, U^*_{\rm GRA} = \diag(-1,-1,1)\,.
\end{align}
Thus, as in the previous cases, the phases  $\xi_i$ are fixed by generalised CP 
symmetry to be either $0$ or $\pi$. 

 It follows from the results derived in the present Section 
that the two phases $\xi_{21} = \xi_2 -\xi_1$ and 
$\xi_{31} = \xi_3 - \xi_1$, present in the matrix $Q_0$ (see eq.~(\ref{PsieQ0})) 
and giving contributions to 
the Majorana phases $\alpha_{21}$ and $\alpha_{31}$ 
in the PMNS matrix, are constrained to be 
either $0$ or $\pi$ for all examples considered.

 Finally, we note that although 
in the cases of the flavour symmetry groups considered~---~%
$A_4$, $T^\prime$, $S_4$ and $A_5$~---~%
we choose to work in specific basis for the generators of 
each symmetry group, the results on the phases $\xi_{1,2,3}$
we have obtained, as we show below, are basis-independent. 
Indeed, let $B$ be a unitary matrix, which realises the change of basis. 
Then, the representation matrices of the group elements in 
the new basis, $\tilde\rho(g)$, are given by
\begin{equation}
\tilde\rho(g) = B \, \rho(g) \, B^\dagger\,, 
\quad g \in G_f\,.
\end{equation}
%
Expressing $\rho(g)$ from this equation and substituting it 
in the consistency condition given in eq.~(\ref{eq:consistency}) 
leads to
\begin{equation}
\tilde X \, \tilde\rho^*(g) \, \tilde X^{-1} = \tilde\rho (g^\prime)\,, 
\quad g, g^\prime \in G_f\,,
\label{eq:consistencynewbasis}
\end{equation}
%
where 
\begin{equation}
\tilde X = B \, X \, B^T
\end{equation}
are the generalised CP transformations in the new basis.
Now we substitute $X$ from this equation in eq.~(\ref{eq:xiphases}) 
and obtain
\begin{equation}
\left(\tilde{\bar U}_\nu\right)^\dagger \, \tilde X_i \, \left(\tilde{\bar U}_\nu\right)^* 
= \bar U_\nu^\dagger \, X_i \, \bar U_\nu^*
= \diag\left(\pm e^{i \xi_1}, \pm e^{i \xi_2}, \pm e^{i \xi_3}\right)\,,
\label{eq:basisindependence}
\end{equation}
%
where $\tilde{\bar U}_\nu = B\,\bar U_\nu$ is the matrix 
which diagonalises the neutrino Majorana mass matrix $\tilde M_\nu$,  
$\tilde M_\nu = B^* M_\nu \, B^\dagger$, in the new basis, i.e.,
\begin{equation}
\tilde{\bar U}_\nu^T\, \tilde M_\nu\, \tilde{\bar U}_\nu = 
\bar U_\nu^T\, M_\nu\, \bar U_\nu = 
\diag\left(m_1 e^{-i \xi_1}, m_2 e^{-i \xi_2}, m_3 e^{-i \xi_3} \right)\,.
\end{equation}
%
What concerns the charged lepton sector, in all cases we consider in the 
present study a flavour symmetry $G_f$ is completely broken 
in the charged lepton sector, i.e.,  
the residual symmetry group $G_e$ consists only of the identity element $E$. 
The change of basis yields 
$\tilde\rho(E) = B \, \rho(E) \, B^\dagger$. 
As can be easily shown, 
the matrix $U_e^\prime = B\,U_e$ diagonalises the hermitian matrix 
$\tilde M_e \, \tilde M_e^\dagger$, 
$\tilde M_e \, \tilde M_e^\dagger = B \, M_e \, M_e \, B^\dagger$, 
in the new basis, 
$M_e$ being the charged lepton mass matrix in the initial basis. 
Namely, 
\begin{equation}
{U_e^\prime}^\dagger \, \tilde M_e \, \tilde M_e^\dagger \, U_e^\prime =
U_e^\dagger \, M_e \, M_e^\dagger \, U_e =
\diag\left(m_e^2, m_\mu^2, m_\tau^2\right)\,.
\end{equation}
%
Taking into account that $U_\nu^\prime = B \, U_\nu = B \, \bar U_\nu \, \bar Q_0$, 
we obtain for the PMNS matrix $U$: 
\begin{equation}
U = {U_e^\prime}^\dagger \, U_\nu^\prime = U_e^\dagger \, U_\nu = 
U_e^\dagger \, \bar U_\nu \, \bar Q_0\,.
\label{eq:UPMNSbasisindependence}
\end{equation}
%
Thus, as eqs.~(\ref{eq:basisindependence}) and  
(\ref{eq:UPMNSbasisindependence})
demonstrate, the results 
for the phases $\xi_i$ are basis-independent.

%
\section{Summary and Conclusions}
\label{sec:summary}
%
%

In the present article we have obtained 
predictions for the Majorana 
phases $\alpha_{21}/2$ and $\alpha_{31}/2$ of
the $3\times 3$ unitary neutrino mixing
matrix $U = U_e^{\dagger} \, U_{\nu} =
(\tilde{U}_{e})^\dagger\, \Psi \tilde{U}_{\nu} \, Q_0\,$, 
$U_e$ ($\tilde{U}_e$)
and $U_{\nu}$  ($\tilde{U}_\nu$)
being $3\times 3$ unitary (CKM-like) 
matrices arising from the diagonalisation 
respectively of the charged lepton and neutrino Majorana 
mass terms. Each of the diagonal phase matrices $\Psi$ and $Q_0$ 
contains, in  general, two physical CPV phases \cite{Frampton:2004ud}. 
The phases in the matrix $Q_0$, $\xi_{21}/2$ and  $\xi_{31}/2$, 
contribute to the Majorana phases in the PMNS matrix. 
Our study employs a method proposed in \cite{Petcov:2014laa} and 
is a natural continuation of the studies performed 
in \cite{Petcov:2014laa,Girardi:2014faa,Girardi:2015vha}.
We have considered forms of $\tilde{U}_e$ and 
$\tilde{U}_{\nu}$, permitting to express $\delta$ as 
a function of the PMNS mixing angles,
$\theta_{12}$, $\theta_{13}$ and $\theta_{23}$, present in $U$, 
and the angles contained in $\tilde{U}_{\nu}$ 
\cite{Petcov:2014laa,Girardi:2015vha}. 
As we have shown, for the same forms,  
the Majorana phases $\alpha_{21}/2$ and $\alpha_{31}/2$
are determined by the values 
of $\theta_{12}$, $\theta_{13}$ and $\theta_{23}$ 
and the phases  $\xi_{21}/2$ and  $\xi_{31}/2$ (see below).
We have derived such sum rules for 
$\alpha_{21}/2$ and $\alpha_{31}/2$ in the following cases: 
\begin{itemize}
\item[i)] $U = R_{12}(\theta^e_{12})\Psi R_{23}(\theta^{\nu}_{23}) 
R_{12}(\theta^{\nu}_{12}) Q_0$ (case A1),
\item[ii)] $U = R_{13}(\theta^e_{13})\Psi R_{23}(\theta^{\nu}_{23}) 
R_{12}(\theta^{\nu}_{12}) Q_0$ (case A2),
\item[iii)] $U =R_{12}(\theta^e_{12})R_{23}(\theta^e_{23})\Psi
R_{23}(\theta^{\nu}_{23}) R_{12}(\theta^{\nu}_{12}) Q_0$ (case B1),
\item[iv)] $U =R_{13}(\theta^e_{13})R_{23}(\theta^e_{23})\Psi
R_{23}(\theta^{\nu}_{23}) R_{12}(\theta^{\nu}_{12}) Q_0$ (case B2),
\item[v)] $U =R_{12}(\theta^e_{12}) R_{13}(\theta^e_{13}) \Psi 
R_{23}(\theta^{\nu}_{23}) R_{12}(\theta^{\nu}_{12}) Q_0$ (case B3),
\item[vi)] $U =R_{12}(\theta^e_{12})\Psi R_{23}(\theta^{\nu}_{23})
R_{13}(\theta^{\nu}_{13}) R_{12}(\theta^{\nu}_{12}) Q_0$ (case C1),
\item[vii)] $U = R_{13}(\theta^e_{13}) \Psi R_{23}(\theta^{\nu}_{23})
R_{13}(\theta^{\nu}_{13}) R_{12}(\theta^{\nu}_{12}) Q_0$ (case C2),
\end{itemize} 
where $R_{ij}$ are real matrices, $R^T = R^{-1}$,
and  $\theta^e_{ij}$ and $\theta^{\nu}_{ij}$ 
denote the rotation angles in 
$\tilde{U}_e$ and $\tilde{U}_{\nu}$, respectively.  
The sum rules are summarised in Section~\ref{sec:sumrules}.
In the sum rules, $\alpha_{21}/2$ and $\alpha_{31}/2$
are expressed, in general, 
in terms of the three measured angles 
of the PMNS matrix, $\theta_{12}$, 
$\theta_{13}$ and $\theta_{23}$, 
the phases $\xi_{21}/2$ and $\xi_{31}/2$ of the matrix $Q_0$, 
and the angles in $\tilde{U}_{\nu}$, which are 
supposed to have known values, determined by symmetries.
In the cases of schemes B1 and B2 (scheme B3),  
$\alpha_{31}/2$ ($\delta$, $\alpha_{21}/2$ and $\alpha_{31}/2$)
depends (depend) on one additional, in general, unknown 
phase $\beta$ ($\omega$), whose value can only be 
fixed in a self-consistent theory of generation of  
neutrino masses and mixing.

 In order to obtain predictions for the Majorana phases 
one has to specify, in particular, the values of the 
angles in the matrix $\tilde{U}_{\nu}$.
In the present study we have considered 
the following symmetry forms of $\tilde{U}_{\nu}$:
tri-bimaximal (TBM), bimaximal (BM), 
golden ratio A (GRA), golden ratio B (GRB),
and hexagonal (HG). 
All these forms are characterised by the same
$\theta^{\nu}_{23} = -\pi/4$ and $\theta^{\nu}_{13} = 0$, 
but differ by the value of the angle $\theta^{\nu}_{12}$.
For the forms cited above and used in the present study 
the values of $\theta^{\nu}_{12}$ are given in the Introduction.
In  schemes  
C1 and C2 we have employed three 
representative fixed values of $\theta^{\nu}_{13}\neq 0$ 
considered in the literature and appearing in models 
with flavour symmetries, $\theta^{\nu}_{13} = \pi/20,~\pi/10$ and 
$\sin^{-1} (1 / 3)$, 
together with certain fixed values of $\theta^{\nu}_{12}$~---~%
in total five different pairs of values 
of $[\theta^{\nu}_{13},\theta^{\nu}_{12}]$ in each of the 
two schemes. The values of the five pairs
are given in Table~\ref{tab:deltaNO}. 

Thus, for the specific symmetry 
forms of $\tilde{U}_{\nu}$ listed above and used 
in our numerical analysis, the phase differences  
a) $(\alpha_{21}/2 - \xi_{21}/2)$ and 
$(\alpha_{31}/2 - \xi_{31}/2)$ in 
 schemes A1, A2, C1 and C2,
b) $(\alpha_{21}/2 - \xi_{21}/2)$ and 
$(\alpha_{31}/2 - \xi_{31}/ - \beta)$
in schemes B1 and B2, and 
c)  $(\alpha_{21}/2 - \xi_{21}/2)$ and 
$(\alpha_{31}/2 - \xi_{31}/2)$ for a fixed $\omega$ 
in scheme B3, 
are determined completely by the values of 
the measured neutrino mixing angles
$\theta_{12}$, $\theta_{13}$ and $\theta_{23}$ 
and the angles in the matrix $\tilde{U}_{\nu}$.
If the value of the Dirac phase 
$\delta$ is measured,  that will allow 
to fix the value of $\omega$ in scheme B3. 
Using the best fit values 
of $\theta_{12}$, $\theta_{13}$ and $\theta_{23}$, 
we have obtained predictions for the  
phase differences listed above, which
are summarised in Tables \ref{tab:alpha21NO} and 
 \ref{tab:alpha31NO}. In the case of scheme B3, 
we have set $\omega =0$. For this value of $\omega$ 
the predicted value of the Dirac phase 
$\delta$ lies in the $2\sigma$ 
interval of allowed values quoted in eq.~(\ref{deltaexp}).
The results reported in  Tables \ref{tab:alpha21NO} and 
\ref{tab:alpha31NO} show that the phase differences 
of interest involving the Majorana phases $\alpha_{21}/2$ and 
$\alpha_{31}/2$ are determined with a two-fold ambiguity 
by the values of  $\theta_{12}$, $\theta_{13}$ and $\theta_{23}$.
This is a consequence of the fact that, as long as the sign 
of $\sin\delta$ is not fixed by the data, 
the Dirac phase $\delta$, 
on which the phase differences under discussion depend, 
is determined by the values of 
$\theta_{12}$, $\theta_{13}$ and $\theta_{23}$ 
in the schemes studied by us with a two-fold ambiguity 
\cite{Petcov:2014laa,Girardi:2014faa,Girardi:2015vha}, 
as Table \ref{tab:deltaNO} also shows. 
It follows from eq.~(\ref{deltaexp}) that  
the current data appear to favour negative values 
of  $\sin\delta$.   
The predictions for the BM (LC) symmetry form 
of $\tilde{U}_{\nu}$ in  Tables \ref{tab:alpha21NO} and 
\ref{tab:alpha31NO} correspond to the 
current $3\sigma$ upper bound of allowed values of 
$\sin^2 \theta_{12} = 0.354$ and the best fit values of 
$\sin^2 \theta_{23}$ and $\sin^2 \theta_{13}$, 
since using the best fit values of 
the three neutrino mixing angles 
one gets unphysical values of $|\cos\delta| > 1$ 
\cite{Marzocca:2013cr,Petcov:2014laa,Girardi:2015vha}. 
Physical values of $\cos\delta$ are found for larger (smaller) 
values of $\sin^2\theta_{12}$ ($\sin^2\theta_{23}$) 
\cite{Petcov:2014laa,Girardi:2014faa,Girardi:2015vha}.
For $\sin^2\theta_{12} = 0.354$ and the best 
fit values of $\sin^2\theta_{23}$ and $\sin^2\theta_{13}$,  
$|\cos\delta|$ has an unphysical value 
greater than one only for schemes 
B1 with the IO spectrum, B2 with the NO spectrum and B3, 
and for these cases we do not present results 
for the relevant phase differences.

 We have investigated also how the predictions 
for the phase differences 
$(\alpha_{21}/2 - \xi_{21}/2)$ and $(\alpha_{31}/2 - \xi_{31}/2)$ 
($(\alpha_{31}/2 - \xi_{31}/2 - \beta)$) presented in 
Tables~\ref{tab:alpha21NO} and \ref{tab:alpha31NO} change when 
the uncertainties in  determination of the neutrino mixing parameters 
are taken into account 
(see Figs.~\ref{fig:MPs3sigmaB1} and \ref{fig:MPs3sigmaB2} and 
the related discussion as well as Appendix~\ref{app:A1A2}).

 Extracting the values of the Majorana phases 
$\alpha_{21}/2$ and $\alpha_{31}/2$ from the 
results presented in Tables \ref{tab:alpha21NO} and 
\ref{tab:alpha31NO} 
for two fixed values of 
each of the phases $\xi_{21}$ and $\xi_{31}$,   
$\xi_{21}=0$ and $\pi$, $\xi_{31} = 0$ and $\pi$ 
(altogether four cases),
and using also the predicted values of the 
Dirac phase $\delta$ from Table \ref{tab:deltaNO}   
and the best fit values of  $\sin^2\theta_{12}$, 
$\sin^2\theta_{23}$ and $\sin^2\theta_{13}$,
we derived (in graphic form) predictions 
for the absolute value of the neutrinoless double beta decay effective Majorana 
mass $\meff$ as a function of the lightest neutrino 
mass $m_{\rm min} \equiv {\rm min}(m_j)$, $j=1,2,3$,
for both the NO and IO neutrino mass spectra 
(Figs. \ref{fig:1}~--~\ref{fig:4}).
For schemes B1 and B2 the predictions are 
obtained by varying the phase $\beta$ 
in the interval $[0,\pi]$.
As a possible justification of the choice of the 
two values of the phases $\xi_{21}$ and $\xi_{31}$ used 
for the predictions of $\meff$, we show that the requirement of 
generalised CP invariance of the neutrino Majorana mass term 
in the cases of the $S_4$, $A_4$, $T^\prime$ and $A_5$ 
lepton flavour symmetries leads to the constraints 
$\xi_{21}=0$ or $\pi$, $\xi_{31}=0$ or $\pi$.

 The results derived in the present article
for the Majorana CPV phases in the PMNS neutrino mixing 
matrix $U$ complement the results obtained in 
\cite{Petcov:2014laa,Girardi:2014faa,Girardi:2015vha} 
on the predictions for the Dirac phase $\delta$ 
in $U$ in schemes in which the underlying 
form of $U$ is determined by, or is associated with, 
in particular, discrete (lepton) flavour symmetries.

%
\section*{Acknowledgements}
%
%

 A.V.T. would like to thank E.V.\,Titov for discussions 
facilitating the making of the figures in the present article. 
This work was supported in part 
by the INFN program on Theoretical Astroparticle Physics (TASP),
by the research grant  2012CPPYP7 ({\sl Theoretical Astroparticle Physics})
under the program  PRIN 2012 funded by the Italian Ministry 
of Education, University and Research (MIUR), 
by the European Union FP7 ITN INVISIBLES 
(Marie Curie Actions, PITN-GA-2011-289442-INVISIBLES), and 
by the World Premier International Research Center
Initiative (WPI Initiative), MEXT, Japan (S.T.P.).

\appendix 
\section{Impact of the $\boldsymbol{\sin^2\theta_{ij}}$ Uncertainties in Cases A1 and A2}
\label{app:A1A2}

\begin{figure}[h]
\centering
\includegraphics[width=0.49\textwidth]{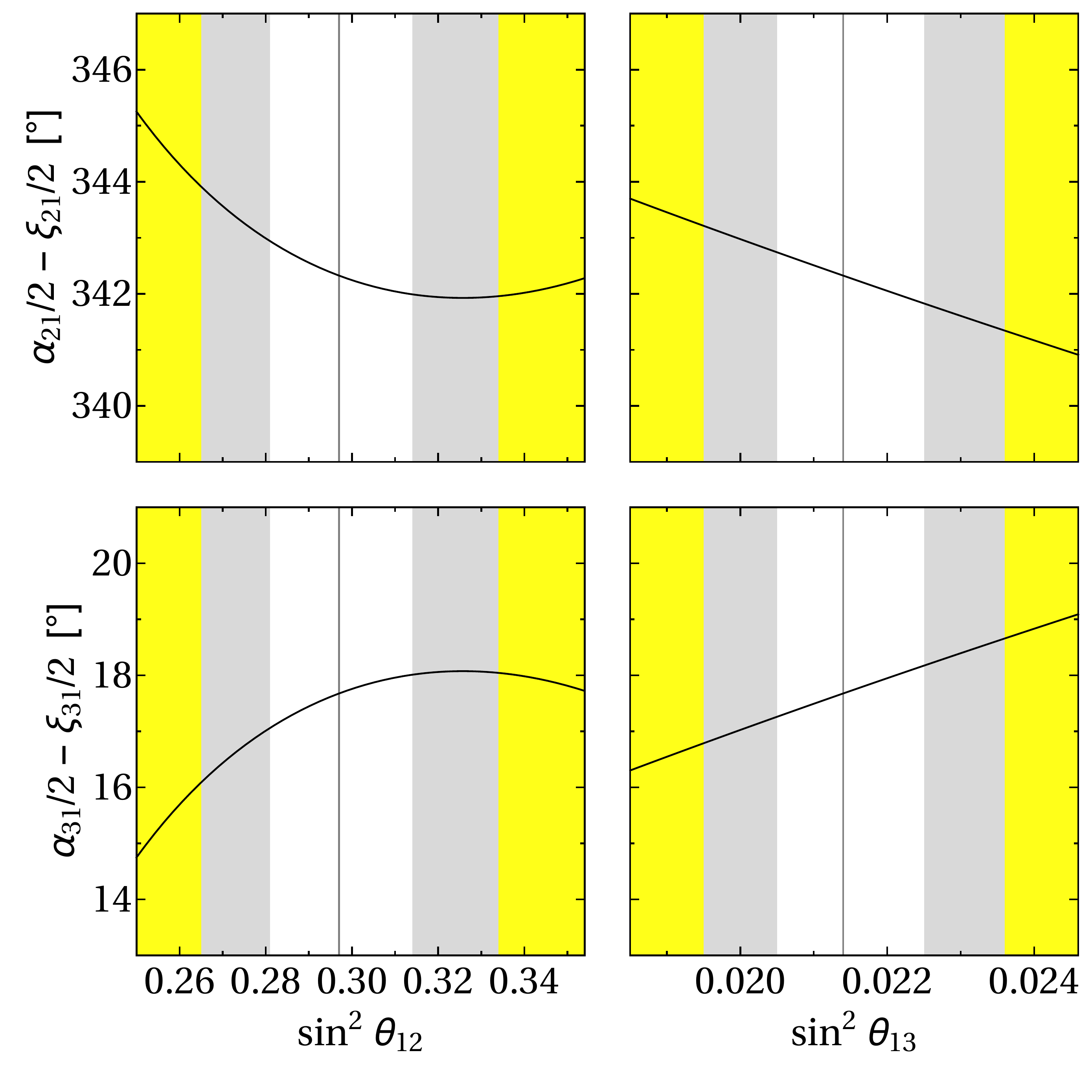}
\includegraphics[width=0.49\textwidth]{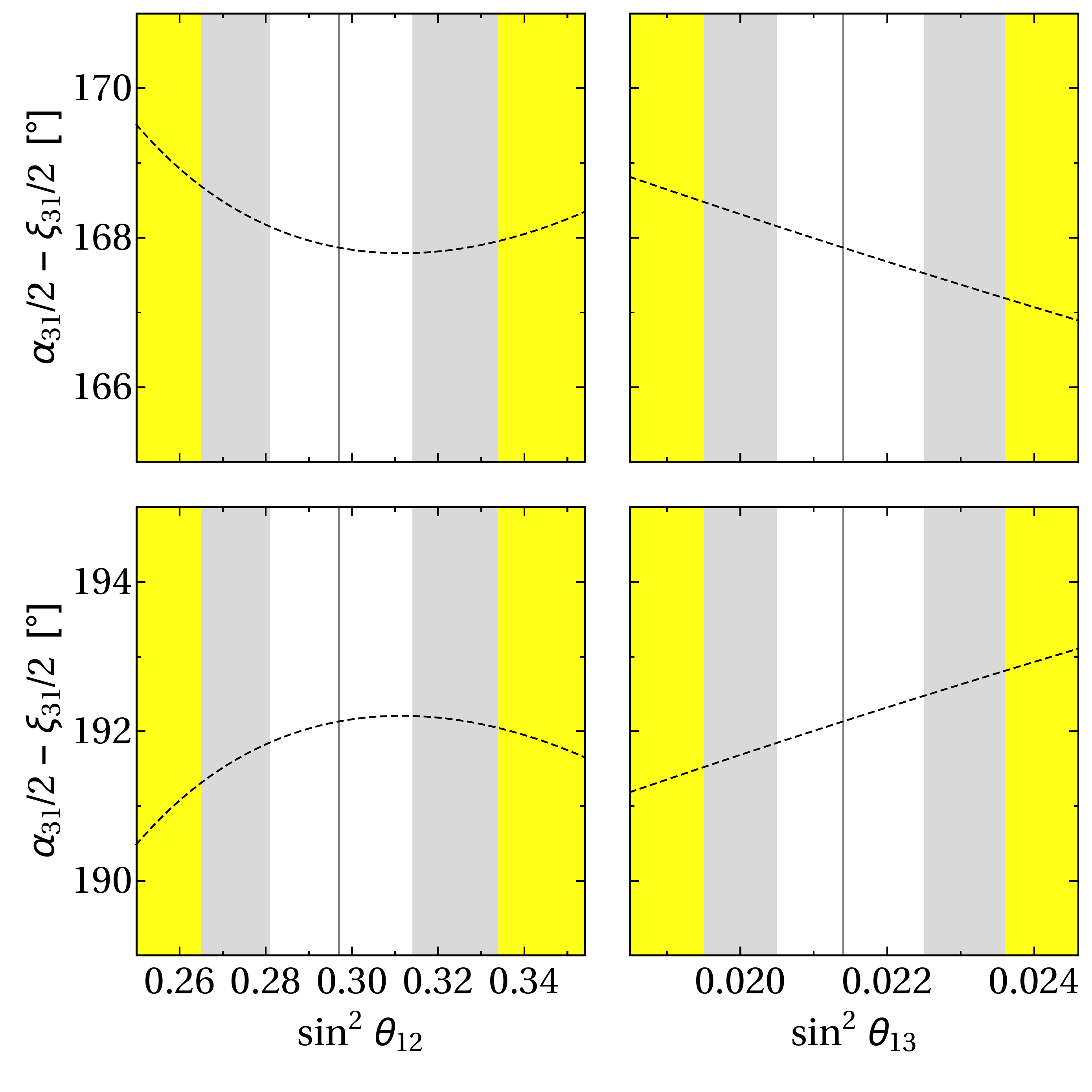}
\caption{\label{fig:MPs3sigmaA1}The phase differences 
$(\alpha_{21(31)}/2 - \xi_{21(31)}/2)$   
as functions of $\sin^2\theta_{12(13)}$
in case A1 and for the TBM form of the matrix $\tilde U_\nu$, 
fixing $\sin^2\theta_{13(12)}$
to its best fit value for the NO spectrum.
The upper panels correspond to $\delta = \cos^{-1}(\cos\delta)$, 
while the lower panels correspond to $\delta = 2\pi - \cos^{-1}(\cos\delta)$.
The vertical line and the three coloured vertical bands 
indicate the best fit value and the $1\sigma$, $2\sigma$ and $3\sigma$ 
allowed ranges of $\sin^2\theta_{12(13)}$.}
\end{figure}

\begin{figure}[t]
\centering
\includegraphics[width=0.49\textwidth]{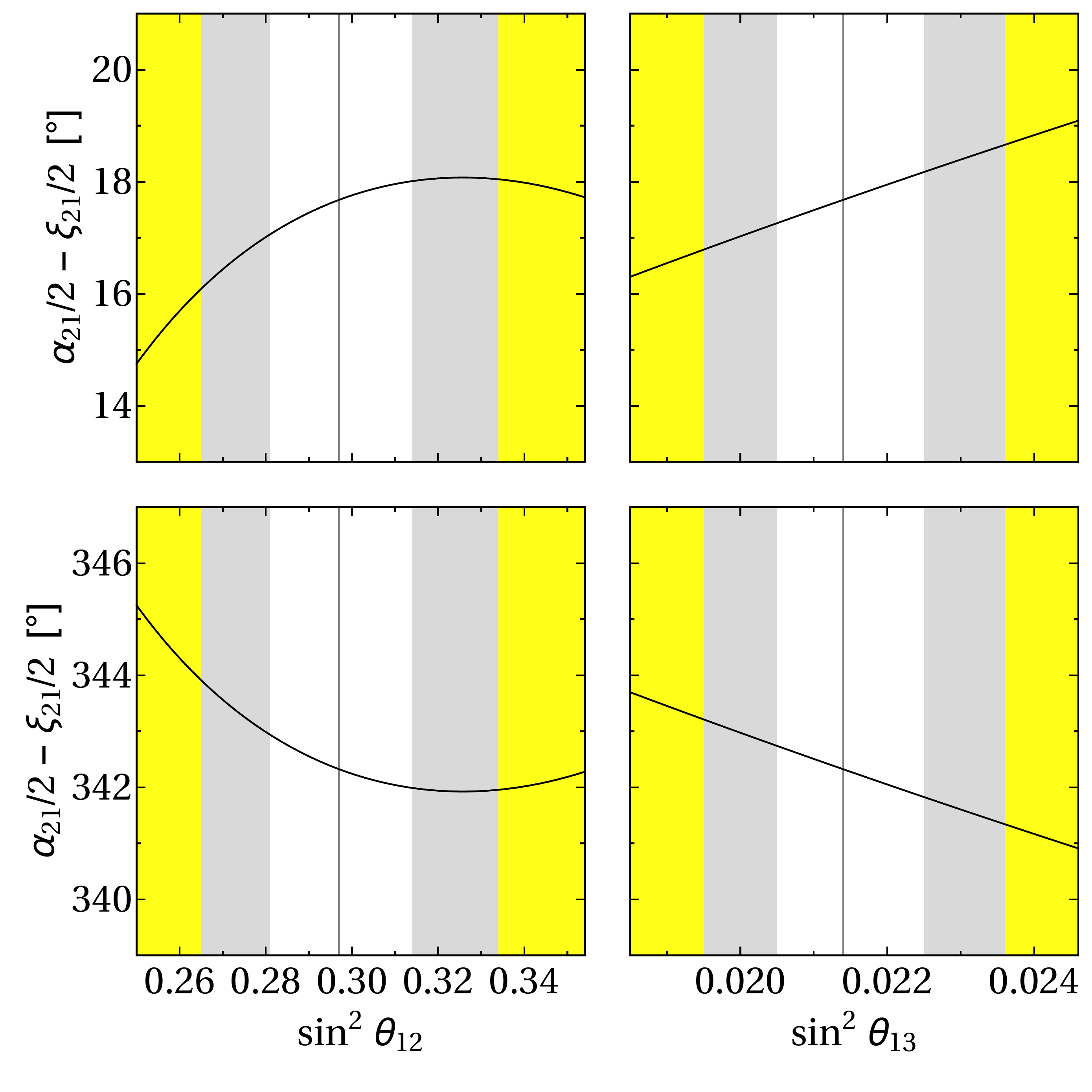}
\includegraphics[width=0.49\textwidth]{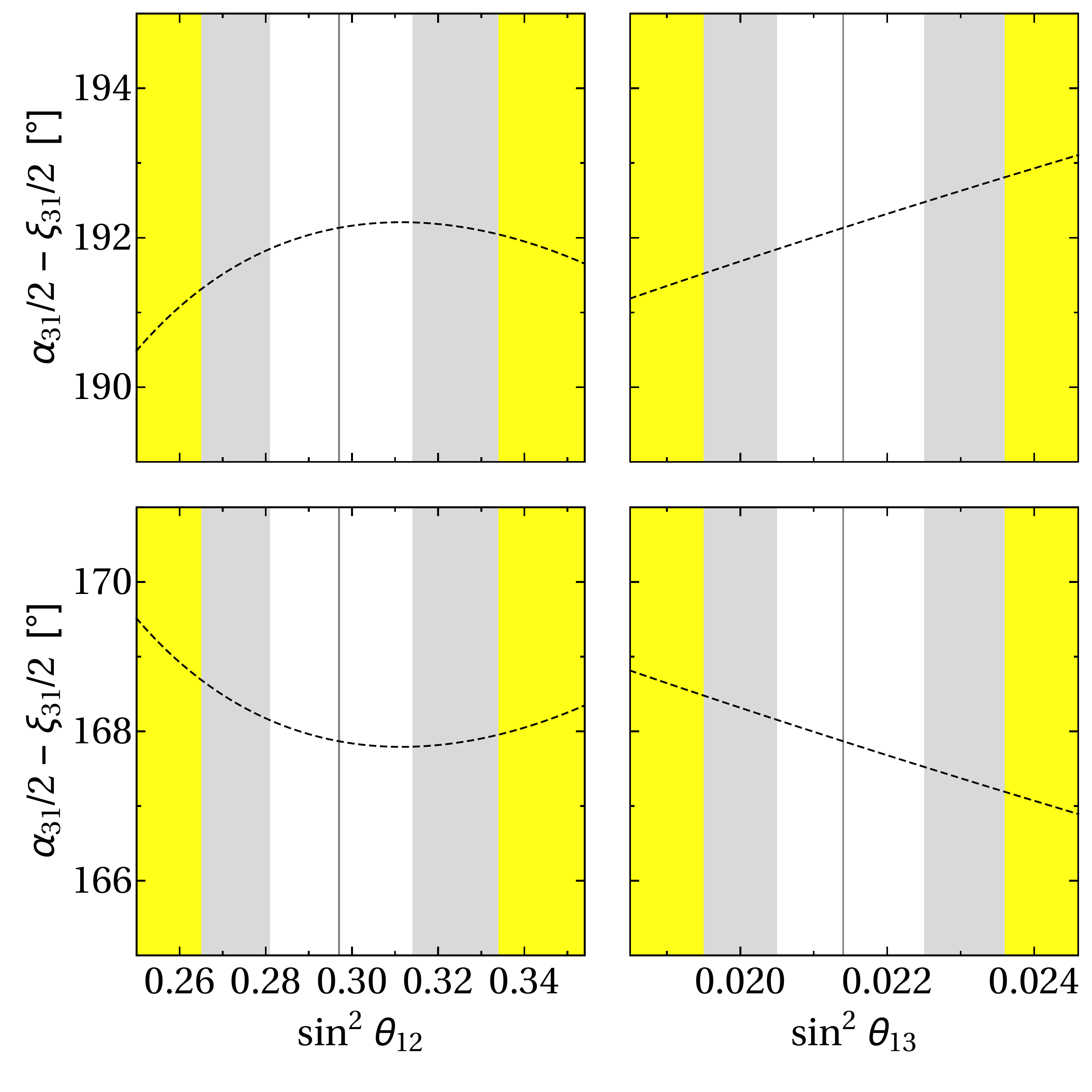}
\caption{\label{fig:MPs3sigmaA2}The same as in Fig.~\ref{fig:MPs3sigmaA1}, but for case A2.}
\end{figure}

 In this Appendix we illustrate the impact of the uncertainties 
in determination of the neutrino mixing parameters on 
the predictions for the phase differences 
$(\alpha_{21}/2 - \xi_{21}/2)$ and $(\alpha_{31}/2 - \xi_{31}/2)$ 
in cases A1 and A2 with the TBM symmetry form of the matrix $\tilde U_\nu$.
In Fig.~\ref{fig:MPs3sigmaA1} we show the dependence of 
$(\alpha_{21}/2 - \xi_{21}/2)$ and $(\alpha_{31}/2 - \xi_{31}/2)$ on 
$\sin^2\theta_{12}$ ($\sin^2\theta_{13}$) 
in case A1, fixing $\sin^2\theta_{13}$ ($\sin^2\theta_{12}$) to its best fit value 
for the NO spectrum. We recall that in this setup 
$\sin^2\theta_{23}$ is correlated with $\sin^2\theta_{13}$ 
by eq.~(\ref{eq:th23A0}) and, hence, is not a free parameter.
In Fig.~\ref{fig:MPs3sigmaA2} we present results for case A2. 
Also in this scheme $\sin^2\theta_{23}$ 
is correlated with $\sin^2\theta_{13}$ 
and is not a free parameter (see eq.~(\ref{eq:th23B0gen})). 
As can be seen from Figs.~\ref{fig:MPs3sigmaA1} and \ref{fig:MPs3sigmaA2}, 
in both cases A1 and A2 
the variation of $(\alpha_{21}/2 - \xi_{21}/2)$ is within $3^\circ$, 
while that of $(\alpha_{31}/2 - \xi_{31}/2)$ is within $2^\circ$. 
%
%
%
%
%
%


\end{document}